\documentclass[onecolumn]{mn2e}  
\usepackage{mnras_cite}  
 \usepackage[dvips]{graphicx}  
\usepackage{amsmath}
\newcommand{\nc}{\newcommand}   

\nc{\citealt}{\cite}
\nc{\de}{\delta}
\nc{\tISW}{\triangle_T^{ISW}}
\nc{\hn}{\hat{n}}
\nc{\bH}{\bar{H}}
\nc{\Ol}{\Om_{\Lambda}}

\nc{\ul}{\underline} \nc{\al}{\alpha} \nc{\g}{\gamma}
\nc{\Del}{\Delta} \nc{\e}{\textrm{e}} \nc{\eps}{\epsilon}
\nc{\lam}{\lambda} \nc{\Om}{\Omega} \nc{\Omm}{\Omega_m}
\nc{\Oml}{\Omega_\Lambda} \nc{\LCDM}{$\Lambda$CDM~} 
\nc{\ve}{\varepsilon} \nc{\mn}{{\mu\nu}} \nc{\vp}{\varphi}

\def\gsim{\; \raise0.3ex\hbox{$>$\kern-0.75em
\raise-1.1ex\hbox{$\sim$}}\; }

\nc{\Section}[2]{\section{#2}\label{#1}}   
\nc{\Bibitem}[1]{\bibitem{#1}}   
\nc{\Label}[1]{\label{#1}}   

\nc{\beq}[1]{\begin{equation}\label{#1}}     
\nc{\eeq}{\end{equation}}

\nc{\hq}{\hat{q}}
\nc{\hw}{\widehat{w}}

\def\l{\ell}
\def\ben{\begin{enumerate}}
\def\een{\end{enumerate}}
\def\bi{\begin{itemize}}
\def\ei{\end{itemize}}
\def\ee{\end{equation}}
\def\bea{\begin{eqnarray}}
\def\eea{\end{eqnarray}}

\nc{\Mpc}{Mpc/h}   
\nc{\vev}[1]{\langle #1 \rangle}   
   
\def\etal{{et al. }}   
   
\def\ltsima{$\; \buildrel < \over \sim \;$}   
\def\gtsima{$\; \buildrel > \over \sim \;$}   
\def\simlt{\lower.5ex\hbox{\ltsima}}   
\def\simgt{\lower.5ex\hbox{\gtsima}}   
\nc{\w}{$w(\theta)$\ }   
\nc{\ie}{i.e.}    
\nc{\eg}{e.g.}   
\def\q{{\hat n} }   
   
\def\Cl{C_{\ell}}

\input epsf

\begin{document}

\title[Error analysis in cross-correlation of sky maps]
{Error analysis in cross-correlation of sky maps: \\
application to the ISW detection }

\author[Cabr\'e \etal]{Anna Cabr\'e, Pablo Fosalba, Enrique Gazta\~{n}aga \&
Marc Manera \\ 
Institut de Ci\`encies de l'Espai, IEEC-CSIC, Campus UAB,
F. de Ci\`encies, Torre C5 par-2,  Barcelona 08193, Spain}

\twocolumn   
\maketitle 

\begin{abstract}

Constraining cosmological parameters from measurements of the Integrated Sachs-Wolfe effect requires developing robust and accurate methods for computing statistical errors in the cross-correlation between maps. This paper presents a detailed comparison of such error estimation applied to the case of cross-correlation of Cosmic Microwave Background (CMB) and large-scale structure data. We compare theoretical models for error estimation with montecarlo simulations where both the galaxy and the CMB maps vary around a  fiducial auto-correlation and cross-correlation model which agrees well with the current concordance \LCDM cosmology. Our analysis compares estimators both in harmonic and configuration (or real) space, quantifies the accuracy of the error analysis and discuss the impact of partial sky survey area and the choice of input fiducial model on dark-energy constraints. We show that purely analytic approaches yield accurate errors even in surveys that cover only 10\% of the sky and that parameter constraints strongly depend on the fiducial model employed. Alternatively, we discuss the advantages and limitations of error estimators that can be directly applied to data.
In particular, we show that errors and covariances from the Jack-Knife method
agree well with the theoretical approaches and simulations.
We also introduce a novel method in real space that is computationally efficient and
can be applied to real data and realistic survey geometries.
Finally,  we present a number of new findings and prescriptions that can be useful
for analysis of real data and forecasts, and present a critical summary
of the analyses done to date.

\end{abstract}   
   

\section{Introduction}   

The ISW  effect  (Sachs \& Wolfe, 1967) has emerged as a new and powerful tool to probe
our universe on the largest physical scales, testing deviations from General Relativity and
the existence of dark-energy independent of other classical probes (e.g, Crittenden and Turok 1996;
Bean and Dore 2004; Lue et al 2004; Cooray et al. 2004; Garriga et al. 2004; Song et al. 2006).
Recently, a number of groups have obtained the first detections of the ISW effect by
cross-correlating low redshift tracers of the large scale structure
(LSS) with the cosmic variance limited cosmic microwave background (CMB)
maps obtained by WMAP  (e.g, Boughn and Crittenden 2004;
Nolta et al. 2004; Fosalba and Gazta\~naga 2004; Fosalba et al. 2003; Scranton et al. 2003;
Afshordi et al 2004, Rassat et al 2006, Cabr\'e et al. 2006).
Although current detections are only claimed at the 2-4 $\sigma$ level,
all analyses coherently favor a flat \LCDM model that is consistent with WMAP observations
(Spergel et al. 2006). Moreover, the redshift evolution of the measured signal already provides
first constraints on alternative cosmological scenarios ( Corasaniti et al. 2005; Gazta\~naga et al. 2006).

However, sample variance from the primary CMB aniso-tropies limits the ability with which
one can detect CMB-LSS correlations. For the observationally favored flat \LCDM model,
even an optimal measurement of the cross-correlation could only achieve a signal-to-noise ratio
of $\sim 10$ (Crittenden and Turok 1996; Peiris and Spergel 2000; Afshordi 2004, see also
Fig.\ref{fig:s2npred} below).
Given the low significance level of ISW detections, a good understanding
of the systematic and statistical errors is crucial to optimally exploit CMB-LSS correlation
data that will be collected in future surveys such as PLANCK, DES, SPT, LSST, etc.,
for cosmological purposes
(see e.g., Pogosian et al. 2005). Recent work has focused on the impact of
known systematics on cross-correlation measurements (Boughn and Crittenden 2005; Afshordi 2004), however
no detailed analysis has been carried out to assess how different error estimates
compare or what is the accuracy delivered by each of them. 
So far, most of the published analyses have implemented one specific error estimator
(primarily in real or configuration space) without justifying the choice of that particular
estimator or quantifying its degree of accuracy.

In particular,
most of the groups that first claimed ISW detections
(Boughn and Crittenden 2004; Nolta et al. 2004; Fosalba and Gazta\~naga 2004;
Fosalba et al. 2003; Scranton et al. 2003, Rassat et a 2006) 
estimated errors from CMB Gaussian montecarlo
(MC) simulations  alone. In this approach statistical errors are obtained from the dispersion
of the cross-correlation between the CMB sky realizations with
a (single) fixed observed map tracing the nearby large-scale structure. This estimator is expected to
be reasonably accurate as long as the cross-correlation signal is weak and
the CMB autocorrelation dominates the total variance of the estimator.
We shall call this error estimator MC1 (see below).

Fosalba, Gazta\~naga \& Castander (2003), Fosalba \& Gazta\~naga (2004), also used
Jack-knife (JK) errors. They found that the JK errors perform
well as compared to the MC1 estimator, but the JK error from the real data seems  up to a factor of two
smaller (on sub-degree scales) than the JK error estimated from simulations.
This discrepancy arise from the fact that the fiducial theoretical model
used in the MC1 simulations does not match the best fit to the data (see conclusions).

Afshordi et al (2004) criticize the MC1 and JK estimators and implement
a purely theoretical Gaussian estimator in harmonic space (which we shall
call TH bellow). However,  they did not show why their choice of estimator
should be more optimal or validate it with simulations.
This criticism to the JK approach has been spread in the literature
without any critical assessment. Vielva, Martinez-Gonzalez \& Tucci (2006)
also point out the apparent limitations  of the JK
method and adopted the MC1 simulations instead. However
they seem to find that the signal-to-noise of their measurement depends on the statistical method used.

Padmanabham etal (2005) use Fisher matrix approach and MC1-type simulations to validate
and calibrate their errors. They also claim that JK errors tend to underestimate errors
because of the small number of uncorrelated JK patches on the sky,
but they provide no proof of that.

Giannantonio et al (2006), use errors from MC simulations
that follow the method put forward by Boughn et al (1997).
In their work the error estimator is built from pairs 
of simulation maps (of the CMB and large-scale structure fields)
including the predicted auto and cross-correlation.
This is the estimator we shall name MC2 below.
They point out that their results are consistent with what is obtained from the
simpler MC1 estimator.

In this paper we develop a systematic approach to compare different error estimators in
cross-correlation analyses. Armed with this machinery, we address some of the open questions
that have been raised in previous work on the ISW effect detection:
how accurate are JK errors? are error estimators different due to the input theoretical models
or the data themselves? how many Montecarlo simulations should one use to get accurate results?
can we safely neglect the cross-correlation signal in the simulations?
do harmonic and real space methods yield compatible results?
What is the uncertainty associated to the different error estimates?

The methodology and results presented here should largely apply
to other cross-correlation analyses of different sky maps such as galaxy-galaxy or lensing-galaxy
studies.

This paper is organized as follows:
Section 2 presents the Montecarlo and the theoretical methods used to compute the errors
for the galaxy-temperature cross-correlation signal. Sections 3 \& 4 shows a comparison
between the normalized covariances and diagonal errors from different estimators.
The impact of the choice of error estimator on cosmological parameters
is discussed in Section 5. Finally, in Section 6 we summarize our main results and conclusions.

\section{Methods}   
\label{sec:methods}

 We consider four methods to estimate errors. The first one is based on
Montecarlo (MC) simulations of the pair of maps we want to correlate.
We consider two variants: MC2, where pairs are  correlated
with a given fiducial model  and MC1, where one map in each pair is fixed and no
cross-correlation signal is included. The next two methods rely on theoretical estimation.
We will use a popular harmonic space prediction, that we shall call
TH (Theory in Harmonic space).
We will also introduce a novel error estimator that is an analytic function
of the auto and cross-correlation of the fields in real space that we shall call  TC
(Theory in Configuration space). 
Finally, we will estimate Jack-Knife (JK) errors which uses sub-regions  of the actual
data map to calculate the dispersion in our estimator.

Once we have errors estimated in one space, it is also possible
to translate them, through Legendre transformation, into the complementary space.
We shall make a clear distinction between the method for the
error calculation (i.e., MC, TH, TC or JK), and the estimator onto which
the errors are propagated: i.e. either $w(\theta)$ or $\Cl$.
For example, $TH-w$ means  errors in $w(\theta)$ propagated from theoretical
errors originally computed in harmonic space. This notation is summarized in Table \ref{notation}.

\begin{table}{notation}
\caption{Notation used in this paper.}
\begin{tabular}{|c|c|}
\hline
TC & Theory in Configuration space \\
TH & Theory in Harmonic space \\
MC & Montecarlo simulations \\
MC2 & MC of the 2 fields, with correlation signal\\
MC1 & MC of 1 field alone (CMB), no correlation signal\\
JK & Jack-knife errors  \\
MC2-w & errors in $w_{TG}$ from MC2 simulations \\
MC1-w & errors in $w_{TG}$ from MC1 simulations \\
MC2-$\Cl$ & errors in $\Cl$ from MC2 simulations \\
TH-w &  errors in $w_{TG}$ from TH theory \\
TC-w &  errors in $w_{TG}$ from TC theory \\
TH-$\Cl$ & errors in $\Cl$ from TH theory \\
JK-w &  errors in $w_{TG}$ from JK simulations \\
\hline
\label{notation}
\end{tabular}
\end{table}

In all cases  (except for the JK) we are assuming Gaussian statistics.
In principle, it is also possible to do all this with non-Gaussian statistics
but this requires particular non-Gaussian models, which are currently not
well motivated by observations. Ultimately, our focus here is on the
comparison of different methods for a well defined set of reasonable assumptions.

\subsection{Montecarlo Simulations (MC)}

We have run 1000 Montecarlo (MC) simulation pairs of the CMB temperature anisotropy and
the dark-matter over-density field, including its cross-correlation, following
the approach presented in Boughn et al. 97 (see Eq.\ref{eq:mc} below).
These simulations are produced using
the {\it synfast} routine of the Healpix package\footnote{http://healpix.jpl.nasa.gov/}.
We assume that both fields are
Gaussian: this is a good approximation for the CMB field on the largest scales
(i.e few degrees on the sky), which are the relevant scales for the ISW effect.
However, the matter density field is weakly non-linear on these scales (eg see Bernardeau
etal 2002) and therefore non-Gaussian, i.e it has non-vanishing higher-order moments.
Therefore our simulations are realistic as long as non-Gaussianity does not
significantly alter the CMB-matter cross-correlation and its associated errors.
\footnote{As a check, we will compare below, in Fig.\ref{fig:varreal}, the results of the MC1
simulations, which have a fix galaxy map with Gaussian statistics, with the results
using the observed SDSS DR5, which is not Gaussian. We find no significant differences,
 indicating that the level of non-Gaussianity in observations does not influence much the error
 estimation.}
We take galaxies to be fair tracers of the underlying spatial distribution of
the matter density field on large-scales:  we assume that a simple linear bias
model relates both fields, so that  $w_{GG} = b^2 w_{MM}$ and $w_{TG} = b ~w_{TM}$.
Therefore, in what follows, we make no
difference between matter and galaxies in our analysis (other than $b$),
without loss of generality.

Decomposing our simulated fields on the sphere, we have
\beq{pablo1}
f(\hat{n}) = \sum_{\l m} a_{\l m} Y_{\l m}(\hat{n})
\eeq
where $a_{\ell m}$ are the amplitudes of the scalar field projected on the
spherical harmonic basis $Y_{\l m}$. In our simulations, the $a_{\l m}$'s
are given by linear combinations of unit variance random Gaussian fields $\psi$ ,
\bea
a^T_{\l m} &=& \sqrt{C^{TT}_{\l}} \, \psi_{1,\l m} , \label{eq:mc} \\ \nonumber
a^G_{\l m} &=& \frac{C^{TG}_{\l}}{\sqrt{C^{TT}_{\l}}} \, \psi_{1,\l m} +
\left(C^{GG}_{\l} - \frac{[C^{TG}_{\l}]^2}{C^{TT}_{\l}}\right)^{1/2} \, \psi_{2,\l m}
\eea
for the CMB temperature (T) and galaxy over-density (G) fields, respectively,
and $C^{XY}_{\l}$ is the (cross) angular power spectrum of the $X$ and $Y$ fields.

Our simulations assume a model that is broadly speaking in agreement with
current observations, although the precise choice of parameter values is not critical
for the purpose of this paper. We assume adiabatic initial conditions and a
spatially flat FRW model with the following fiducial cosmological parameters:
$\Omega_{DE} = 0.7$, $\Omega_B = 0.05$,  $\Omega_{\nu}=0$,
$n = 1$, $h = 0.7$, $\sigma_8 = 0.9$.
Although we will base most of our analyses on this fiducial model, we
have also run a set of 1000 MC simulations for a more strongly dark-energy dominated 
\LCDM model with $\Omega_{DE}=0.8$ (other parameters remain as in our fiducial model).
This will allow us to test how robust are our main results to changes around our
fiducial model.

Galaxies are distributed in redshift according to an analytic selection function,
\beq{dNdz}
\frac{dN}{dz} = \frac{3}{2} \frac{z^2}{z_0^3} e^{-(z/z_0)^{3/2}}
\eeq
where $z_m = 1.412\,z_0 $ is the median redshift of the source distribution, and by definition,
$\int dN/dz = 1$. Note that for such selection function, one can show that its
width simply scales with its median value,
$\sigma_z \simeq z_m/2$. For convenience we shall take $z_m = 0.33$ as our fiducial model.
For all the sky we have set the monopole ($\l=0$) and
dipole ($\l=1$) contribution to zero in order to be consistent
with the WMAP data.

We have run simulations for surveys covering different areas, ranging from
an all-sky survey ($f_{sky}=1$) to a survey that covers only 10\% of the sky ($f_{sky}=0.1$).
The latter is realized by
intersecting a cone with an opening angle of $37^{\circ}$ from the north pole with the sphere.
Larger survey areas are obtained by taking larger opening angles.
For $f_{sky}=0.1$ we have done the same analysis taking a compact square in the equator
(with galactic coordinates $l=0^{o}$ to $l=66^{o}$ and $b=-33^{o}$ to $b=33^{o}$)
and have found similar results.
We note that the wide $f_{sky}=0.1$ survey is comparable in area and depth to the distribution of main
sample galaxies in the SDSS DR2-DR3.

\subsubsection{Clustering in the simulations}

\begin{figure}
	\centering{\epsfysize=8cm \epsfbox{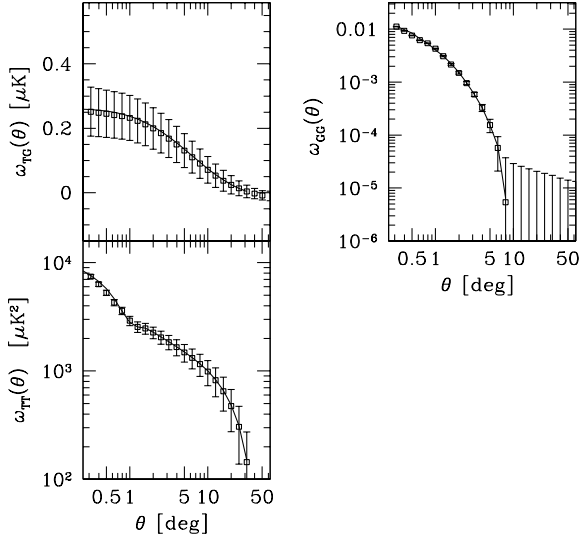}}
        \centering{\epsfysize=8cm \epsfbox{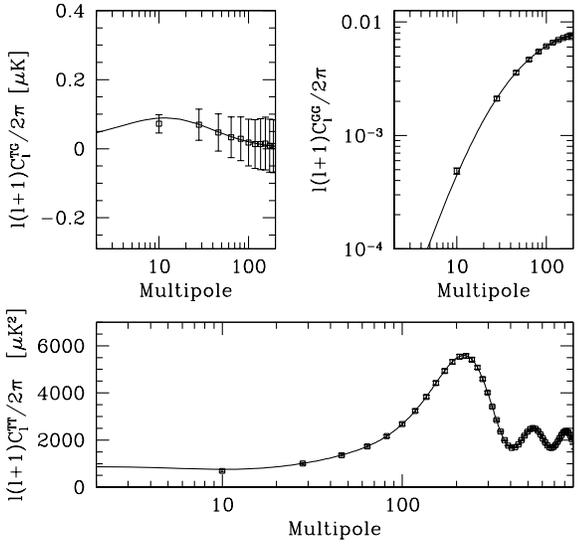}}
	\caption{2-point correlation function (top panels) and angular power spectra (bottom panels)
for all-sky surveys with median depth $z_m=0.3$. Different panels correspond to TT
(Temperature-Temperature), GG (Galaxy-Galaxy)
and TG (Temperature-Galaxy) cross-correlation. Errors shown correspond to dispersion over
Montecarlo MC2-type simulations binned with $\Delta_{\l}=18$ (see text for details)}
	\label{fig:wsim1}
\end{figure}

\begin{figure}
	\centering{\epsfysize=8cm \epsfbox{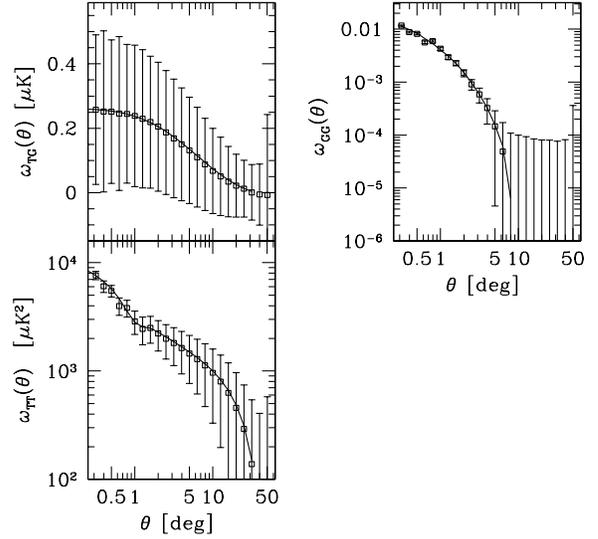}}
        \centering{\epsfysize=8cm \epsfbox{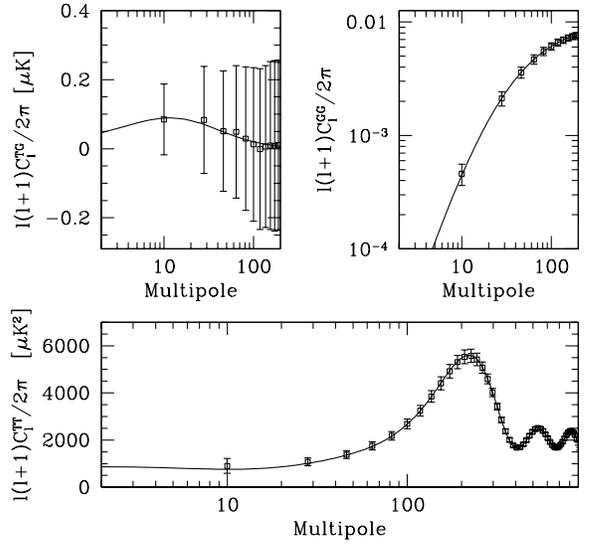}}
	\caption{Same as Fig. \ref{fig:wsim1} but for a wide field survey $f_{sky}=0.1$}
	\label{fig:wsim6}
\end{figure}

We have computed the angular 2-point correlation function for the galaxy over-density $w_{GG}$,
the temperature $w_{TT}$ and their cross-correlation $w_{TG}$, as well as their (inverse)
Legendre transforms, i.e, the angular power spectra,
\bea
w(\theta) &=& \sum_{\l} \frac{2\l+1}{4\pi} C_{\l} P_{\l}(cos\theta) \\
C_{\l} &=& 2\pi \int_{-1}^{1} dcos\theta \, w(\theta)\, P_{\l}(cos\theta)
\eea
where we denote by $P_{\l}$ the Legendre polynomial of order ${\l}$. \

In real space, we define the cross-correlation function as the expectation value of galaxy number
density $\delta_G$  and temperature $\Delta_T$ fluctuations:
\bea
\delta_G &=& {{N_G}\over{<N_G>}}-1 \\
\Delta_T &=& T-T_0 ~~(in ~~\mu K)   
\eea
at two positions $\q_1$ and $\q_2$ in the sky:
\beq{eqn:wtg}  
w_{TG}(\theta) \equiv  \vev{ \Delta_T({\bf\q_1}) \delta_G({\bf\q_2}) },
\eeq   
where $\theta = |\bf{\q_2}-\bf{\q_1}|$, assuming that   
the distribution is statistically isotropic.   
To estimate $w_{TG}(\theta)$ from the pixel maps we use:   
\beq{eqn:ctg}   
w_{TG}(\theta) = {\sum_{i,j} \Delta_T({\bf\q_i})~    
\delta_G({\bf\q_j})\over Npairs},   
\ee   
where the sum extends to all pairs $i,j$ separated by    
$\theta \pm \Delta\theta$.    
Survey mask and pixel window function effects have been appropriately taken into account using
SpICE (Szapudi et al. 2001a,b). This code has been probed to yield correct results not only on simulations
but also on real data from surveys with partial sky coverage and complex survey geometries
(Fosalba and Szapudi 2004).
In Fig.\ref{fig:wsim1} we show results from the all-sky MC2 simulations, whereas Fig.\ref{fig:wsim6} 
displays the same for a survey covering 10\% of the sky alone ($f_{sky}=0.1$).
Errorbars are computed as the
rms dispersion over the MC2 simulations. For the $C_{\l}$'s, we use linear bins with $\Delta\l = 18$,
to get approximately uncorrelated errobars for $f_{sky}=0.1$ (see Fig.\ref{fig:covCl} and \S\ref{sec:cov}).
As shown in the plots, our all-sky simulations are unbiased with respect to the input fiducial
model (continuous lines):  the mean over 1000 simulations lies on
top of the theoretical (input model) curve.  For finite area surveys, sample variance makes
measurements on the largest scales (i.e, lower $\l$'s) fluctuate around the input theoretical model
\footnote{When we calculate the cross-correlation in a fraction of the
sky, there is a residual monopole in the galaxy and temperature maps,
which changes the normalization of both fluctuations. In a real survey
we are limited by the survey area covered by galaxies and we need to normalize
the fluctuations using the local  mean, which is in general different from the 
mean in all sky (because of sampling
variance). We find that the cross-correlation calculated with the local normalization
with $f_{sky}=0.1$ is about 10\% lower for our fiducial $\Lambda CDM$ model,
but the value can vary for others models and different $f_{sky}$.}.


\subsubsection{Convergence in simulations}
\label{sec:convergence}

\begin{figure}
	\centering
{\epsfysize=2.5cm \epsfbox{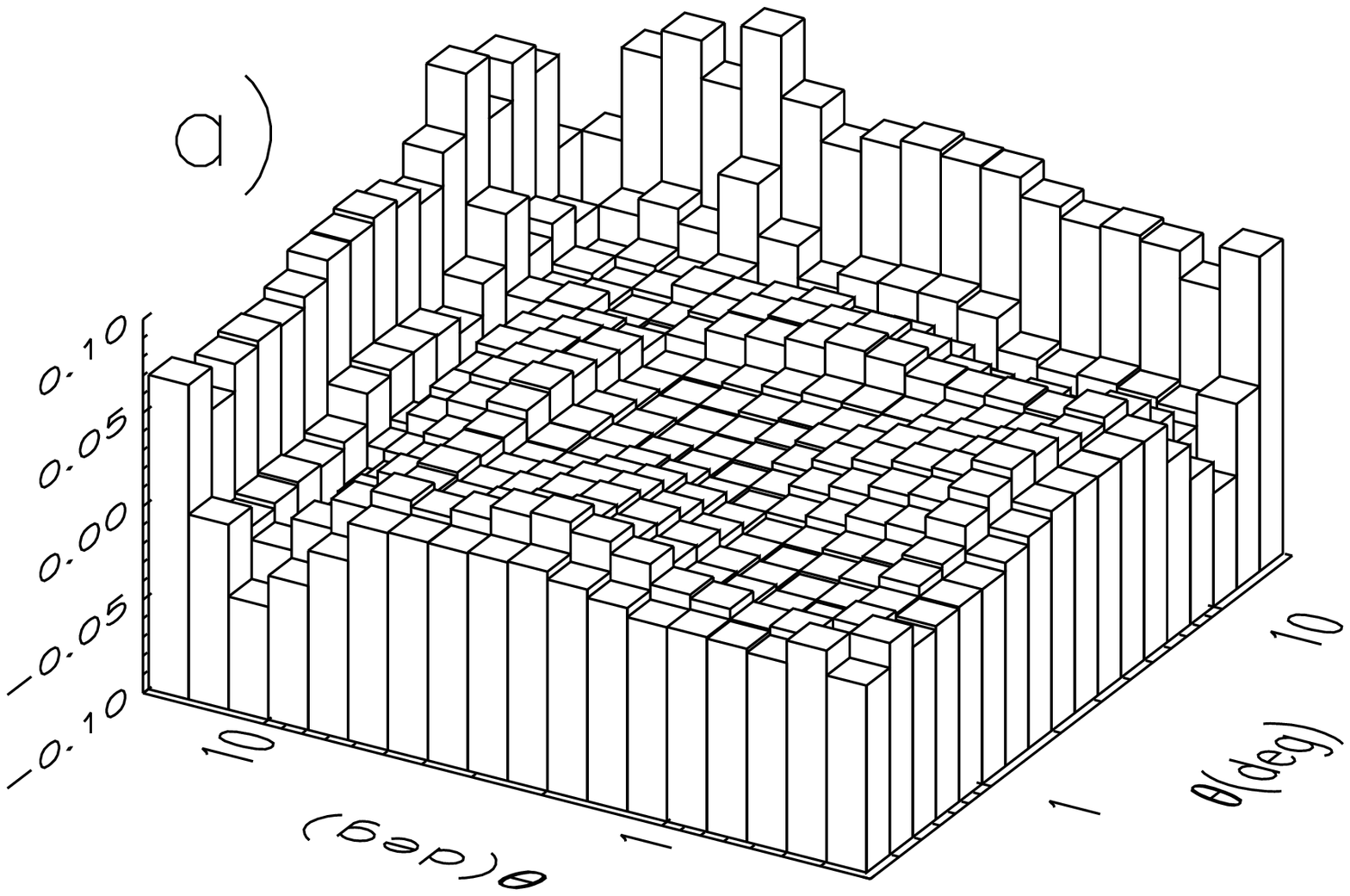}}
  \centering
{\epsfysize=2.5cm \epsfbox{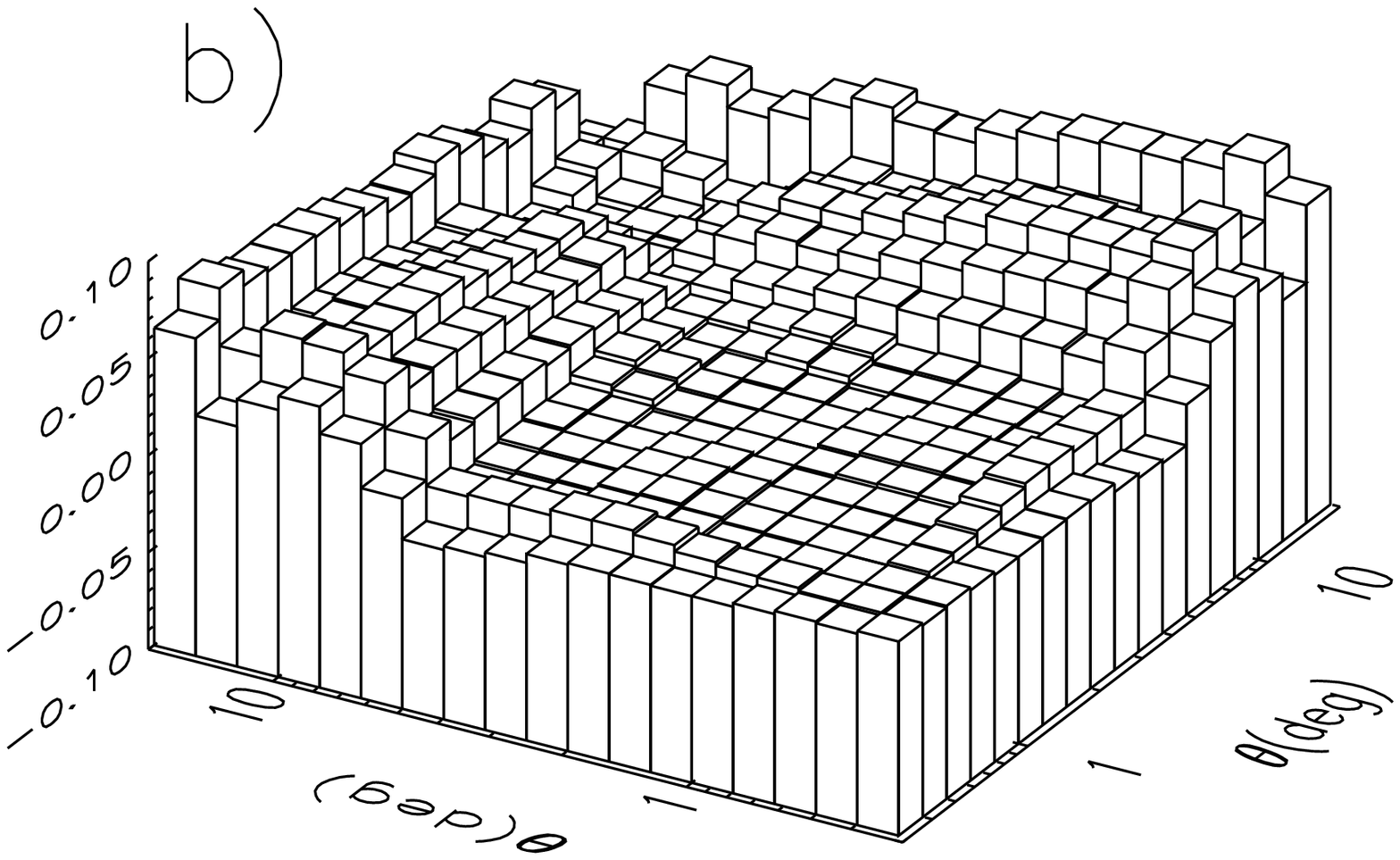}}
  \centering
{\epsfysize=2.5cm \epsfbox{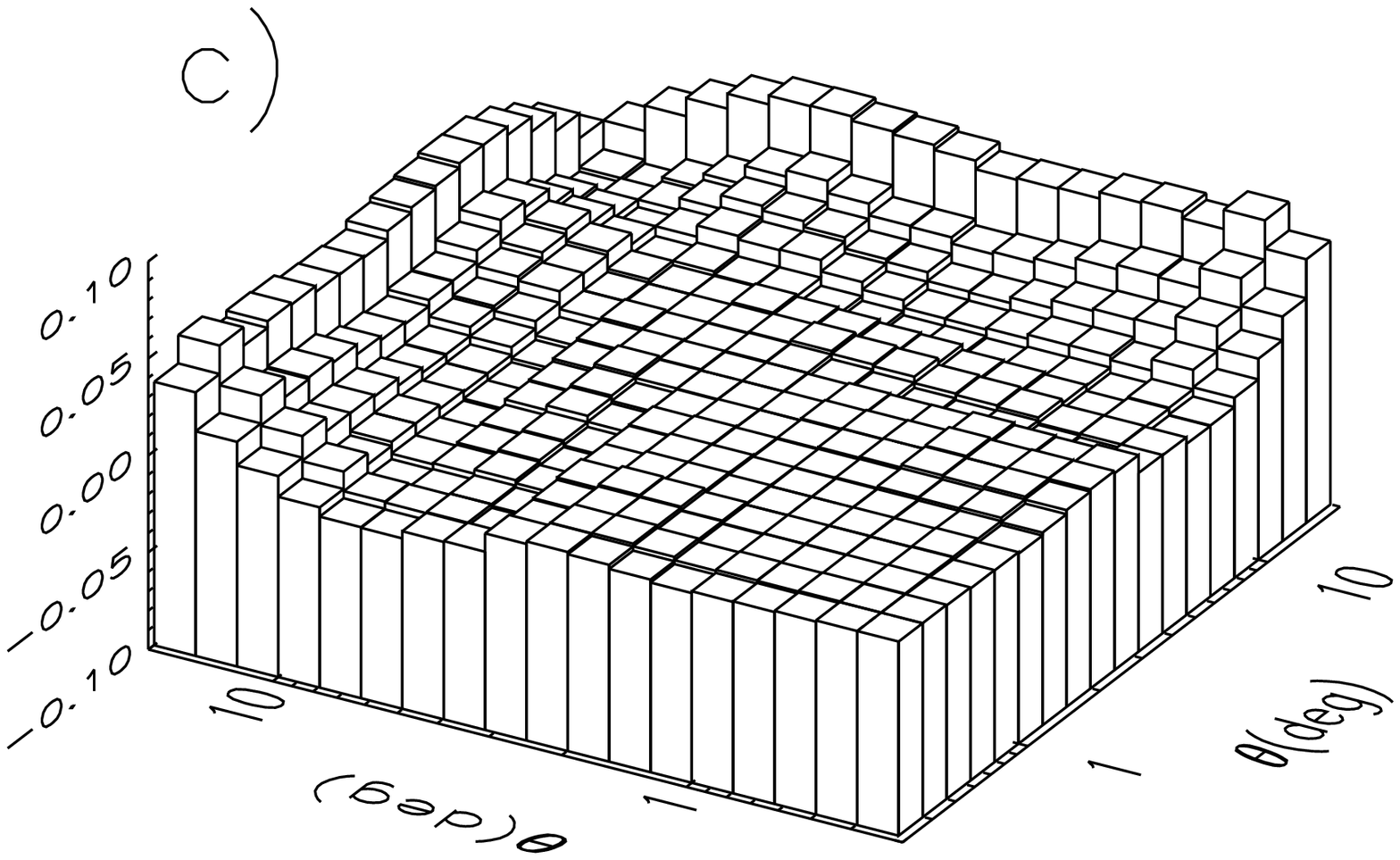}}
   \centering
{\epsfysize=2.5cm \epsfbox{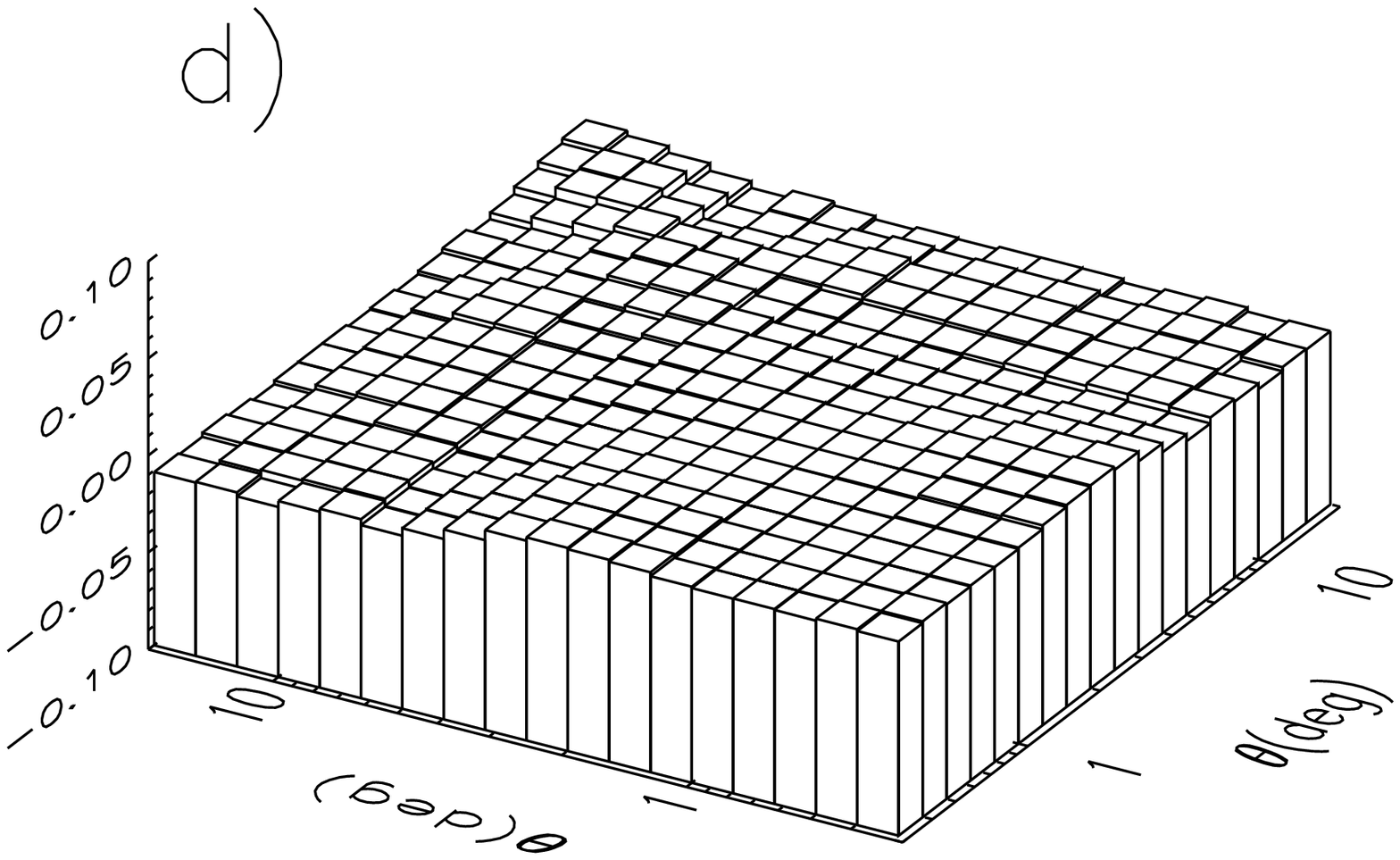}}
	\caption{Convergence in normalized covariance matrix for Monte-Carlo simulations (MC2). We 
compare  covariance when we increase the number of simulations. Here we show the difference between
 a) the first 100 with respect to the first 200 simulations (100-200),  
 b) 200-400, c)400-700, d)700-1000. Results show that one needs at least
700 simulations for the normalized covariance to converge (we use 1000 in our analysis).}
	\label{fig:convergence}
\end{figure}

The MC covariance is defined as:

\begin{equation}\label{eq:covMC}
C_{ij} =  \frac{1}{M} \sum_{k=1}^{M}{\Delta{w}}_{TG}^{k}(\theta_{i}){\Delta{w}}_{TG}^{k}(\theta_{j})
\end{equation}
\begin{equation}\label{eq:covMC2}
\Delta{w}_{TG}^{k}(\theta_{i})=w_{TG}^{k}(\theta_{i})-\widehat{w}_{TG}(\theta_{i})
\end{equation}
where $w_{TG}^{k}(\theta_{i})$ is the measure in the k-th simulation (k=1,...M) and $\widehat{w}_{TG}(\theta_{i})$ is the mean over M realizations. The case i=j gives the diagonal error (i.e, variance).

In order to check the numerical convergence in the computation of the covariance matrix,
we compare the results using
all 1000 (MC2) simulations with the ones using the first 100, 200, 400 or 700 simulations.
For clarity we separate our converge analysis into the diagonal elements (the variance)
and normalized covariance, where we divide the covariance by the diagonal
elements (see Eq.\ref{eq:covnorm}).
As shown in Fig.\ref{fig:convergence}, we find that there is no noticeable difference
($\simeq 5\%$ acuracy)
in the normalized covariance from 700 and 1000 simulations. This suggests that 700 simulations
are enough for our purposes. To be safe, we shall use all 1000 simulations to derive our main results.

\begin{figure}
	\centering
{\epsfysize=6cm \epsfbox{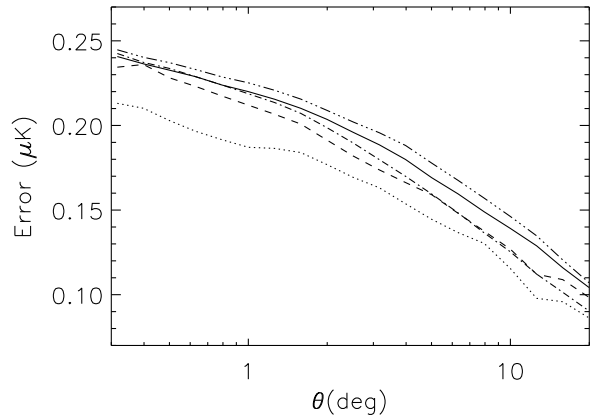}}
	\caption{Convergence in diagonal error for Monte-Carlo simulations. It is shown the error for the first 100 simulations (dotted), 200 (dashed), 400 (dash-dot), 700 (dash-dot-dot) and 999 (solid line).
 With 200 simulations the accuracy in errors is about  $20\%$.}
	\label{fig:errorconv}
\end{figure}

On the other hand, Fig.\ref{fig:errorconv}
shows the convergence on the variance estimation (diagonal elements of the covariance matrix)
for an increasing number of simulations. One needs about 200 simulations to converge
within $~20\%$ accuracy. This is similar  to the dispersion in the errors for a
given realization due to sampling variance, see Fig.\ref{fig:varrealw} below.
 We will use 1000 simulations which will give us better than $5\%$ accuracy in the error
 estimation from these  simulations.

\subsubsection{Simulations with a fixed galaxy map (MC1)}

We can also calculate montecarlo errors by cross-correlating 1000
simulations of CMB with a fixed sky for galaxies (MC1). This is
a common practice because it is quite easy to simulate CMB maps
and not so easy to simulate galaxy maps. In this case the common (and easiest) thing
to do is not to include any cross-correlation signal in the simulated CMB maps.
Thus, this  approach represents two levels of additional approximations: no variance coming from the
galaxy maps and no cross-correlation within the maps. Despite these approximations one
expects MC1 errors to be reasonably accurate because most of
the variance should come from the large scale primary CMB anisotropies, and the
cross-correlation signal is small in comparison.

Here we want to test in detail what is the accuracy of this approach.
We have taken the mean of 20 different cases. Each case has a different
fix galaxy map which is paired with 999 CMB maps, which are not correlated.
For each fixed galaxy case we obtain a MC1 error, so we can calculate the dispersion of
this error with the 20 different galaxy maps. This will be discussed in more
detail in \S\ref{sec:errinerr}.

\subsection{Jack-knife errors (JK)}

The JK method is closely related to the Boostrap method (Press etal 1992) 
which under certain circumstances can provide
accurate errors. The idea is that the data is grouped in sub-regions or zones which are more or less independent.\footnote{It is not adequate here to consider individual points or pixels as the units (sub-regions) to boostrap because they are highly correlated.}
We then use the fair sample hypothesis (ie ergodicity)
to estimate the error (variance between zones) for  the quantity under study.
 In the Boostrap methods one defines new sub-samples (which approach statistically independent realizations) by a random
selection of sub-regions. In the Jack-knife method each new sub-sample
contains all sub-regions but one.  A
potential disadvantage of the JK error is that one may think that it can not be used
on scales that are  comparable to the sub-regions size. This is not necessarily so.
Rare events (such as superclusters) can dominate sampling errors on all scales even
if they only extend over small regions (see Baugh et al . 2001). If JK sub-regions are large enough to
encompass these rare events, they can reproduce well errors on all scales. Nevertheless
it is clear that a danger with JK errors is that the result could in principle depend
on the size and shape of the sub-regions. So this needs to be tested in each
situation.

\begin{figure}
	\centering
{\epsfysize=4.5cm \epsfbox{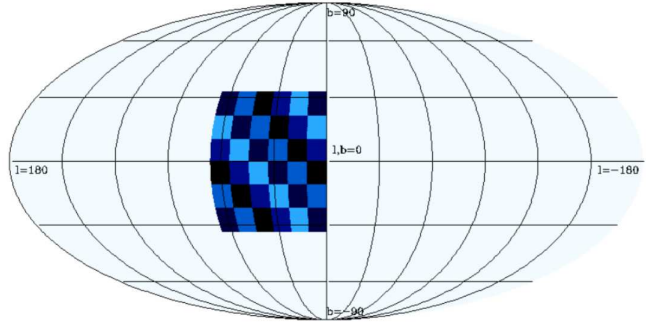}}
	\caption{Compact square with 36 zones, covering a 10/
used to calculate JK error in galactic coordinates
($l=0^{o}$ to $l=66^{o}$ and $b=-33^{o}$ to $b=33^{o}$).
We see that the shapes and sizes of the zones are similar.}
	\label{fig:JK10}
\end{figure}

We can therefore calculate the error from each single map  using the JK method.
To study the JK error in a fraction of the sky of 10\%, we divide a
compact square area in $M$ zones or sub-regions. Fig.\ref{fig:JK10} shows the case $M=36$,
but we
have tried different values for $M=20-80$, and find similar results.
The JK regions have roughly equal area and shape. This is important; we have found
that the JK method could give unrealistic errors when the areas or shapes are not
even. To calculate the covariance, we take  a JK sub-sample to be all the data
removing one of this JK zones, this means that we remove
all the pairs that fall completely or  partially in the JK zone that is removed.
To compensate for the correlation between the JK sub-samples,
we multiply the resulting covariance by $M-1$. The
covariance for this case is thus:
\begin{equation}\label{eq:covJK}
C_{ij} =  \frac{M-1}{M} \sum_{k=1}^{M}{\Delta{w}}_{TG}^k(\theta_{i}){\Delta{w}}_{TG}^k(\theta_{j})
\end{equation}
\begin{equation}\label{eq:covJK2}
\Delta{w}_{TG}^k(\theta_{i})=w_{TG}^{k}(\theta_{i})-\widehat{w}_{TG}(\theta_{i})
\end{equation}
where $w_{TG}^{k}(\theta_{i})$ is the measurement in the k-th sub-sample ($k=1,...M$)
and $\widehat{w}_{TG}(\theta_{i})$ is the mean for the $M$ sub-samples. For
each of the MC2 pair of simulated maps we have a JK estimation of $C_{ij}$. We can therefore
calculate a JK mean and its dispersion (and distribution) to
compare to the true MC2 covariance in the maps.

\subsection{Errors in harmonic space (TH)}
\label{sec:TH}

Theoretical expectations for the errors are the simplest in
Harmonic space where the covariance matrix is diagonal in the all-sky limit.
In particular, for Gaussian fields, one can easily see that the variance (or diagonal error) is,
\begin{equation}\label{eq:errorCl}
\Delta^{2} C^{TG}_{\l} = \frac{1}{f_{sky} (2\l+1)} \left[({C^{TG}_\l})^2 +
C^{TT}_{\l} C^{GG}_{\l} \right].
\end{equation}
This  indicates that the variance of the power spectrum estimator results
from quadratic combinations of the auto and cross power, with an amplitude that
depends on the number of independent $m$-modes available to estimate the power at a
scale ${\l}$,
which is approximately given by $f_{sky} (2\l+1)$.
We shall emphasize that this is only approximate and rigorously it is only
expected to yield accurate predictions for azimuthal sky cuts. However, as we shall see
later, this result is of more general applicability.
We note that the dominant contribution to the error and covariance comes from
the auto-power of the fields $C^{TT}_{\l} C^{GG}_{\l}$ involved
in the cross-correlation, whereas the cross-correlation signal $(C^{TG}_\l)^2$
only gives a few percent contribution, depending
on cosmology and survey selection function.

\begin{figure*}
	\centering{\epsfysize=4cm \epsfbox{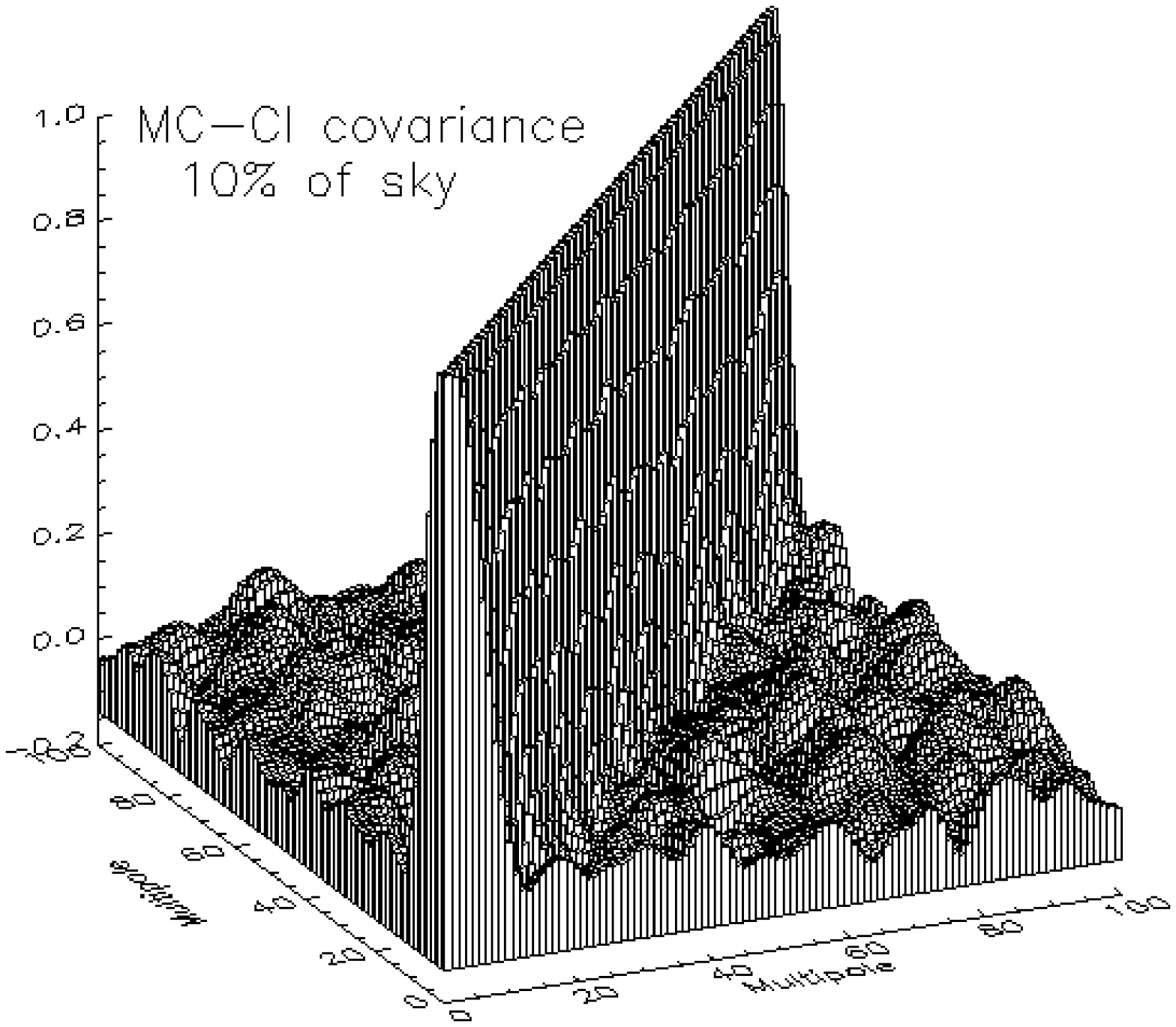}}
	\centering{\epsfysize=4cm \epsfbox{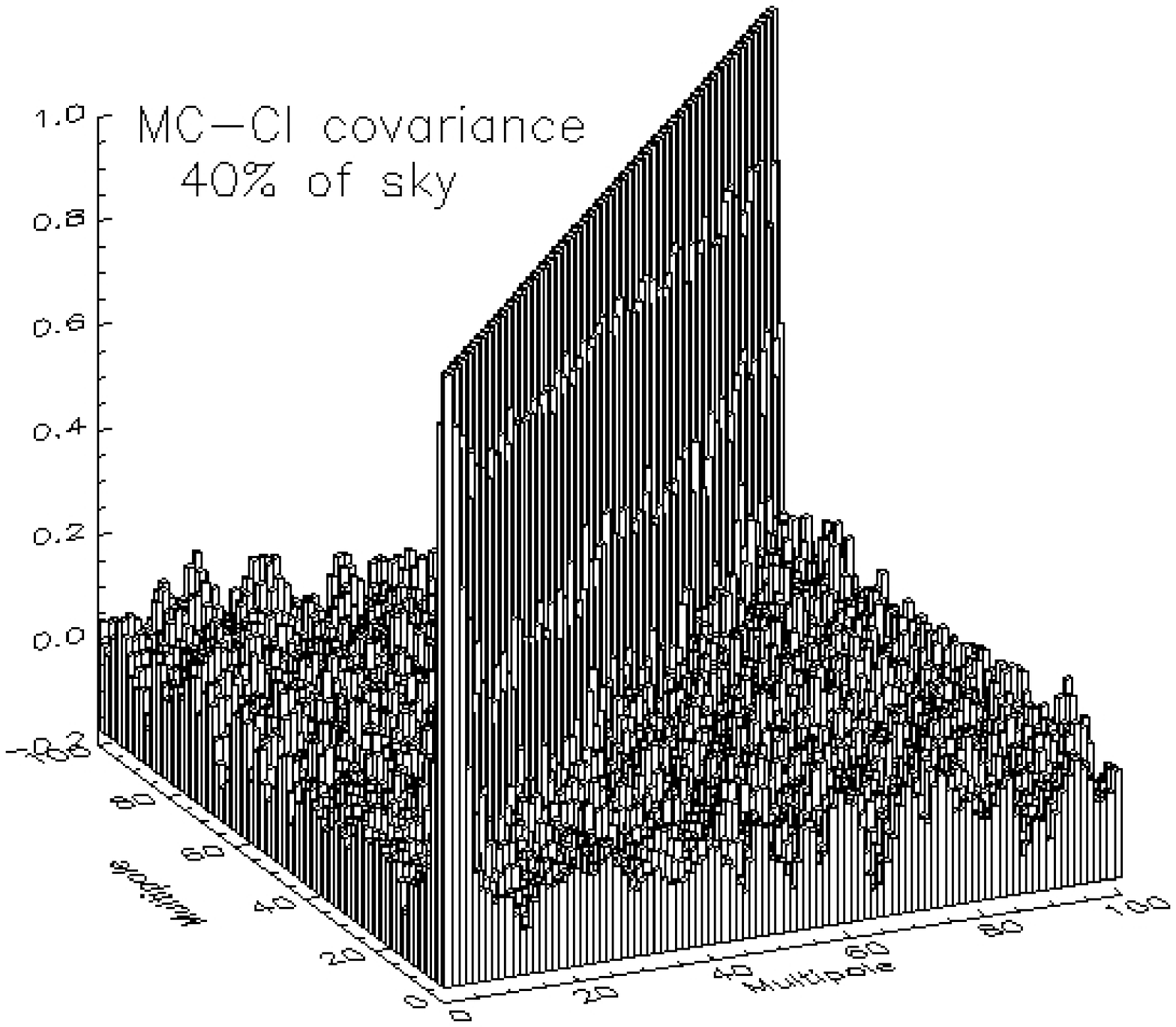}}	
	\centering{\epsfysize=4cm \epsfbox{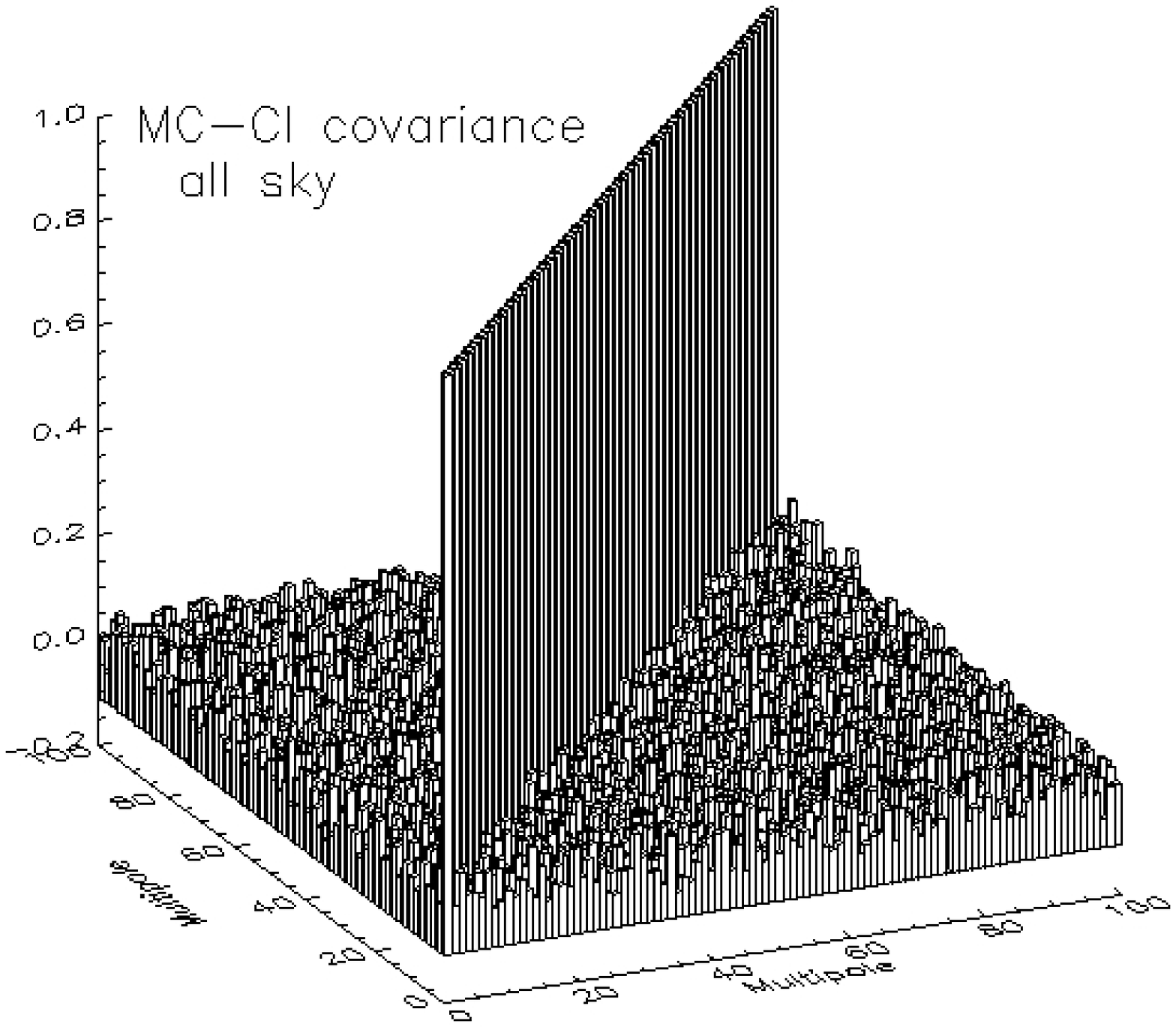}}	
\caption{Normalized covariances in $\Cl$ space from montecarlo (MC2) method.
The covariance becomes progressively dominated by its diagonal elements as we increase the sky fraction
$f_{sky}$.}
	\label{fig:covCl}
\end{figure*}

Partial sky coverage introduces a boundary
 which results in the coupling (or correlation) of
different ${\l}$ modes: the spherical harmonic basis in no longer orthonormal
on an imcomplete sky. Thus
the covariance matrix between different modes,
\begin{equation}\label{eq:covCl}
Cov(C_{\l},C_{\l'}) = <(C_{\l} - <C_{\l}>)(C_{\l'} - <C_{\l'}>)>
\end{equation}
is no longer diagonal (ie see Fig.\ref{fig:covCl}). Because of the partial sky
coverage there is less power on the smaller multipoles. This results in a systematic bias on the low multipoles  of $\Cl$ that can sometimes be modeled with the appropriate  window correction of the survey mask.

Using the Legendre transform one can propagate the error $\Delta C_{\l}$
in Eq.\ref{eq:errorCl} above  to configuration space,
\begin{equation}\label{eq:errorW}
\Delta^2 w(\theta)  = \sum_l \left(\frac{2\l+1}{4\pi}\right)^2
~P_{\l}^2(\mu) ~ \Delta^{2} C_{\l},
\end{equation}
where $\mu \equiv cos{\theta}$. For
the covariance matrix, we find:
\begin{eqnarray}\label{eq:covW}
C_{ij} &\equiv&  Cov(w(\theta_i), w(\theta_j))  \\
&  =& \sum_l  \left(\frac{2\l+1}{4\pi}\right)^2~
P_{\l}(\mu_i)~ P_{\l}(\mu_j) ~ \Delta^{2} C_{\l},
  \nonumber
\end{eqnarray}
where $\mu_i \equiv cos{\theta_i}$.
Eq(\ref{eq:covW}) and Eq(\ref{eq:errorW}) assumes that different ${\l}$ multipoles are
uncorrelated which is only strictly true
for all-sky surveys. We shall see below that this approximation is quite
accurate anyway even for
surveys that cover only $10 \%$ of the sky, i.e, cosmological parameter
contours derived from this expression
do not significantly differ from those computed with simulations that take into
account the exact covariance matrix.

\subsection{Errors in configuration space (TC)}

The cross-correlation function in configuration space is
estimated by averaging over all pairs of points separated an angle $\theta$
in the survey,
\begin{equation}\label{estimatorA}
w_{TG}(\theta)= <\Delta T(q)\delta_g(q') ~\arrowvert_{\hat{qq'}=\theta}>_{survey}.
\end{equation}

We have derived a formula for the covariance of the estimator in an
ensemble of sky realizations. Details of this derivation can be found
in Appendix A,
\begin{equation}
\label{result}
 C_{ij} = {1\over 8 \pi^2 P(\theta_i) P(\theta_j)}
\int_0^{\pi} \frac{ K[\theta_i,\theta_j,\psi]}{P(\psi)} \sin\psi d\psi
\end{equation}
where the kernel $K$ is given by:
\begin{eqnarray}
\label{kernelK}
& K[\theta,\theta',\psi]= {1\over 2}\left[ W_{TT}(\theta,\psi) W_{GG}(\theta',\psi)+ \right.  & \\  \nonumber
& \left.  W_{TT}(\theta',\psi) W_{GG}(\theta,\psi)\right]
+ W_{TG}(\theta,\psi) W_{TG}(\theta',\psi) &
\label{www}
\end{eqnarray}
and $W_X$ is a mean over the corresponding  correlation $w_X$, with
$X=TT, GG$ or $TG$:
\begin{equation}
\label{result2}
W_X(\theta,\psi) = 2 \int_0^{\pi} d\varphi P(\psi,\theta,\phi) ~w_X(\phi)
\end{equation}
where $cos\phi=cos\theta~ cos\psi+sin\theta ~sin\psi~ cos\varphi$. 
Survey geometry is encoded in $P(\theta)$ and  $P(\psi,\theta,\phi)$ probabilities. These are the
probabilities for two points separated by an angle $\theta$ or for a triangle of sides $\psi$,
$\theta$,$\phi$ to fall completely into the survey area if they are thrown randomly on
the full sky. For partial sky surveys these probabilities depend mainly on the survey
area and can be well approximated by the formula provided in Appendix A. Particularly simple
analytic expressions can be obtained for a ``polar cap'' survey (area obtained by intersecting a cone
with the sphere) and are given in the Appendix B.

This new method of computing errors in real space has several
advantages. Since it takes into account the survey geometry, it can provide
more accurate errors at large angles where both the jackknife errors and the
harmonic-space errors become more inaccurate. Compared to montecarlo errors this method is
faster because one does not need to generate a large number of sky realizations.
What is more, this estimator does not need to rely on any theoretical/fiducial model and
one can readily apply it to correlation functions measured on the real data to estimate the errors.
\footnote{A FORTRAN code (named TC-ERROR) which takes as input $w_{TG}$,$w_{TT}$,and $w_{GG}$
and compute the covariance matrix and errors for the cross-correlation function,
can be obtained upon request from the authors (please contact Marc Manera). 
Of course, this code can also be used to estimate the
autocorrelation error in a single map by just placing $w_{TG}=w_{TT}=0$.}

\begin{figure*}
	{\centering{\epsfysize=4cm \epsfbox{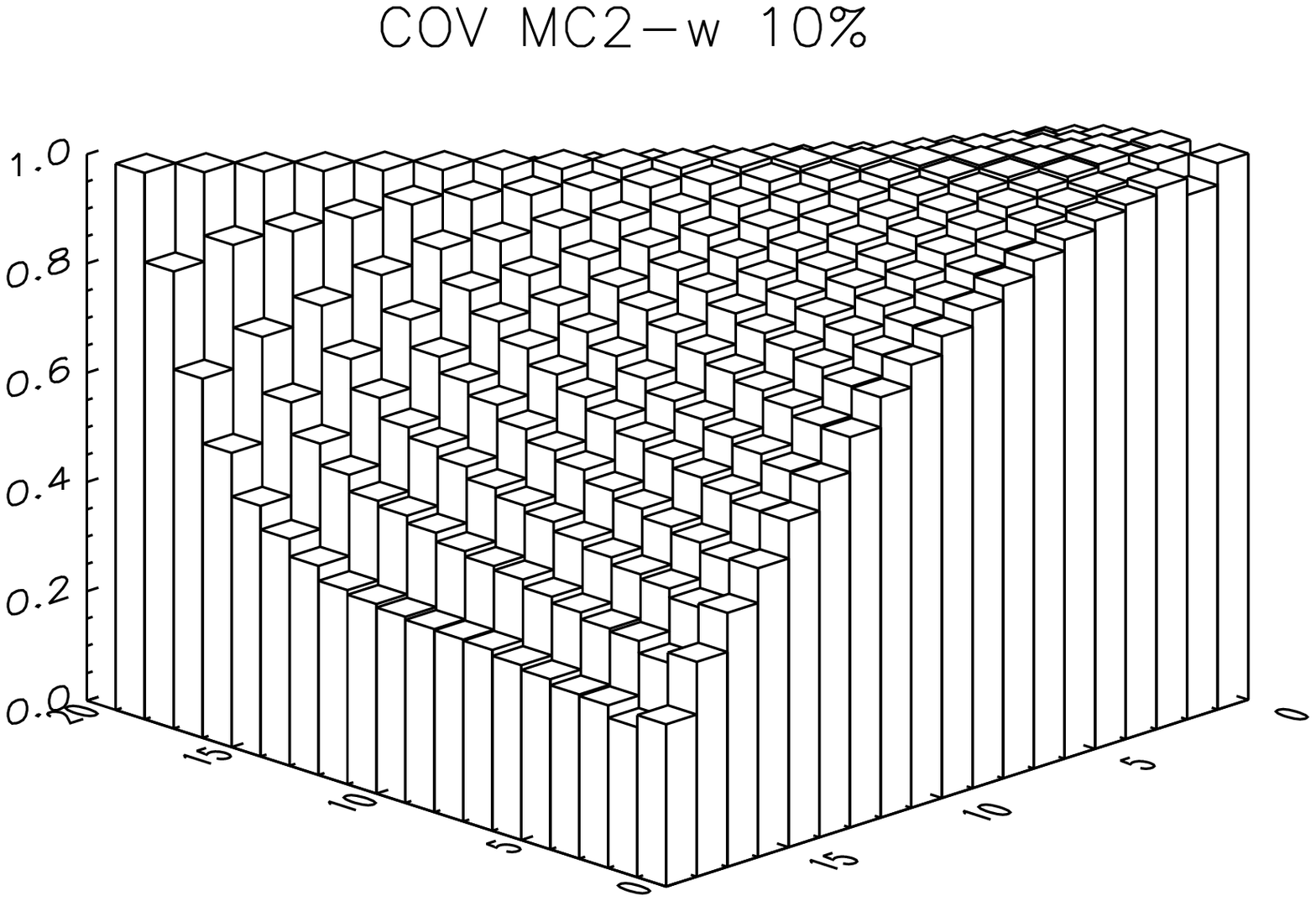}}
	\centering{\epsfysize=4cm \epsfbox{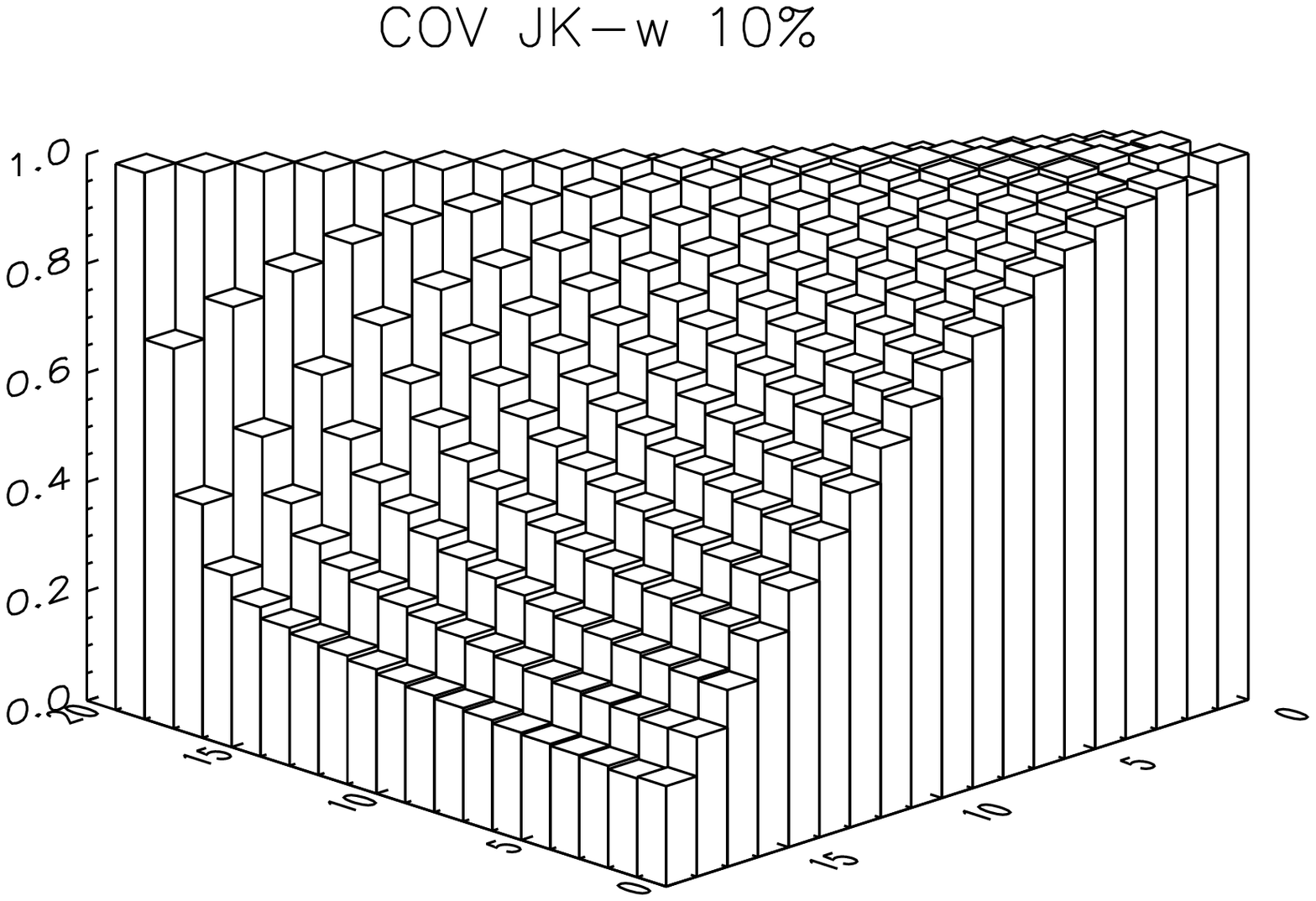}}	
	\centering{\epsfysize=4cm \epsfbox{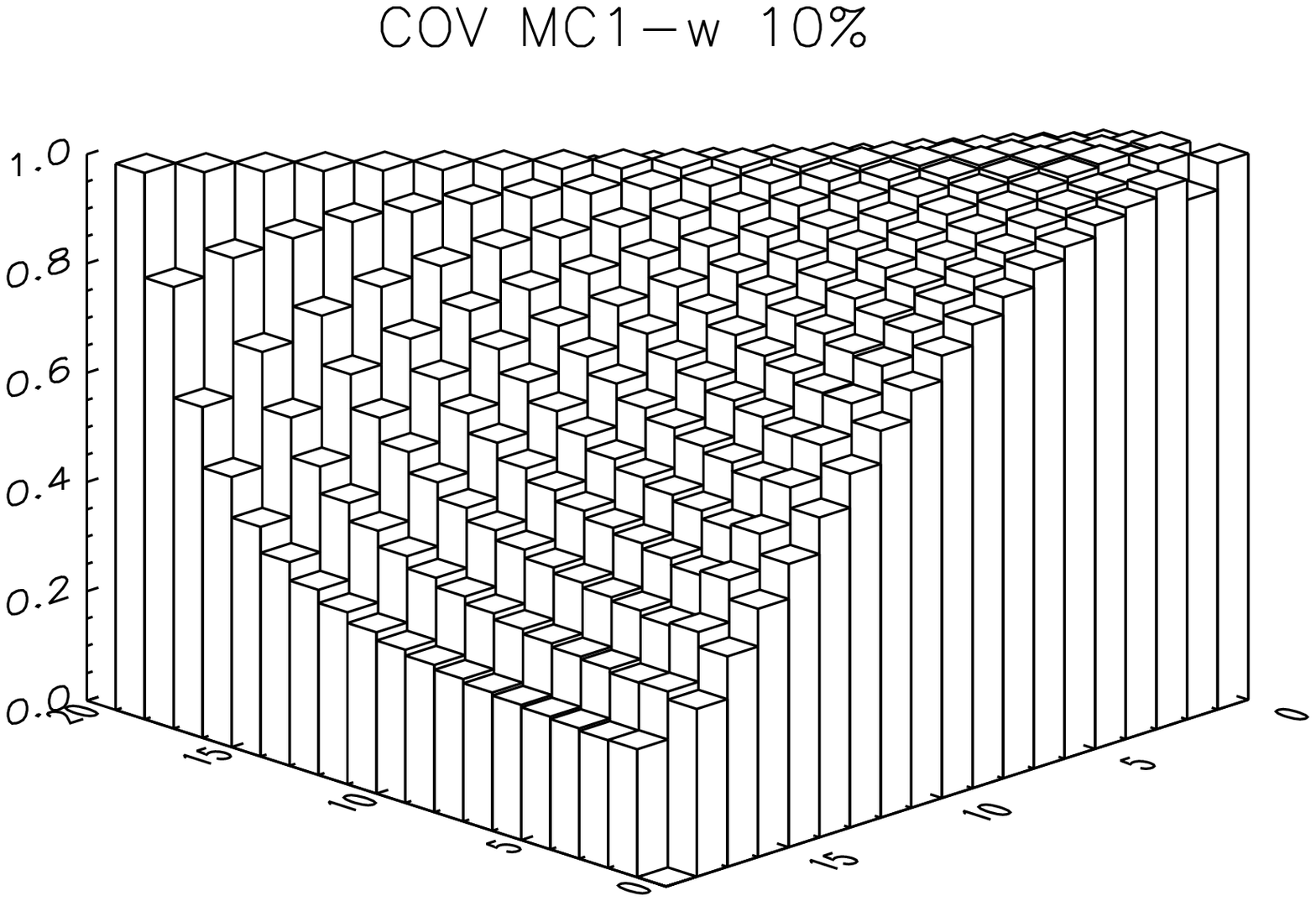}}	
	\centering{\epsfysize=4cm \epsfbox{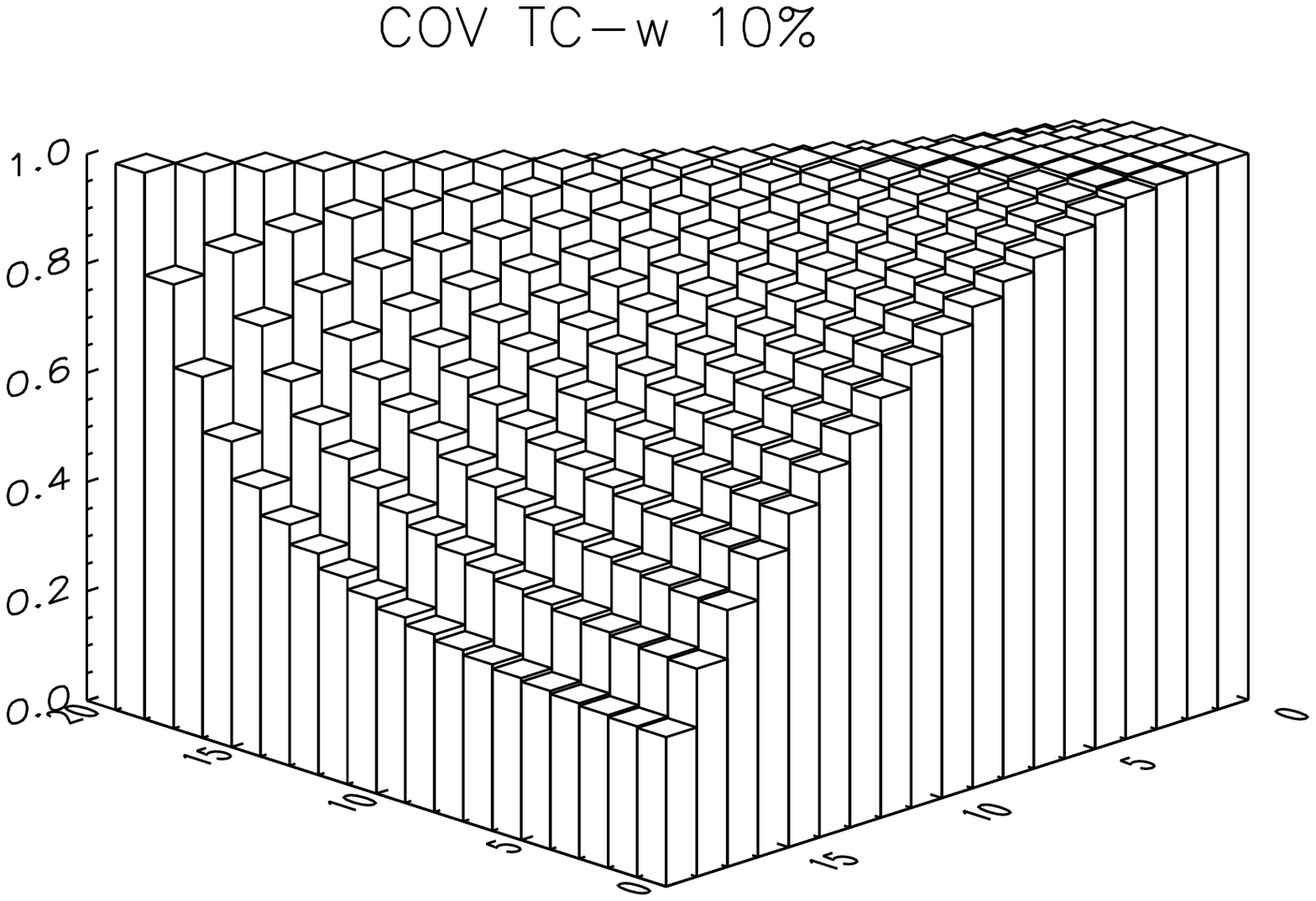}}	
	\centering{\epsfysize=4cm \epsfbox{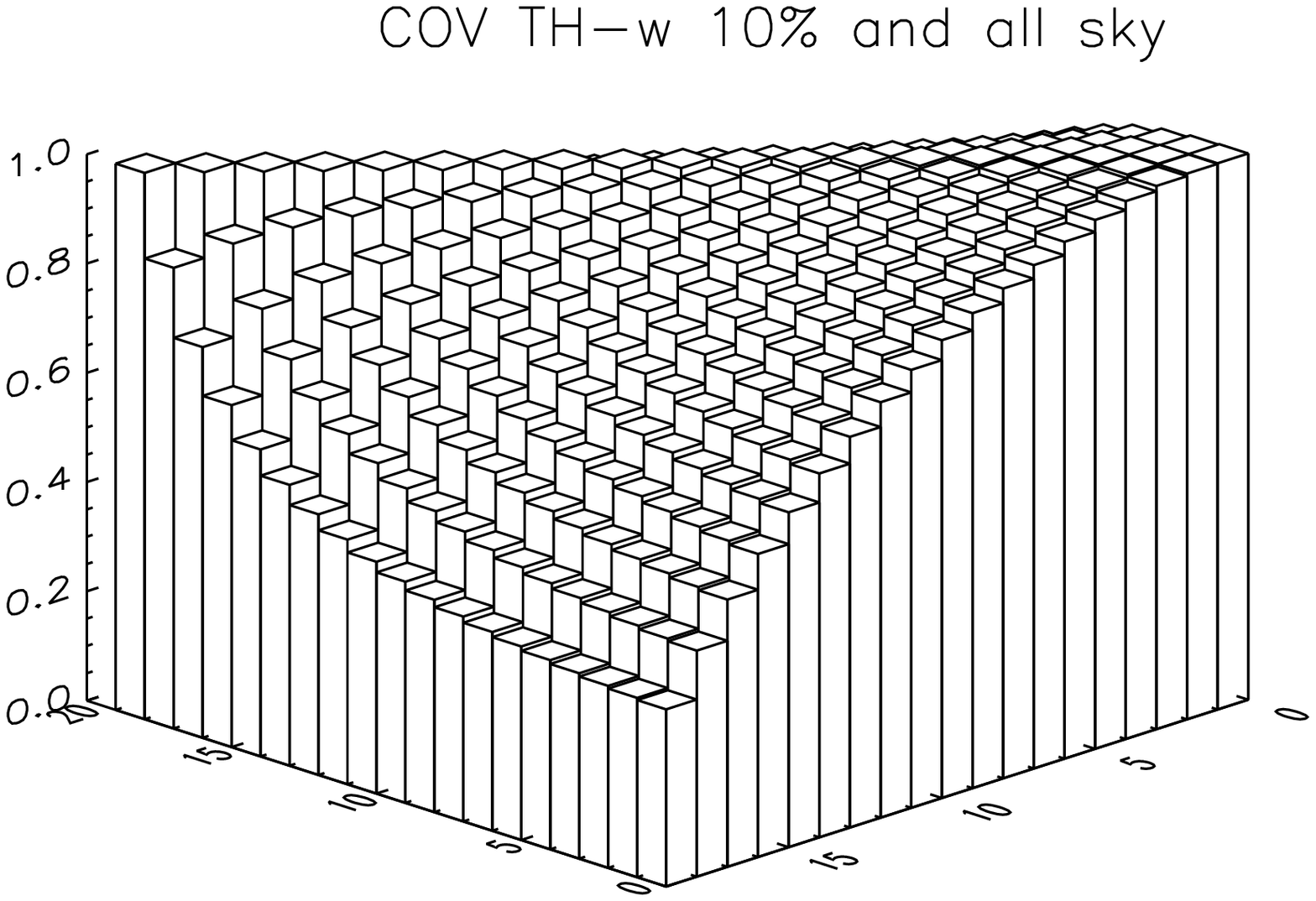}}
	\centering{\epsfysize=4cm \epsfbox{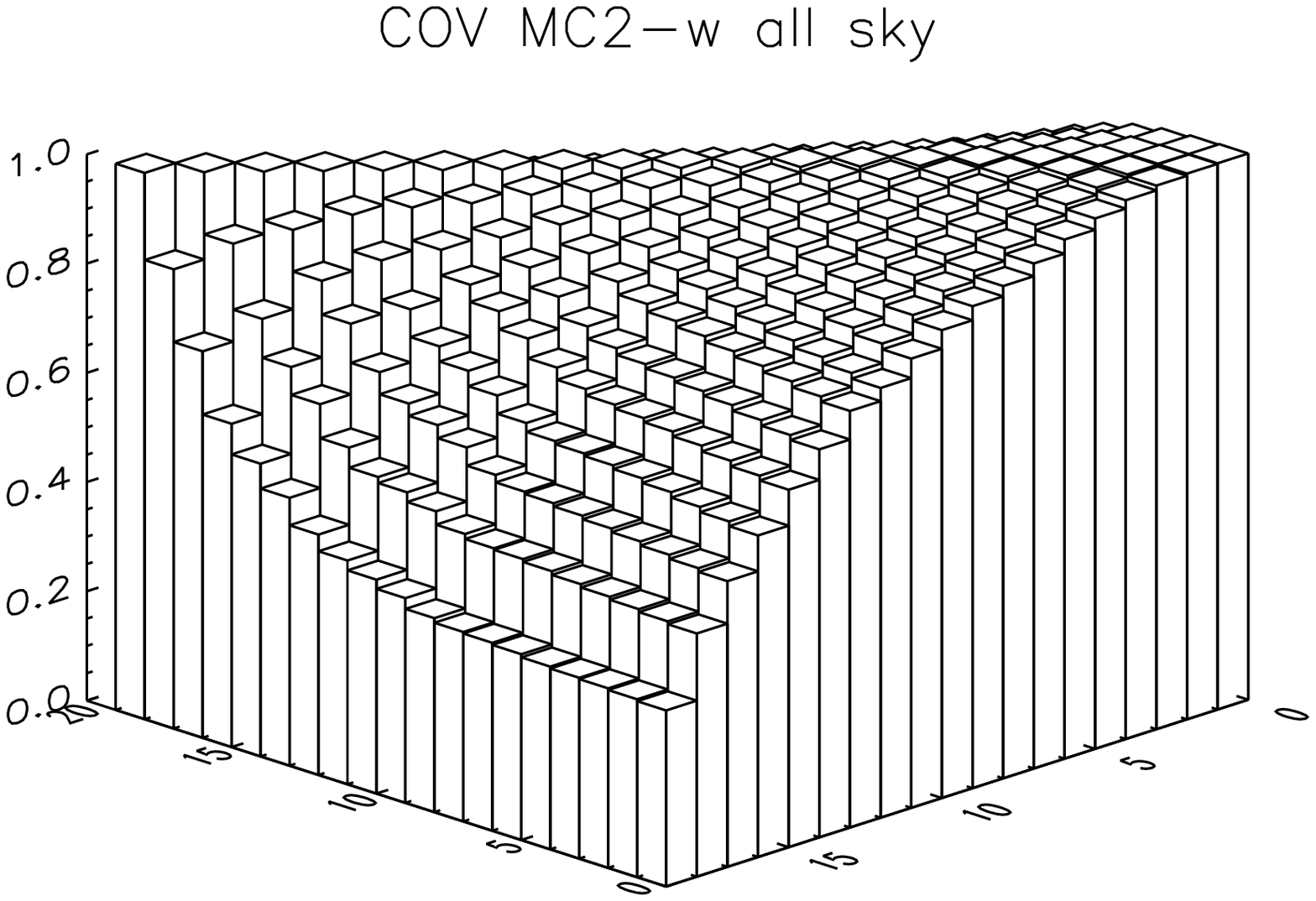}}}	

\caption{Normalized covariances in real space for different methods as labeled in the figure.
No significant changes are found for different methods and sky fractions used.}
\label{fig:cov}
\end{figure*}

\newpage

\begin{figure*}
	{\centering{\epsfysize=4cm \epsfbox{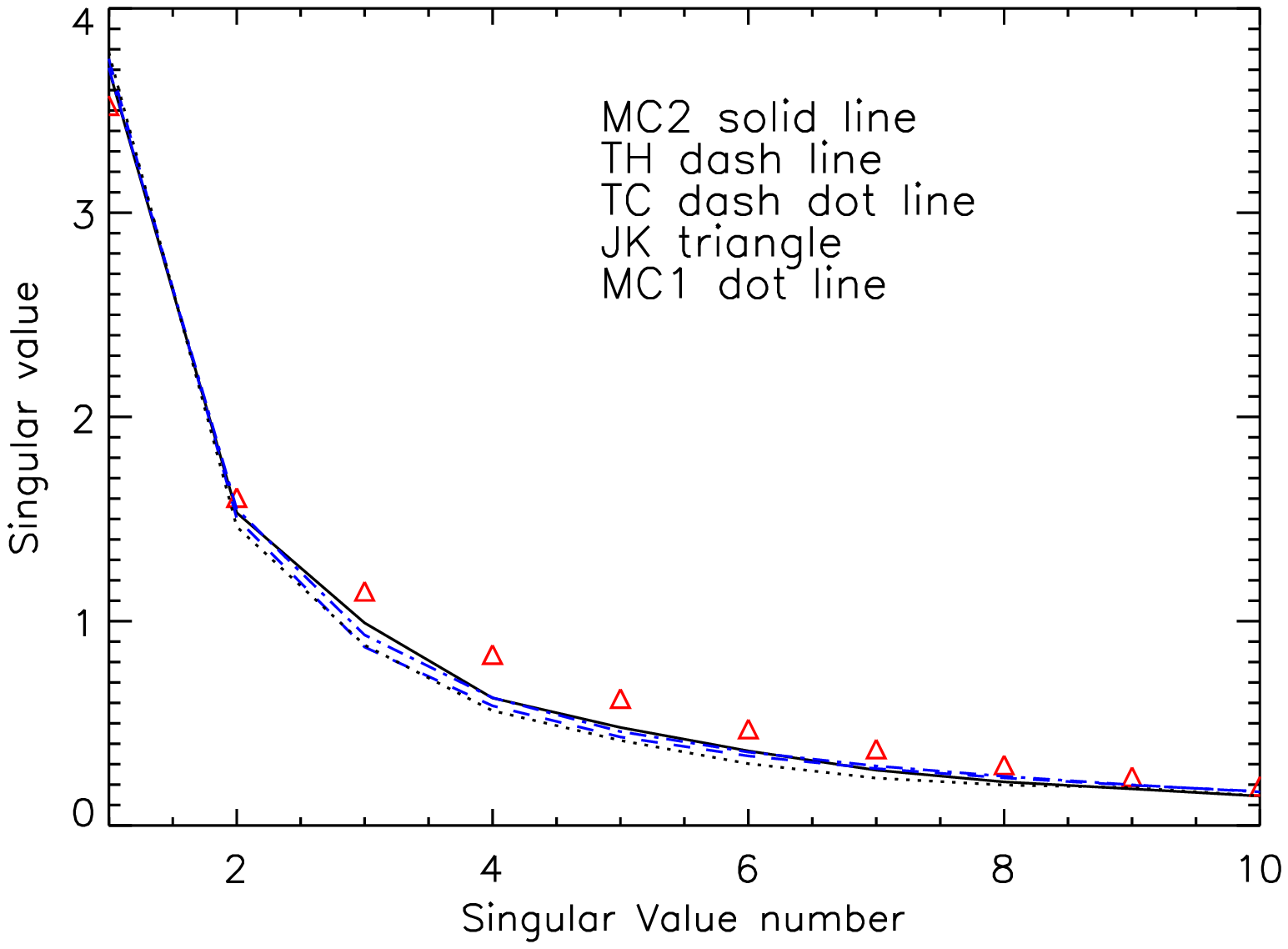}}}
	{\centering{\epsfysize=4cm \epsfbox{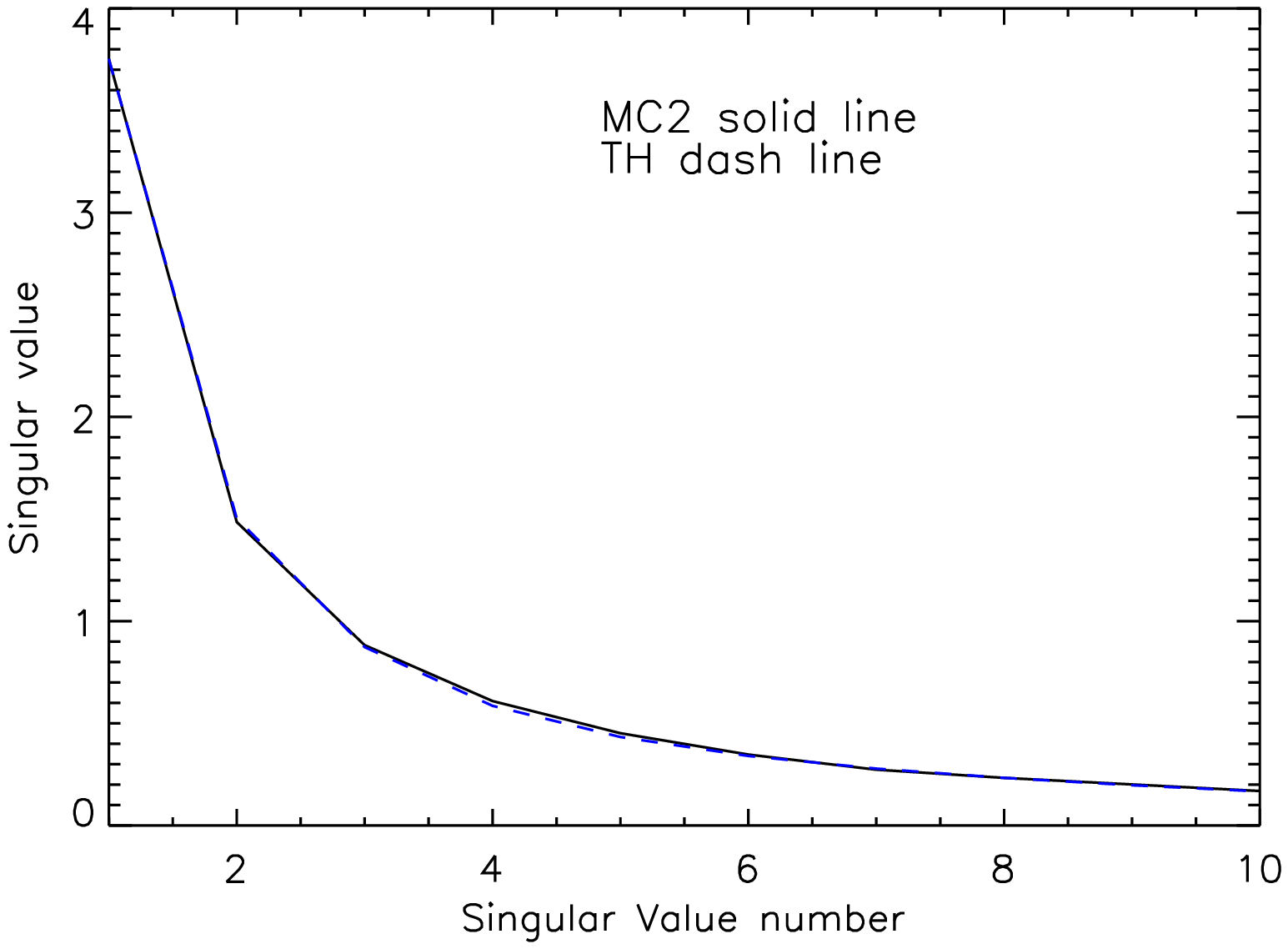}}}
\caption{Eigenvalues in real space for 10\% sky (left panel) and all sky (right).}
\label{fig:svw}
\end{figure*}

\begin{figure*}
	{\centering{\epsfysize=4cm \epsfbox{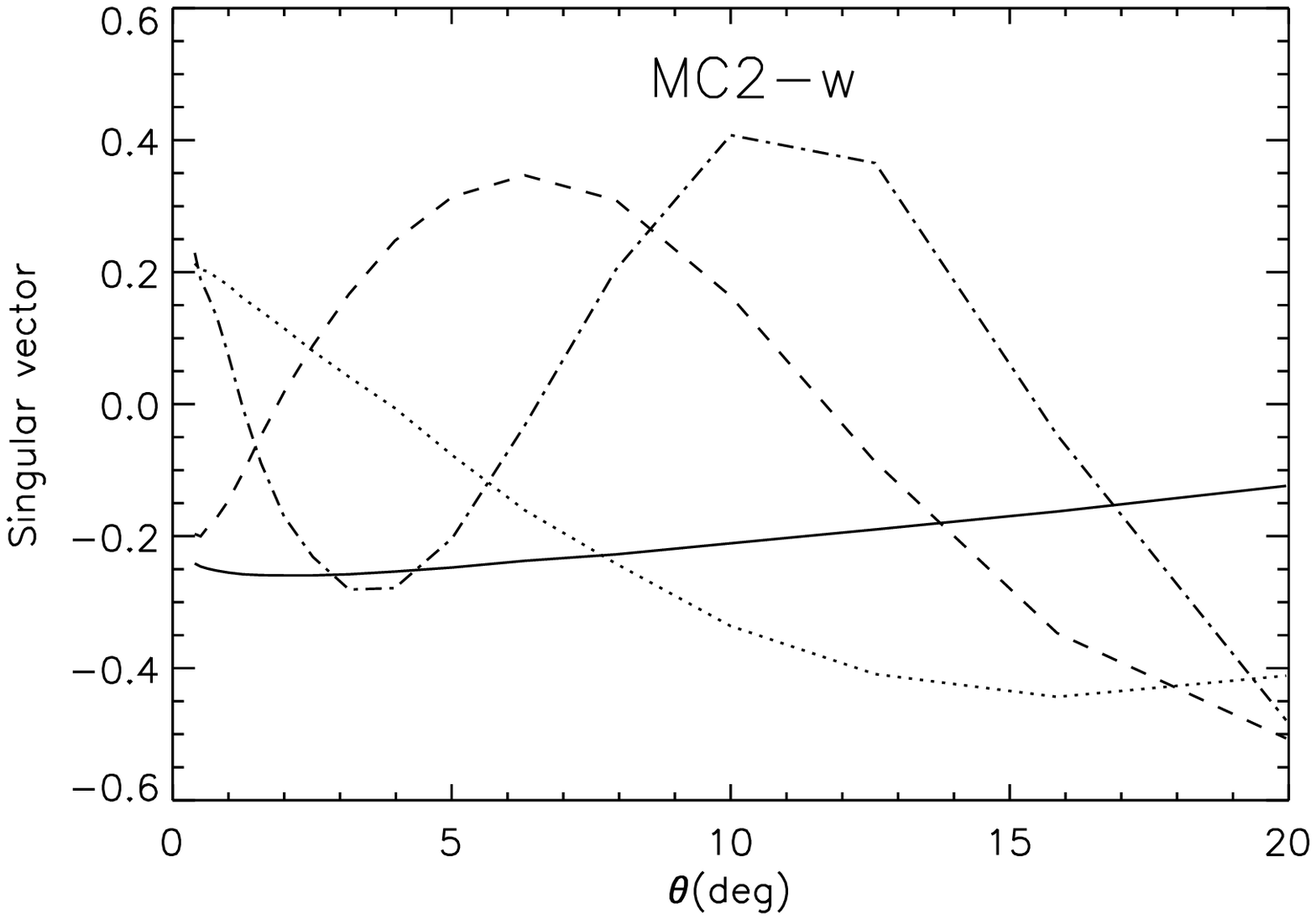}}
	\centering{\epsfysize=4cm \epsfbox{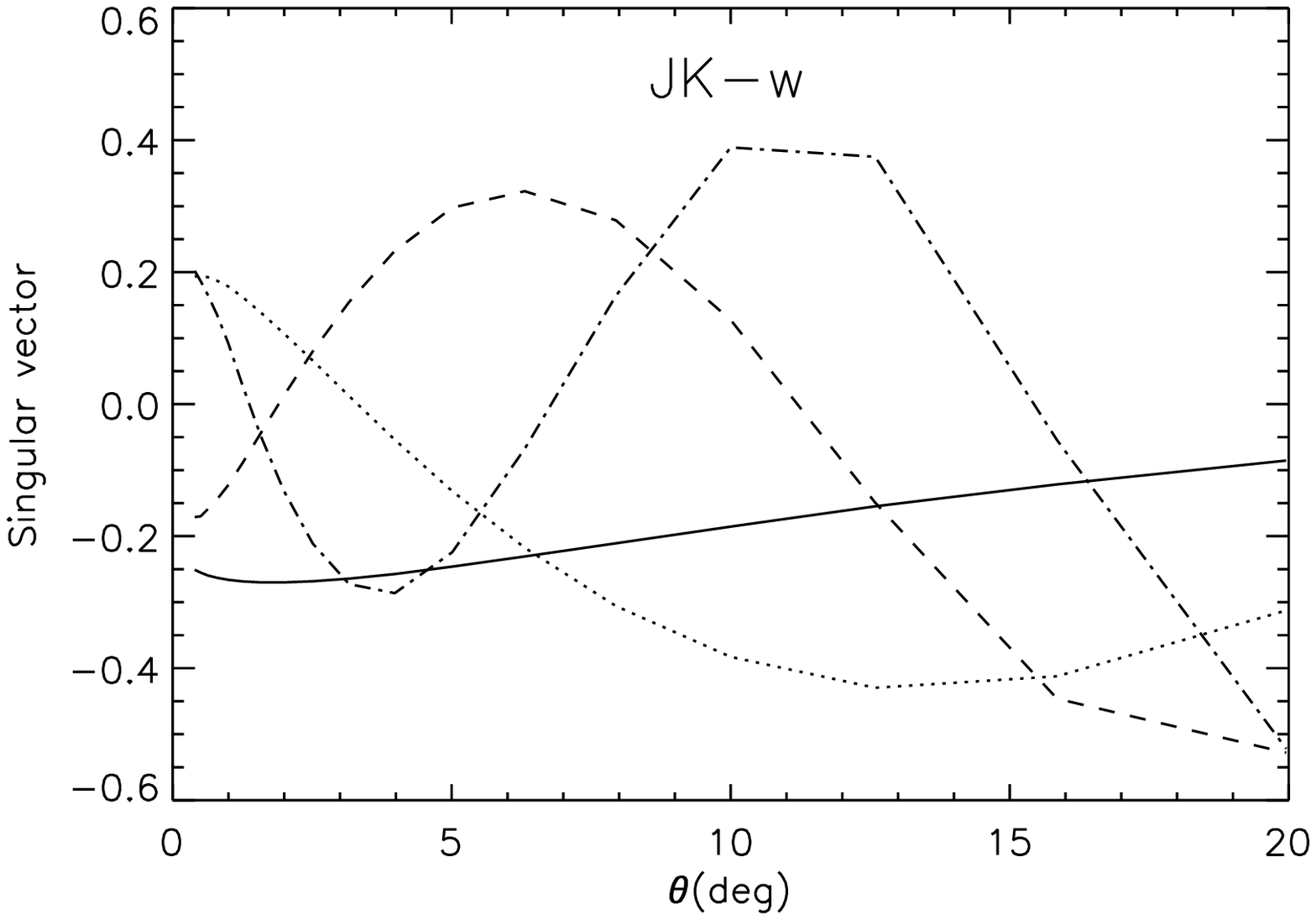}}	
	\centering{\epsfysize=4cm \epsfbox{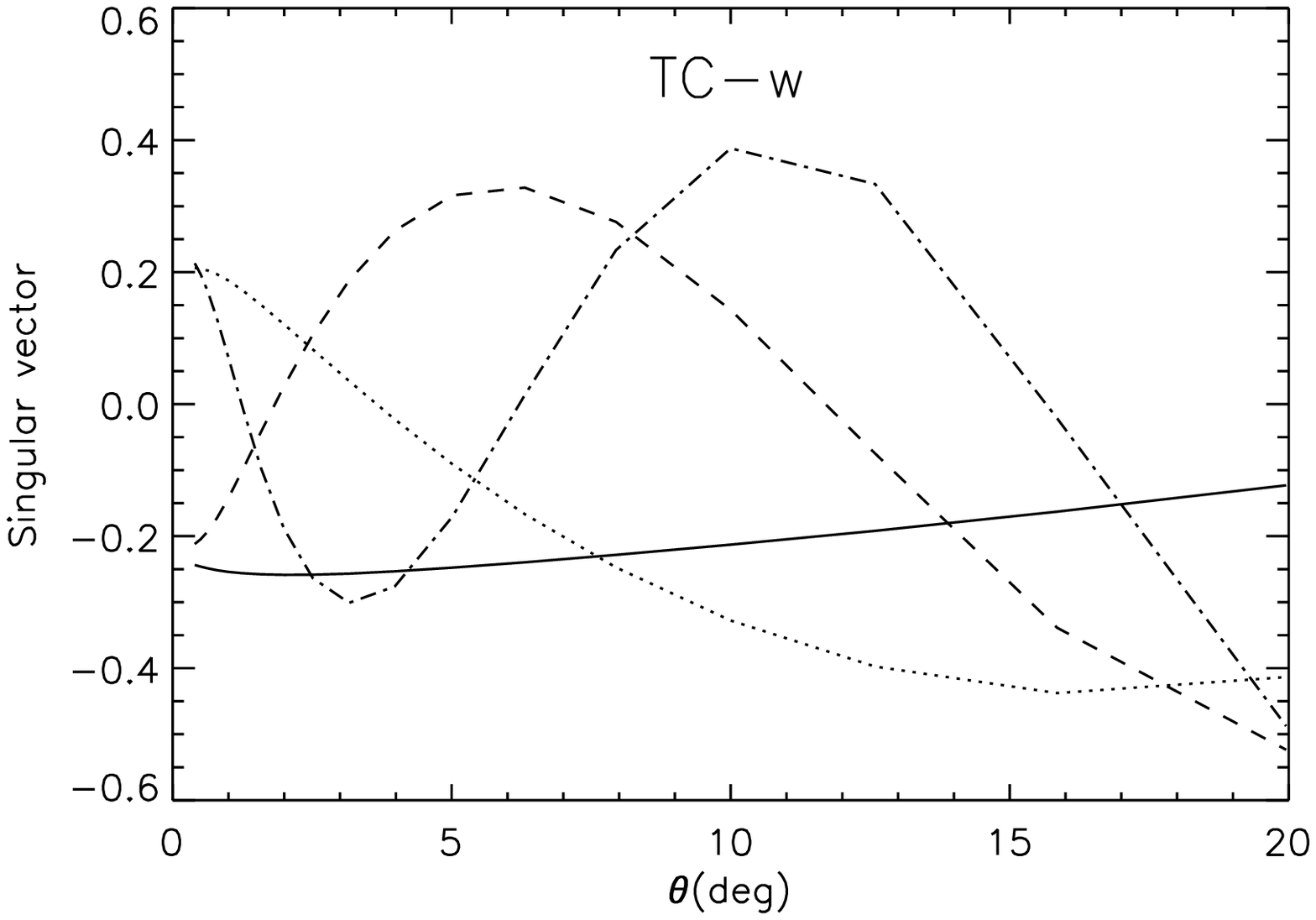}}	
	\centering{\epsfysize=4cm \epsfbox{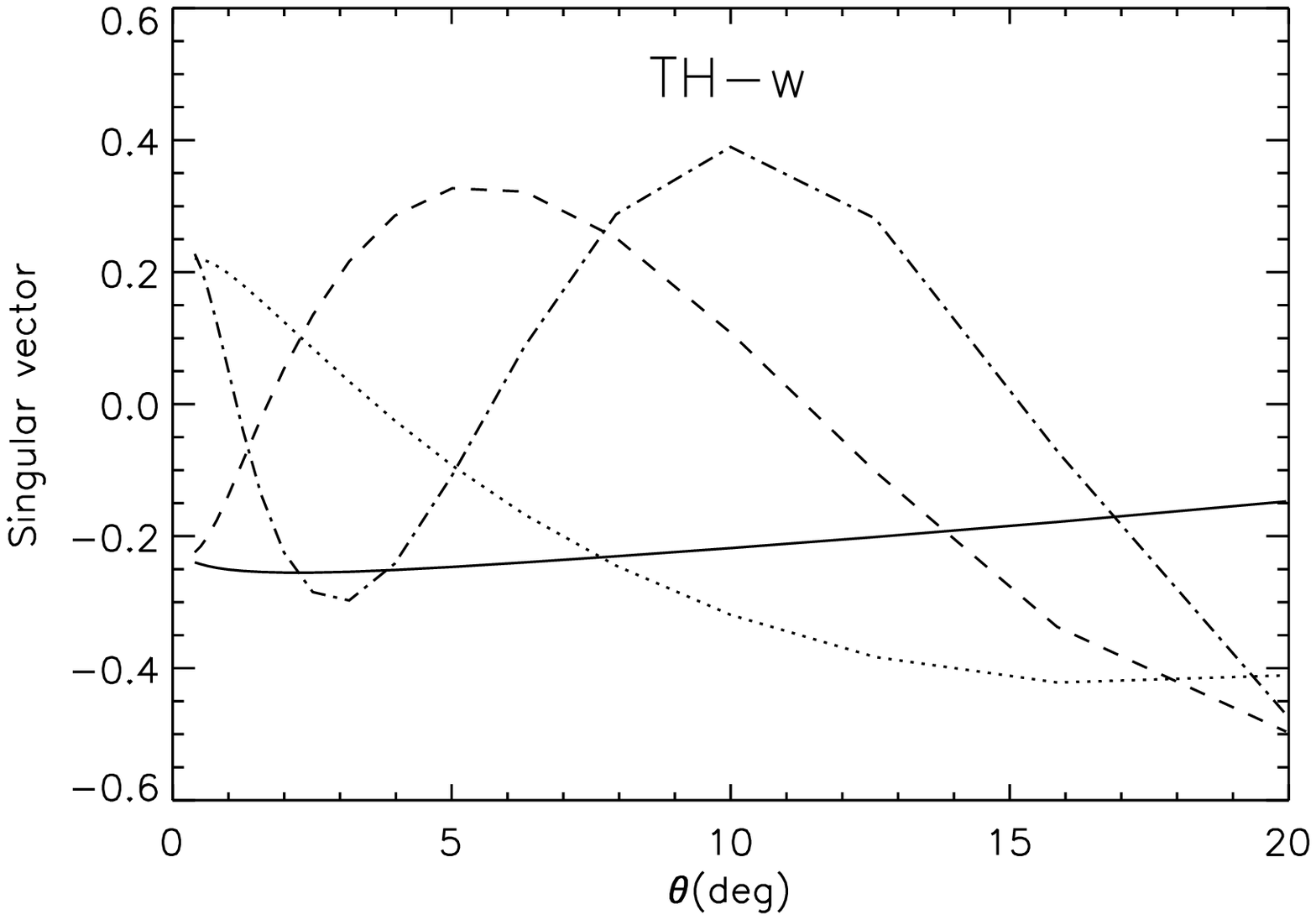}}	}
	\centering{\epsfysize=4cm \epsfbox{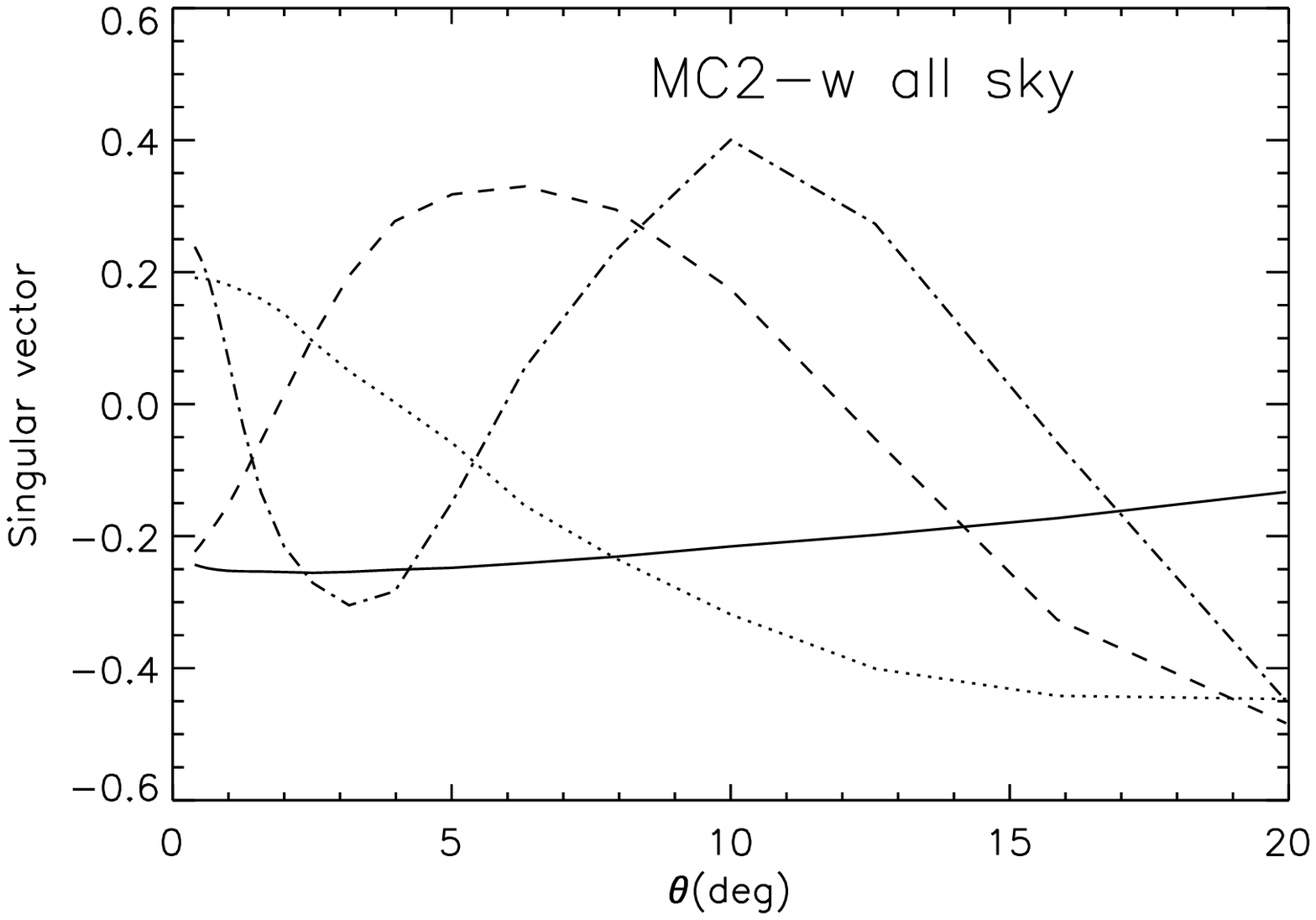}}	
	\centering{\epsfysize=4cm \epsfbox{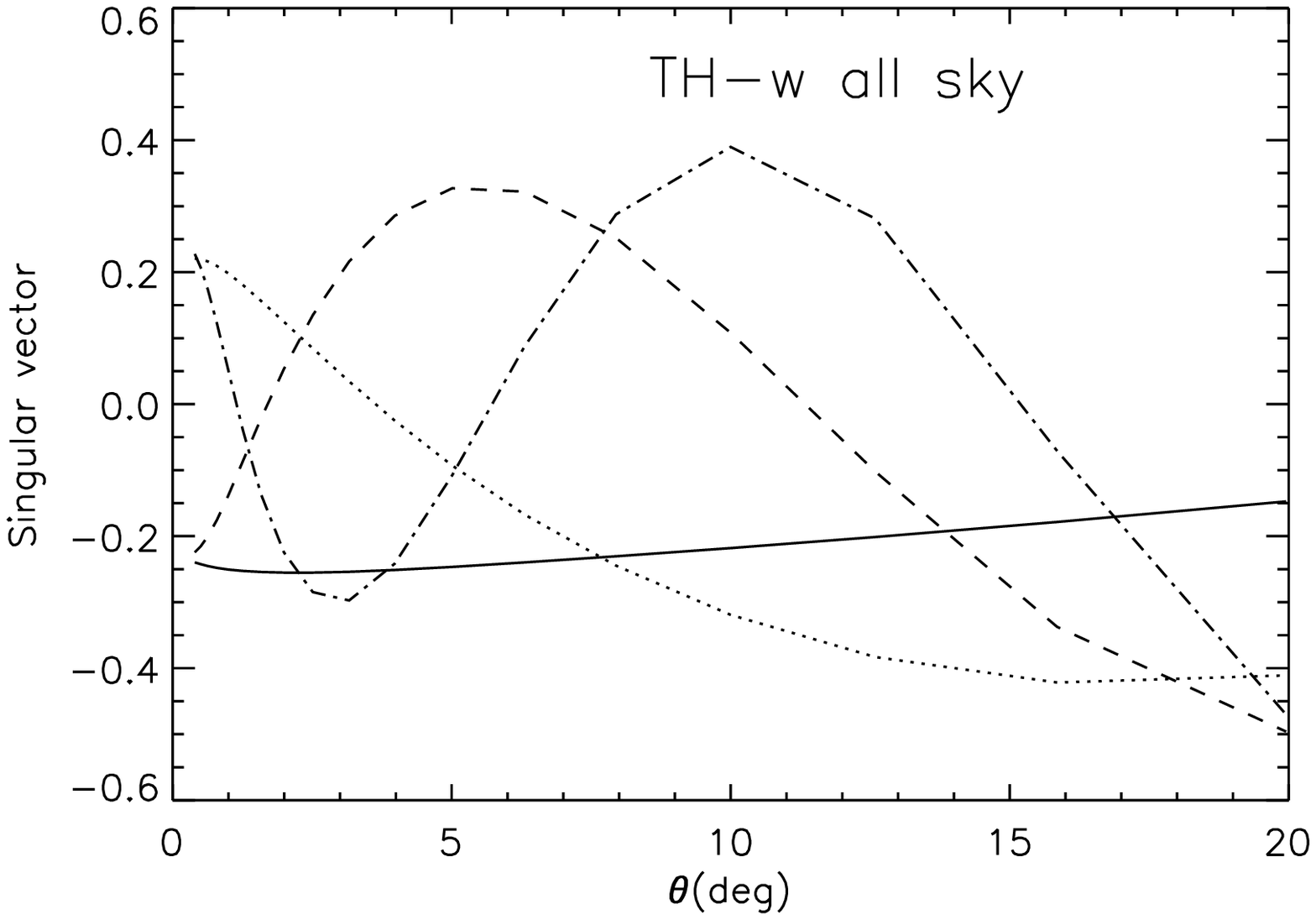}}	
\caption{Eigenvectors in real space for 10\% of the sky and for all sky, as labeled in each panel.
First eigenvector is shown as solid lines, second as dotted,
third dashed and fourth dot-dash. Results for $MC1-w$ and $TC-w$, not shown here, are very similar.}
\label{fig:eigenw}
\end{figure*}

\section{Normalized Covariance Matrix}   
\label{sec:cov}

For each one of the methods presented in the previous section,
we next compare the normalized covariance:
\begin{equation}\label{eq:covnorm}
\widehat{C}_{ij}=\frac{C_{ij}}{\sqrt{(C_{ii}C_{jj})}}
\end{equation}
The diagonal values (the variance) and associated dispersion will be investigated in
\S\ref{sec:variance}.

\subsection{Configuration space}

As shown in Fig.\ref{fig:cov} all the normalized covariances $\widehat{C}_{ij}$
in real space are very similar. The appearance of the  plots does not seem to
 depend strongly
 on the method we use to estimate them, or the survey area $f_{sky}$.
Here we only show  results for 10\% and all the sky, but intermediate values
yield similar results. However,  we want to question if
slight differences in the covariance could have a non-negligible
impact on cosmological paramter estimation. We will discuss this in detail in \S\ref{sec:chi2}.

\subsection{Harmonic space}

In $\Cl$ space, there is no correlation between different $l$-modes  (bins $\Delta{\l}=1$)
for  the case of all sky (MC2)  maps. The normalized covariance matrix is diagonal,
as can be seen in the right panel of Fig.\ref{fig:covCl}.
Also shown, in the left and central panels,  are the results for 10\% and 40\% of the sky,
where the covariance between modes gives rise to large amplitude off-diagonal elements. This is in sharp contrast to the
results in configuration space (in Fig.\ref{fig:cov}) where
there is no significant difference between normalized covariances in real space when
we decrease the area. This is because the main effect of increasing the area
 in configuration space is the reduction
of diagonal errors (which are shown in next section), while in harmonic space there
is  a transfer of power from diagonal to off-diagonal elements.

\subsection{Eigenvalues and Eigenvectors from SVD}
\label{sec:svd}

To calculate the distribution $\chi^2$ and the signal to noise we need to invert
the covariance matrix. We use the Singular Value Decomposition method to
decompose the covariance in two orthogonal matrices $U$ and $V$ and a diagonal
matrix $W$ which contains the singular values $\lambda_i$ squared on the
diagonal (eg see Press etal 1992). This method is adequate to separate the
signal  from the noise:

\begin{equation}\label{eq:coveq}
\widehat{C}_{ij}=(U_{ik}^T)W_{kl}V_{lj}
\end{equation}
where $W_{ij}= \lambda_i^2\delta_{ij}$ and
$\widehat{C}_{ij}$ is the normalized covariance in Eq.\ref{eq:covnorm}.
By doing this decomposition, we can choose the number of modes that we wish to
include in the analysis. This SVD is effectively a decomposition in
different modes ordered in decreasing amplitude.

We obtain very similar singular values for each mode and for each method,
as show in Fig.\ref{fig:svw}  for some of the cases (other cases give very similar
results).

We can understand the effect of modal decomposition looking at the eigenvectors
shown in Fig.\ref{fig:eigenw}, where we have plotted
the four dominant eigenvectors as a function of angle:
first mode (solid) affects only the amplitude, second mode (dotted) shows a bimodal
pattern. The following modes,
third (dashed) and fourth (dot-dash), correspond to modulations on smaller
angular scales. As can be seen in the Figure,  we obtain nearly the same
eigenvectors in all the
cases, in agreement to what was found by direct comparison of the covariance
matrices in Fig.\ref{fig:cov}. Again, we can ask: are the small differences significant?
We will study this in detail in \S\ref{sec:significance}.

\section{Variance \& errors}
\label{sec:variance}

\subsection{Variance in $w(\theta)$}

\begin{figure}
	\centering
{\epsfysize=6.3cm \epsfbox{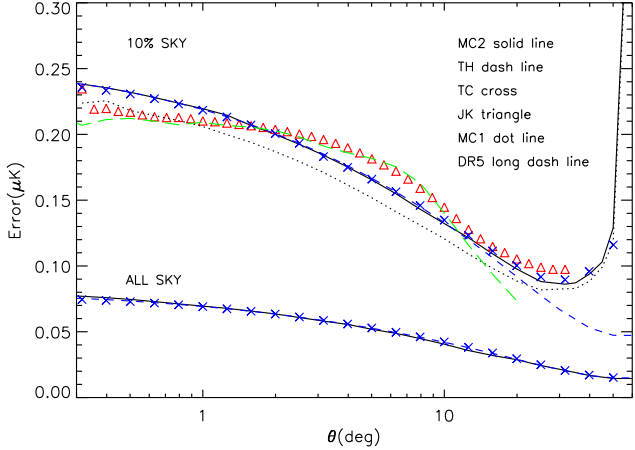}}
	\caption{Error calculated with different methods (as labeled in the figure) in real space
for $\Omega_{DE}=0.7$. For a map covering $10\%$ of the sky (top lines and symbols),
the TC (crosses) and TH (dashed line) theoretical errors
work well compared to the montecarlo MC2 simulations (solid line),
while MC1 simulations (dotted line) seems to underestimate the errors by
10\%. The JK method (triangles) seems slightly biased up/down on large/small scales,
although all the errors are compatible
given the sampling variance dispersion we expect (see Fig. \ref{fig:varrealw}).
For all sky maps (bottom lines and symbols), we show how results for
MC2 (solid), TH (dashed) and TC-w (cross) agree very well.}
	\label{fig:varreal}
\end{figure}

Fig.\ref{fig:varreal} is one of the main result of this paper.
We compare the variance for the different methods, which is the diagonal part
of the covariance matrix. By construction, in the limit of infinite number of realizations, 
the MC2 error from simulations should provide the best approximation to the errors.
We have demonstrated (in section \S\ref{sec:convergence})
that 1000 simulations are enough for convergence within $5\%$ accuracy.
For all sky maps (lower lines in the Figure) we can see that the three methods used:
MC2-w, TH-w and TC-w, yield identical results, as expected. For smaller survey areas we do
 expect some deviations, because of the different approximations on dealing with the survey boundary.
For a survey covering 10\% of the sky these 3 methods also agree well up to 10 degrees. At larger
scales TH-w  (dashed lines) starts to deviate, because  boundary effects are in fact not
taken into account in this method.  The JK error (triangles) has a  slope as a function
of $\theta$ that seems less steep than the other methods, 
but still gives a reasonably good approximation given
that the dispersion in the errors is about $20\%$ (as discuss in
\S\ref{sec:errinerr} below). Note how on scales larger than 10 degrees the JK
method performs better (ie it is closer to MC2)  than the TH-w error. The
TC method seems to account well for the boundary effects, as it reproduces the
MC2 errors all the way to 50 degrees, where all other methods fail.

If we only use one single realization for the galaxies
(MC1) the error seems to be systematically underestimated by about $10\%$ on all scales.
 This bias is expected as we have neglected the variance in the galaxy field and the
 cross-correlation signal.
A particular case of MC1 is done with real
data from SDSS DR5 (shown as long dashed line in Fig.\ref{fig:varreal}).
We have used here a compact square
of 10\% of the sky from the  SDSS $r$ magnitude slice of 20-21, which has a redshift selection
 function similar to the one in our
simulations ($z_m=0.33$). This case works surprisingly well once scaled with linear
bias $b$ (estimated by comparing  the measured galaxy auto-correlation function with 
the one in our fiducial \LCDM model). It happens to closely follow the JK prediction,
rather than the MC1 prediction,
but we believe this is just a fluke, given  the dispersion in the errors
(see \S\ref{sec:errinerr}) and the uncertainties in the fiducial model.

\subsection{Effect of partial sky coverage}

\begin{figure}
	\centering
{\epsfysize=6.5cm \epsfbox{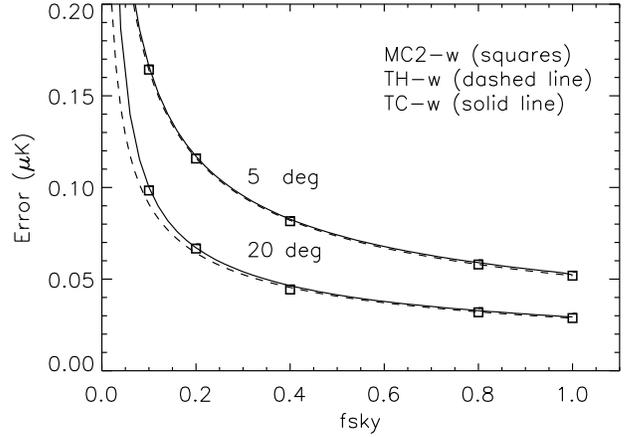}}
	\caption{Error  at a two fix angles: 5 deg (upper lines) and 20 deg (lower lines)
	as a function of  $f_{sky}$, the fraction of the sky covered by the map.
	As predicted,  errors decrease as $1/\sqrt{f_{sky}}$ in both cases. Note how for small
	areas, the TC-w prediction (continuous line) performs better than the TH-w model (dashed line)
	as it better reproduces the MC2 simulations (squares). }
	\label{fig:plotfsky}
\end{figure}

We have tested MC2-w, TH-w and TC-w for different partial sky survey areas $f_{sky}$
and obtained similar results. In Fig.\ref{fig:plotfsky} we have plotted the error
 for a fix angle of 5 degrees (top) and 20 degrees (bottom)
 for the different values of $f_{sky}$.  The three
 methods coincide for large areas. The error scales by a factor $1/\sqrt{f_{sky}}$, as expected.

Notice that errors at angles comparable to the width of the survey are difficult
to estimate theoretically because one needs to take into account the survey geometry.
Even for a map as wide as $10\%$ of the sky, the survey geometry starts to be important
for errors in the cross-correlation above 10 degrees. This is shown in simulations
as a sharp inflection that begins at 30 degrees in Fig. \ref{fig:varreal} (solid line) .
Our new TC-w method predicts well
this inflection, while the more traditional TH method totally misses this feature.
This can also be seen in Fig.\ref{fig:plotfsky} for 20 degrees when we approach
small values of  $f_{sky}$.

\subsection{Uncertainty in $w(\theta)$ errors}
\label{sec:errinerr}

\begin{figure}
	\centering	
{\epsfysize=6.1cm \epsfbox{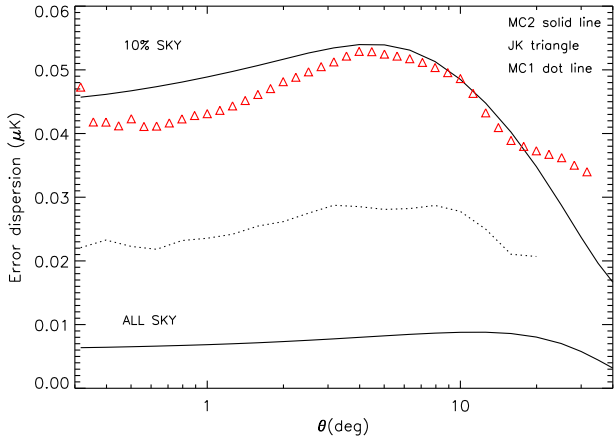}}
	\caption{Dispersion of the error calculated with different methods in real space for $\Omega_{DE}=0.7$. For MC2-w (solid line) or MC1-w (dotted lines), we take each pair of MC2 or MC1 simulations
 as the input for the error in the TH-w calculation  in Eq.\ref{eq:errorW} .
 For JK (triangles), we have one error for each MC2 simulation.  We can see how the error in the error is of other $~20\%$ for MC2 or JK and is lower for MC1
 (mainly because one of the maps in each pair is kept fixed). The lower line corresponds to
the dispersion in all sky maps.}
	\label{fig:varrealw}
\end{figure}

To assess the significance of the differences in the error estimation
that we find using different methods, we will compute here the sampling uncertainties
associated with error estimation.
Fig.\ref{fig:varrealw} shows the sampling dispersion in the error estimates.
This can be calculated from the TH and TC approaches by using $\Cl$ or $w(\theta)$
measure in each realization as the input model for theoretical predictions
(Eq.\ref{eq:errorW} or Eq.\ref{result}). In Fig.\ref{fig:varrealw} solid (or dotted) line
shows the result of using Eq.\ref{eq:errorW}  for each of the MC2 (or MC1) simulations.
This produces an
error for each realization and we can therefore study the error distribution. The
uncertainty in the error (or error in the error) correspond to the rms dispersion
of this distribution. We need all the
multipoles to compute this Legendre transformation (Eq.\ref{eq:errorW})
 although we lose some
information for low multipoles when we use only a fraction of the sky.
The error propagation Eq.\ref{eq:errorW} is not linear and 
we find that this produces
a bias of $3\%$ when we compare the mean of the propagated errors in each
simulation with the propagation of the mean error in all simulations.

We can also calculate the JK-w dispersion of the error, because we have the JK error for each MC2
simulation (remember that we only need one realization to obtain the JK error).
The JK-w dispersion (triangles) in Fig.\ref{fig:varrealw} is quite close to the
MC2-w values. They are both of the order of
$20\%$ relative to the mean error. This uncertainty can be interepreted as
the result of the uncertainties in our input model; typically the model is only
known to the accuracy given by the data and a given sky realization will
deviate from the 'true' model (i.e, the mean over realizations).
Thus, if one chooses to use the estimated
values from the data (or it's bets fit model) as input to the error
estimation, this produces an uncertainty in the error which is of the order
of this scatter. This is always the case with the JK errors, which do not
use any model, but the uncertainty is similar if we use direct measurements
as input to the other error estimations, as shown in Fig.\ref{fig:varrealw}.

For completeness,
Fig.\ref{fig:varrealw} also shows the dispersion for the MC1 error (dotted). There
is less dispersion in the MC1 method because one of the maps is always fixed and this reduces
both the error and, more strongly, its dispersion.

\subsection{Error distribution for JK}

\begin{figure}
	\centering
{\epsfysize=6.2cm \epsfbox{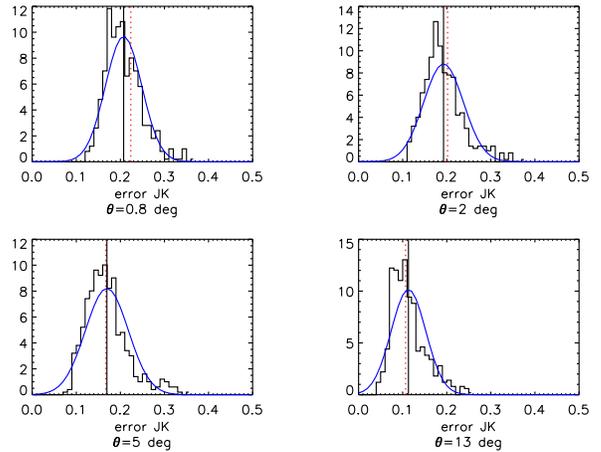}}   	
	\caption{Histograms show the error distribution in JK errors from 1000 simulations at different angles.
Solid line shows a Gaussian with the same mean and dispersion. Vertical lines correspond to  mean JK error (solid)  and the true mean MC2 error (dotted).}
	\label{fig:JKdistr}
\end{figure}

Fig.\ref{fig:JKdistr} shows the distribution of JK errors in the MC2 simulations as
compared to  a Gaussian fit
with the same mean and dispersion. Each panel shows the
distribution of $w(\theta)$ errors at a given fixed angle.
The mean MC2-w error (shown as solid line in Fig.\ref{fig:varreal})
 is shown here by a  dotted vertical line,
while the mean of the JK errors  (shown as triangles in Fig.\ref{fig:varreal})
 corresponds here to the continuous vertical line.
We can see here how the MC2-w error and the mean JK-w error are quite similar.
The variance in the distribution  agrees with the results  in \S\ref{sec:errinerr} above.
Note also that the JK distribution of $w(\theta)$ errors can be well fitted by a Gaussian.
This is important for two reasons. First it shows that there are no important outliers or systematic bias
when one uses a JK estimator in a single realization, as is the case with real data. Second,
it indicates that the error in the error (ie the rms dispersion of this distribution) entails all
relevant  information needed to asses in more detail the accuracy of  the JK error
analysis. One could for example fold the uncertainties in this distribution to asses the significant of a
detection.

\subsection{Variance in $\Cl$}
\label{sec:clvar}

\begin{figure}
	\centering
{\epsfysize=7.5cm \epsfbox{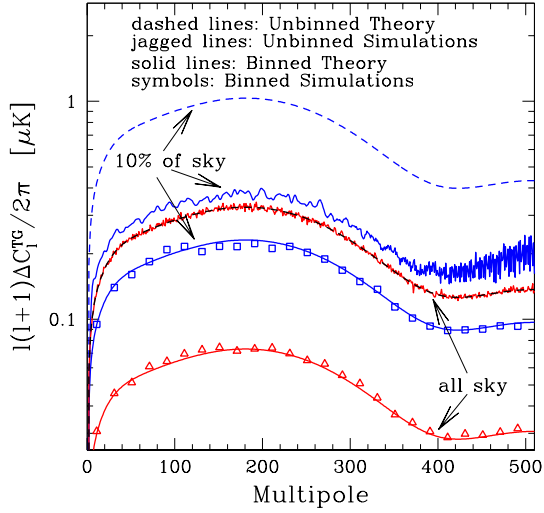}}
	\caption{Errors in $\Cl$ space calculated with (MC2) simulations as compare to
	(TH-$\Cl$) theory.
	 For all sky maps, the theoretical prediction works well, but for 10\% of the sky we see a big  discrepancy between theory (dashed lines) and simuations (jagged lines).
 This is due to covariance between modes and can be solved by binning the $\Cl$ spectrum, as shown by the symbols (simulations) and sold line (predictions
 in Eq.\ref{eq:errorCl2}).}
\label{fig:varcl}
\end{figure}

In $\Cl$ space, we have compared MC2 errors to TH theory.
 Fig.\ref{fig:varcl} shows how both
errors are hard to distinguish for the case of all sky (middle dashed line matches closely
the jagged line). Note the shape of the $C_{\l}$ errors exhibits a broad peak around
$\l=200$ illustrating the fact that errors are dominated by the
$C_{\l}^TT$ term.
For 10\% of the sky the TH error (upper dashed line)  obtained theoretically
from Eq.\ref{eq:errorCl}  (with a factor $1/\sqrt{f_{sky}}$
respect to all the sky) is much larger than the MC2 error (upper jagged line) in the simulations.
 As we have shown in
Fig.\ref{fig:covCl}, there is a strong
 covariance between different bins when $f_{sky}<1$, this is in contrast with
 the TH estimation in Eq.\ref{eq:errorCl}  which assumes a diagonal
covariance matrix. We understand this discrepancy in the variance prediction
as a transfer of power from  the diagonal  to off-diagonal errors.

We can get a better diagonal error estimation by binning $\Cl$
in a $\Delta\l$ that makes the covariance approximately diagonal.
\footnote{This is clear in Fig.\ref{fig:covCl}
which shows that the covariance is confined to a finite number of $\Delta\l$
of off-diagonal elements. It is also apparent in  Fig.\ref{fig:varcl} where the
jagged line for10\% of the sky is clearly correlated on scales
of $\Delta\l \simeq 20$, in contrast to the all-sky jagged line which shows
no correlation from bin to bin. }
When binning by $\Delta\l$,  the theoretical error (TH-Cl) in Eq.\ref{eq:errorCl}
is reduced in quadrature to:
\begin{equation}\label{eq:errorCl2}
\Delta^{2} C^{TG}_{\l} = \frac{1}{\Delta\l  f_{sky} (2\l+1)} \left[({C^{TG}_\l})^2 +
C^{TT}_{\l} C^{GG}_{\l} \right],
\end{equation}
This assumes that the bins are independent.
Because of the partial sky coverage, the bins are not independent
and the above formula will
only be valid in the limit of large $\Delta\l$.

We have tested the above formula for different sky fractions by binning the
$\Cl$ spectrum in the simulations and estimating the error from the scatter in
different realizations. We find that the formula works above some minimum
 $\Delta \l$ which roughly agrees with  the width of off-diagonal coupling in the
 covariance matrix estimated from simulations (Fig.\ref{fig:covCl}). 
We find that $\Delta\l$=20,16,8,1 for $f_{sky}$=0.1,0.2,0.4,0.8 respectively, diagonalize the covariance matrix and provide a good fit to the above theoretical error for binned spectra. In Fig.\ref{fig:varcl} we show the results for $\Delta \l=20$ for both all sky
(triangles) and 10\% of the sky (squares).
The theoretical prediction in Eq.\ref{eq:errorCl2} (solid lines)
works very well in both cases, because the covariance with
 this binning is approximately diagonal.

%

\subsection{Dependence on $\Omega_{DE}$}
\label{sec:OmegaDE}

\begin{figure}
{\epsfysize=6.cm \epsfbox{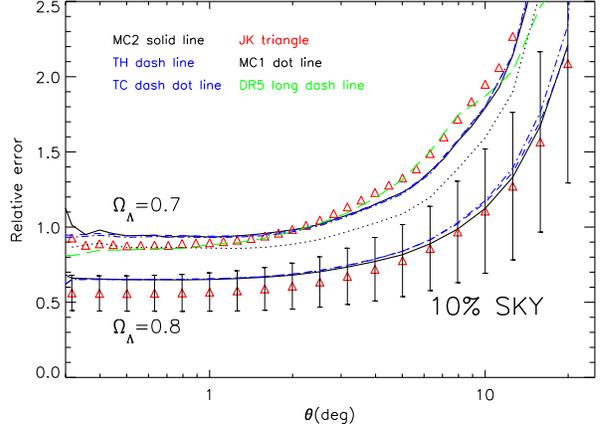}}
	\caption{Relative error for the two fiducial models $\Omega_{DE}=0.7$ (top)
and $\Omega_{DE}=0.8$ (bottom). The different methods are labeled in the figure. We see that the relative error depends on the model, and that in the case
 $\Omega_{DE}=0.8$ there is also  a good agreement within the errors.}
	\label{fig:varreal3}
\end{figure}

Fig.\ref{fig:varreal3} shows a relative comparison of
how our error estimation changes for a different cosmology with
$\Omega_{DE}=0.8$ instead of $\Omega_{DE}=0.7$. The MC error still fits well the TH and TC
predictions, but the JK errors seem to underestimate the errors more than in the
$\Omega_{DE}=0.7$ case. This effect is not
large given the dispersion in the errors from realization to realization (errorbars in
Fig.\ref{fig:varreal3}).

\section{Constraints and significance}
\label{sec:significance}

ISW measurements can directly constrain
dark-energy parameters independent of other cosmological probes.
Here we shall use the covariance analysis presented in the previous
section to derive significance levels for the cosmological parameter constraints
obtained from a cross-correlation analysis.

\subsection{Signal-to-noise from $w(\theta)$}

\begin{figure}
	\centering{\epsfysize=6cm \epsfbox{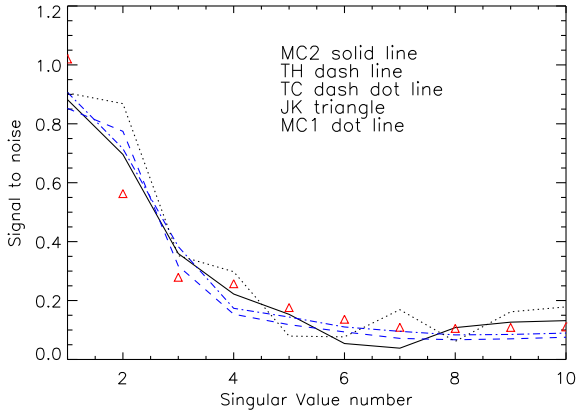}}
	\caption{Signal-to-noise
	for 10\% of the sky for each singular value. Different lines
correspond to methods as labeled.}
	\label{fig:s2n}
\end{figure}

\begin{table}
\begin{tabular}{|r|r|r|r|r|r|}
\hline
$S/N $&10\% & 20\% &40\% &80\% &all sky\\
\hline

Simulations MC2-w & 1.2 & 1.7& 2.5 & 3.5 & 3.8 \\
Theory TH-Cl & 1.2 &1.7 & 2.4 & 3.4 &3.8\\
Theory TH-w & 1.2 &1.7 & 2.4 & 3.4 &3.8\\

\hline
\end{tabular}
\caption{Signal-to-noise
as a function of $f_{sky}$ covered in a survey with a
broad distribution of sources with median redshift
$z_m=0.33$ for the $\Omega_{DE}=0.7$ flat \LCDM model with different error assumptions
 (see Table\ref{notation}). Similar results are found for TC-w and JK-w methods.}
\label{tb:ts2n}
\end{table}

The signal-to-noise ($S/N$ hereafter) depends on both the input
fiducial model used in the simulations and the covariance matrix method we implement.
In this paper we shall invert the covariance matrix using the standard method of
singular value decomposition (SVD), see \S\ref{sec:svd}. In this approach
one projects the signal to the eigenvector space of the thus diagonalized matrix and
only the most significant eigenvalues are kept for the analysis,
\begin{equation}\label{eq:s2n}
\Big(\frac{S}{N}\Big)_i =  \Big| \frac{\widehat{w}_{TG}(i)}{\lambda_i} \Big|=
 \Big| \frac{1}{\lambda_i}\sum_{j=1}^{N_b} U_{ji} \, \frac{w_{TG}(j)}
{\sigma_w(j)} \Big|.
\end{equation}
Fig.\ref{fig:s2n} shows the S/N for each singular value. All methods
agree well even for 10\% of the sky. We get excellent agreement
for all sky maps.

Because eigenvectors are orthogonal the total S/N is just added in
quadrature:
\begin{equation}\label{eq:s2nt}
 \Big(\frac{S}{N}\Big)_T^2  =  \sum_i ~\Big(\frac{S}{N}\Big)_i^2
\end{equation}

Table \ref{tb:ts2n} compares total S/N values from simulations and theory for different survey areas.
Here by {\it Simulations} we
mean the MC2-w method where we have used 6 singular values and {\it theory} refers to
the different methods, including the TH-$\Cl$ approach (see below).
We note that we find apparently lower values than quoted in the literature (see e.g, Afshordi 2004).
This is due to the low value adopted for $\Omega_{DE}$ (i.e $\Omega_{DE}=0.8$ models yield a S/N ratio
$\sim 2$ larger than our fiducial value $\Omega_{DE}=0.7$), and the
fact that these are predictions for a single broad redshift bin (similar to the selection function for
SDSS main sample galaxies), with median redshift $z_m= 0.33$.
A combination of several narrow bins at different redshifts will also increase the
S/N (see  Fig.\ref{fig:s2npred} and Table \ref{tb:ts2n2} below).

\subsection{Signal-to-noise forecast from $\Cl$}
\label{sec:forecast}

In harmonic space the $S/N$ is estimated as

\begin{equation}\label{eq:s2nforecast}
 \Big(\frac{S}{N}\Big)_T^2  = \sum_{\l}
\Big( \frac{C^{TG}_{\l}}{\Delta C^{TG}_{\l}}\Big)^2
\end{equation}
using Eq. \ref{eq:errorCl} in the denominator. Note in particular that the dominant
contribution to $\Delta C_{TG}$ in  Eq. \ref{eq:errorCl}, comes from
the $C_{TT}~ C_{GG}$ term and not from $C_{TG}$ which is an order of
magnitude smaller. This means that the $(S/N)^2$ approximately
scales as:
\begin{equation}\label{eq:s2nforecast2}
 \Big(\frac{S}{N}\Big)_T^2  \simeq  \sum_l
 \frac{C^{TG}_{\l} ~C^{TG}_{\l} }{ C^{GG}_{\l}~C^{TT}_{\l}} \propto  \sigma_8^2
\end{equation}
and therefore depends strongly on the normalization of the dark matter power spectrum $P(k)$,
and is independent of the galaxias bias $b$.

\begin{figure}
	\centering{\epsfysize=5.2cm \epsfbox{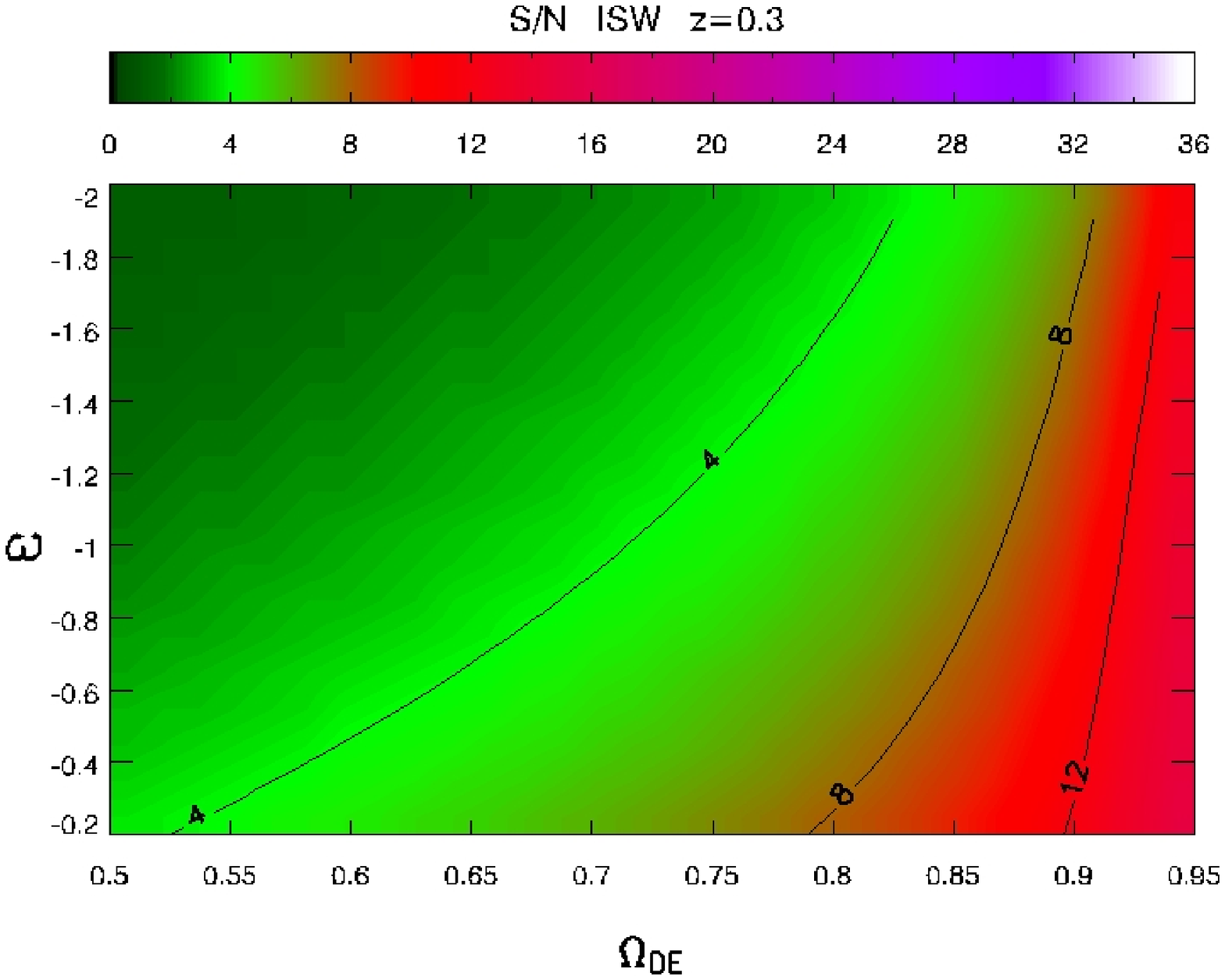}}
	\centering{\epsfysize=5.2cm \epsfbox{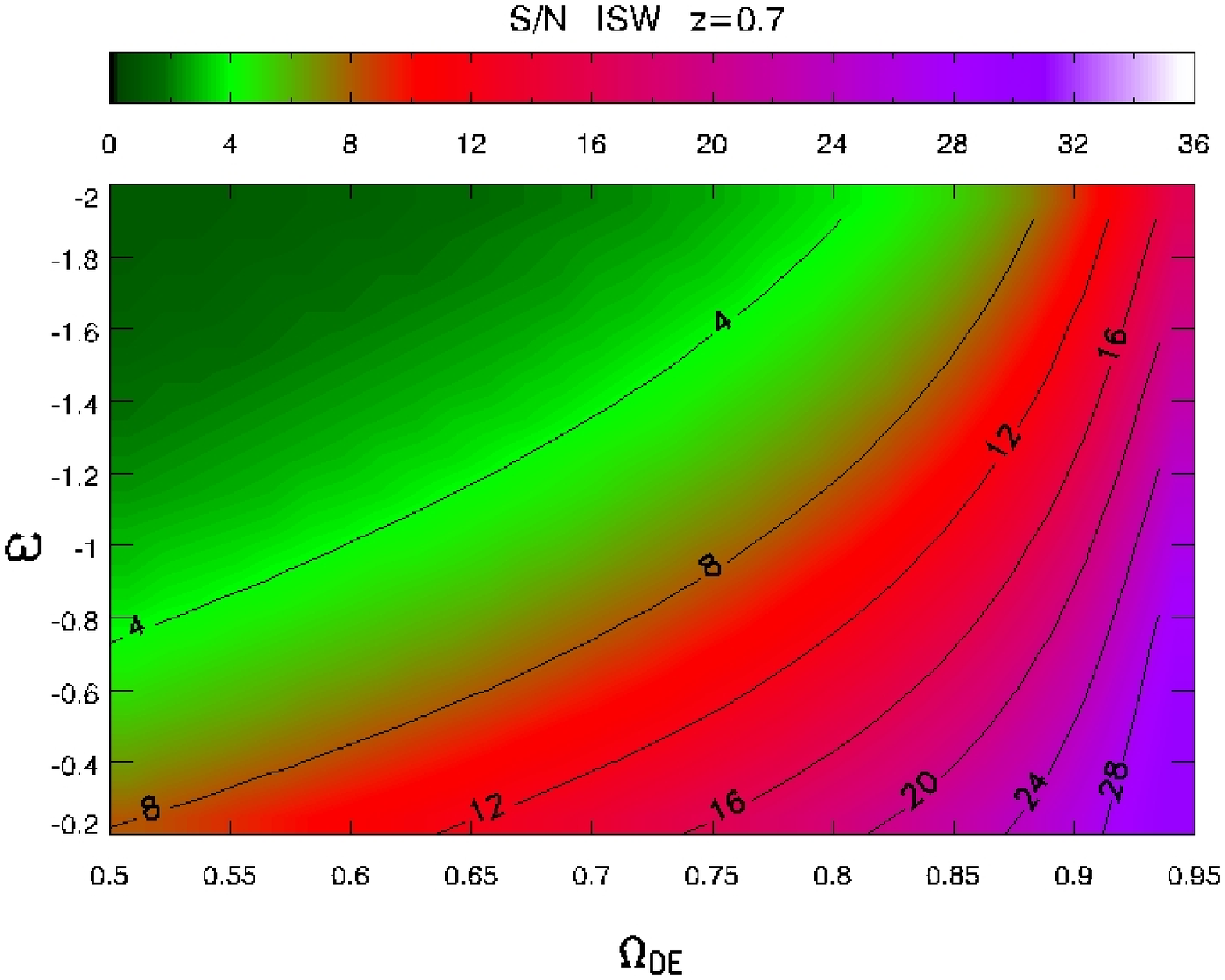}}	
	\centering{\epsfysize=5.2cm \epsfbox{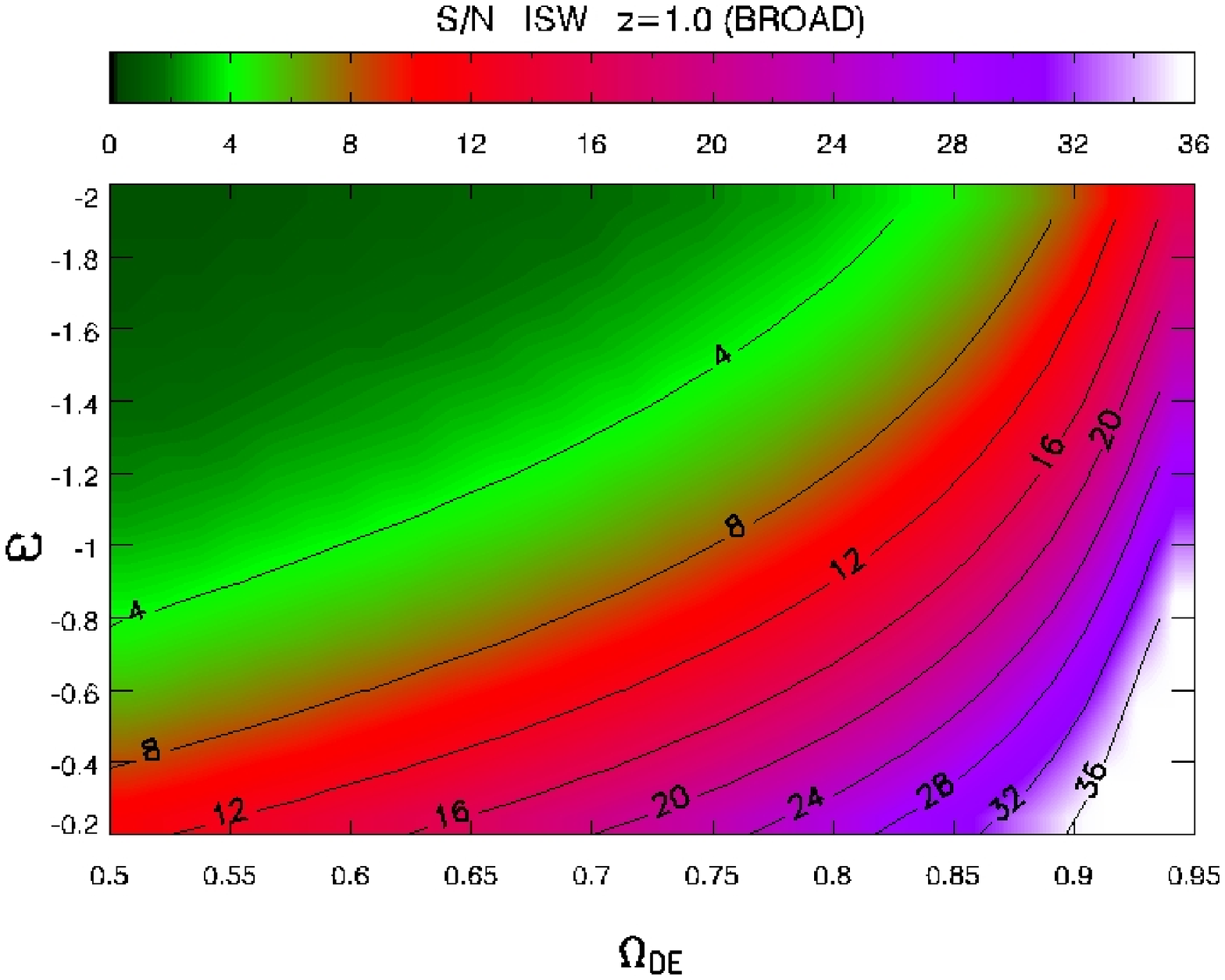}}
	\centering{\epsfysize=5.2cm \epsfbox{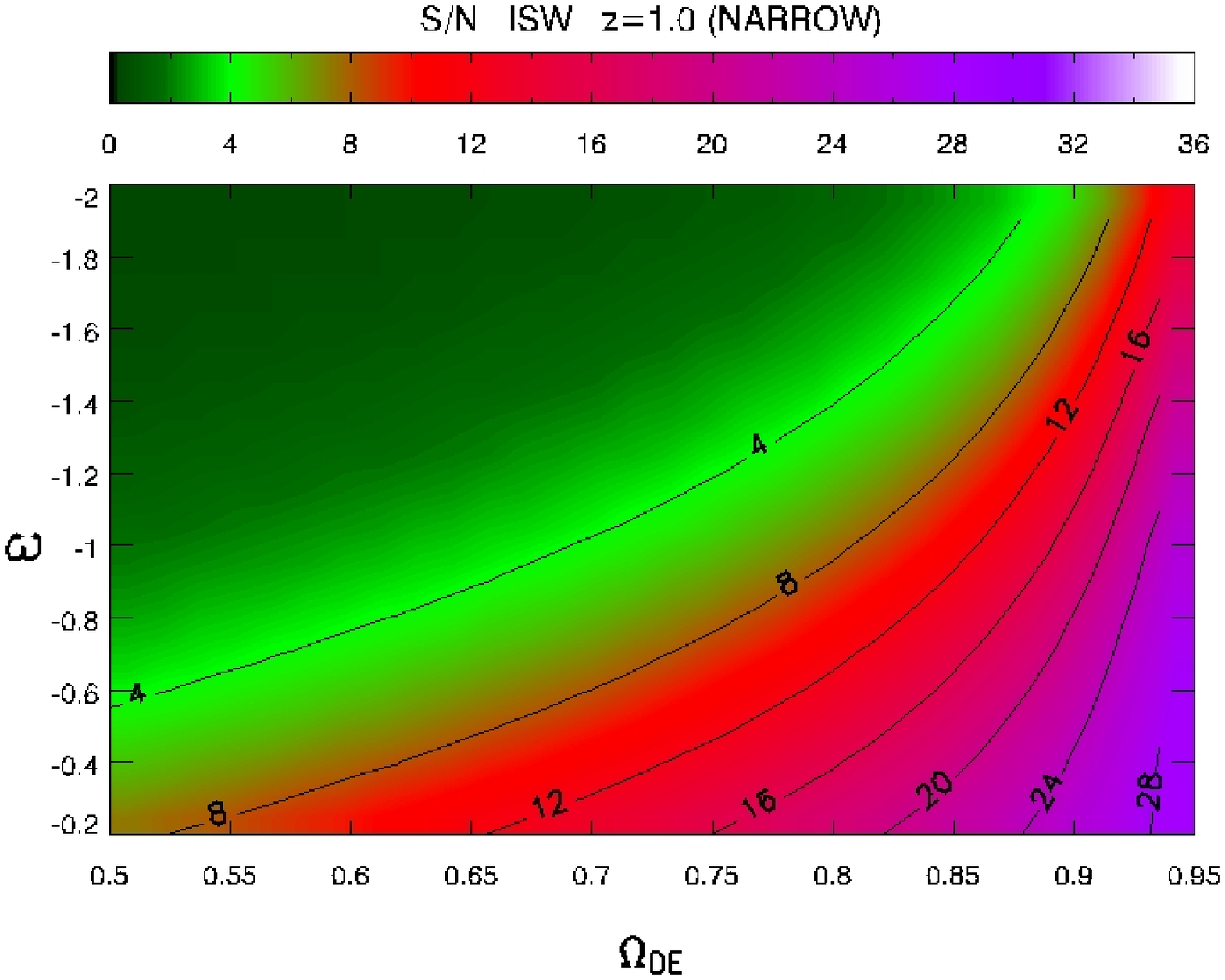}}
	\caption{All sky values of the S/N for different models. Each panel corresponds
	to different redshift distribution: SDSS
	 ($z_m=0.3$), DES($z_m=0.7$), DES+VISTA ($z_m=1$) and DES+VISTA NARROW ($z_m=1$).
Expectations for smaller survey areas are obtained by scaling the displayed values by
$\sqrt{f_{sky}}$, where $f_{sky}$ is the sky fraction covered.}
\label{fig:s2npred}	
\end{figure}

\begin{table}
\begin{tabular}{|c|c|c|c|c|}
\hline
$S/N $ & $z_m=0.33$ & $z_m=0.7$ & $z_m=1$ & $z_m=1$ (N) \\
\hline
$\Omega_{DE}=0.7$ & 3.8 & 6.0 & 6.3 & 4.2 \\
$\Omega_{DE}=0.8$ & 5.5 & 9.5 & 10.6 & 7.6\\
\hline
\end{tabular}
\caption{Signal-to-noise for all sky maps in harmonic space  for two
different values of $\Omega_{DE}$ and for
galaxy maps with different mean depths
$z_m$. The width of the redshift distribution is given by $\sigma_z  \simeq z_m/2$ (see Eq.\ref{dNdz})
except in the last case ($z_m=1$ Narrow), where $\sigma_z \simeq 0.17$}
\label{tb:ts2n2}
\end{table}

Clearly, the S/N will change depending on the fiducial model used.
Fig. \ref{fig:s2npred} shows this dependence on the
plane DE density vs. equation of state, $w$.
Each panel corresponds to
different smooth redshift distributions that closely match current or planned surveys.
The upper panels show predictions for
SDSS main sample ($z_m=0.33$), that anticipated for the DES ($z_m=0.7$),
and a combined DES+VISTA survey ($z_m=1$), respectively. For these 3 surveys we
use broad distributions as given by Eq.(\ref{dNdz}),
with a width that grows linearly with depth, $\sigma_z \simeq z_m/2$.
For this rather generic parametrization of the selection function,
the S/N monotonically increases with $z_m$ as shown by the 3 upper panels in Fig.(\ref{fig:s2npred}),
although the differential contribution, $d(S/N)/dz$, drops for sources at $z \simgt 0.4$
(see Afshordi 2004 for an analytic account of this effect).

In particular, for our baseline survey, SDSS, and our fiducial \LCDM model,
we estimate $S/N = 3.8$, what is in good
agreement with simulations in configuration space (see Table \ref{tb:ts2n}).
 As we sample a wider range of the ISW signal
in redshift, the S/N raises by $\sim 60\%$ when we increase the survey depth by a factor $\sim 2$ to match
the depth of the DES-like survey. However, there is little gain in ISW detection significance when
combining DES+VISTA, as the S/N only increases by an additional $5\%$ with respect to the DES survey.
For comparison, we also show the case of what we shall call DES+VISTA NARROW survey. This survey
has a Gaussian distribution of sources around $z_m=1$,
but with a narrow width, similar to that of SDSS above ($\sigma_z = 0.17$).
In this case, the high redshift population of sources brings a poor added value to the baseline survey (SDSS) by improving the S/N by only $10\%$. As shown in Table \ref{tb:ts2n2}
these conclusions vary somewhat for different values of $\Omega_{DE}$.

We point out that in these estimations we have ignored the lensing magnification
bias contribution (see Loverde, Hui \& Gazta\~naga 2006) which could be important for $z>1$.

\subsection{$\chi^2$ estimation}
\label{sec:chi2}

\begin{figure*}
	\centering
{\epsfysize=4cm \epsfbox{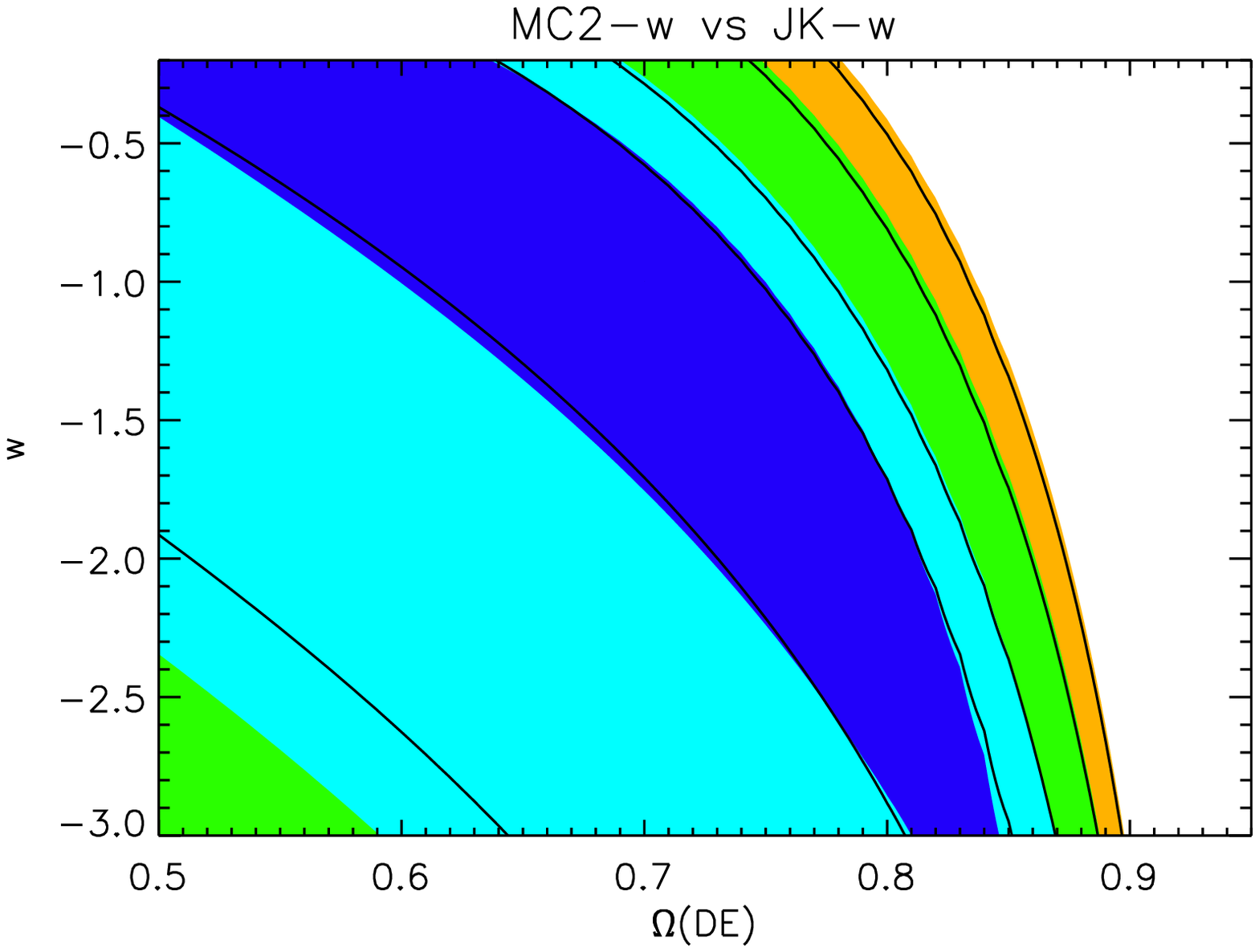}}
	\centering
{\epsfysize=4cm \epsfbox{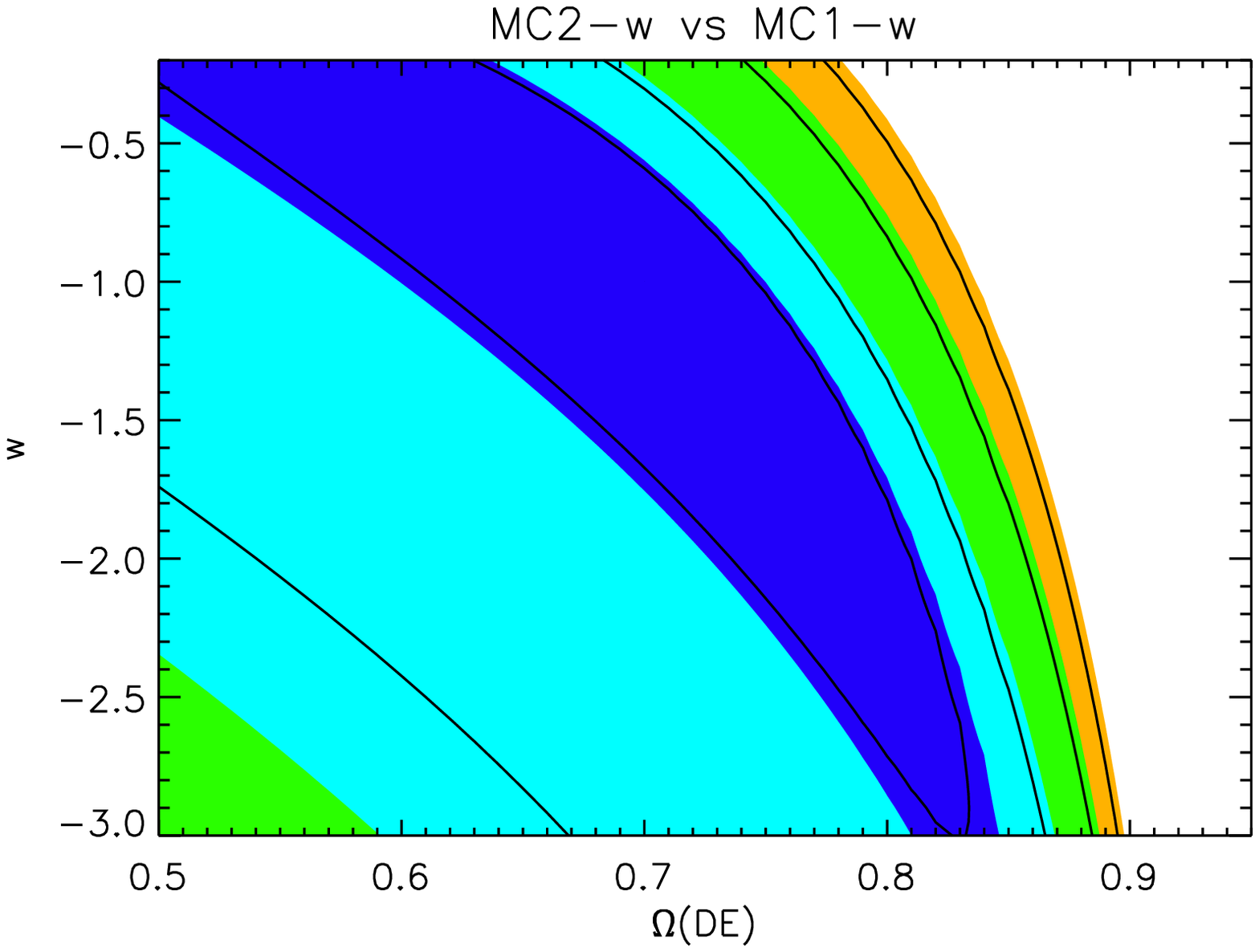}}
	\centering
{\epsfysize=4cm \epsfbox{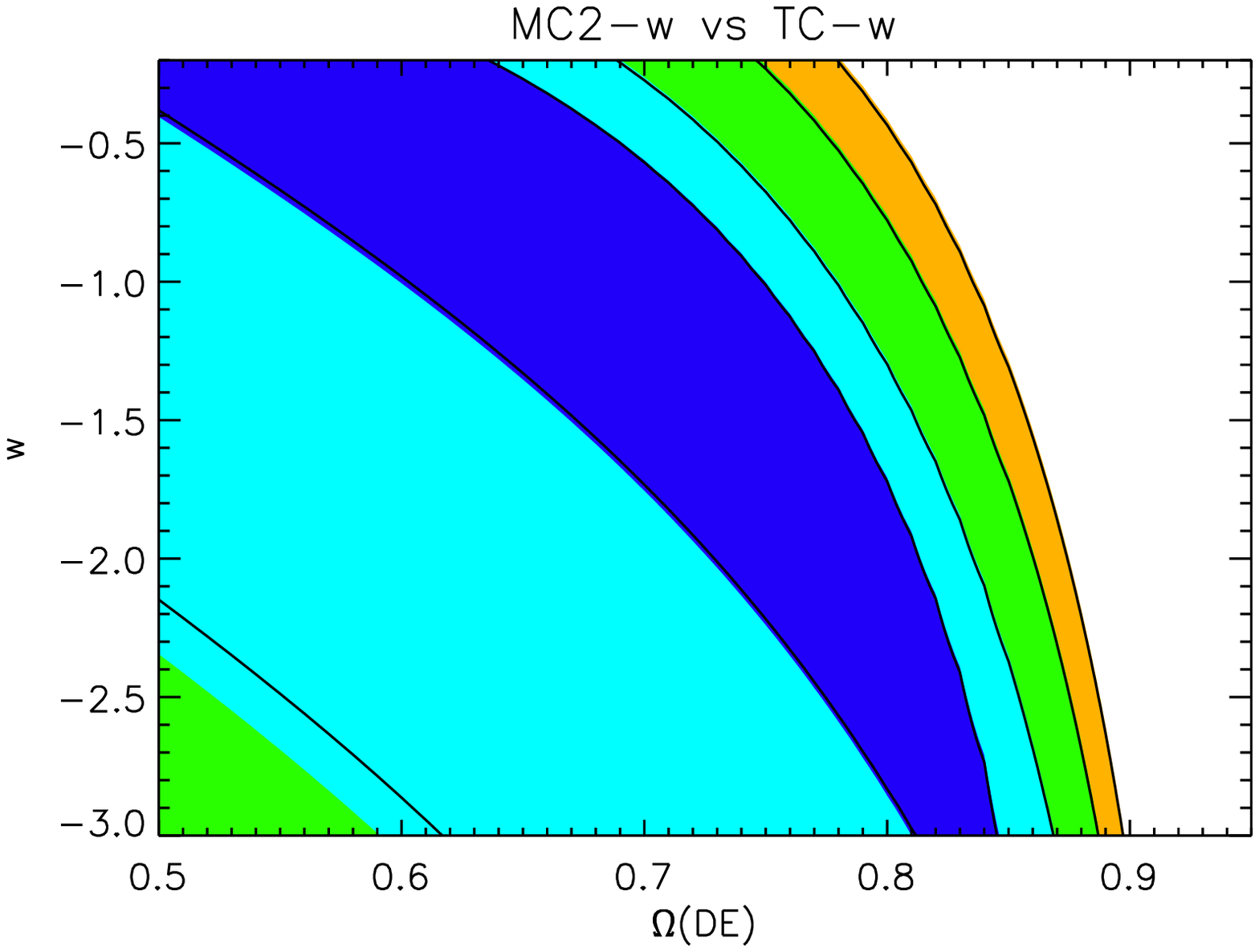}}	
	\centering
{\epsfysize=4cm \epsfbox{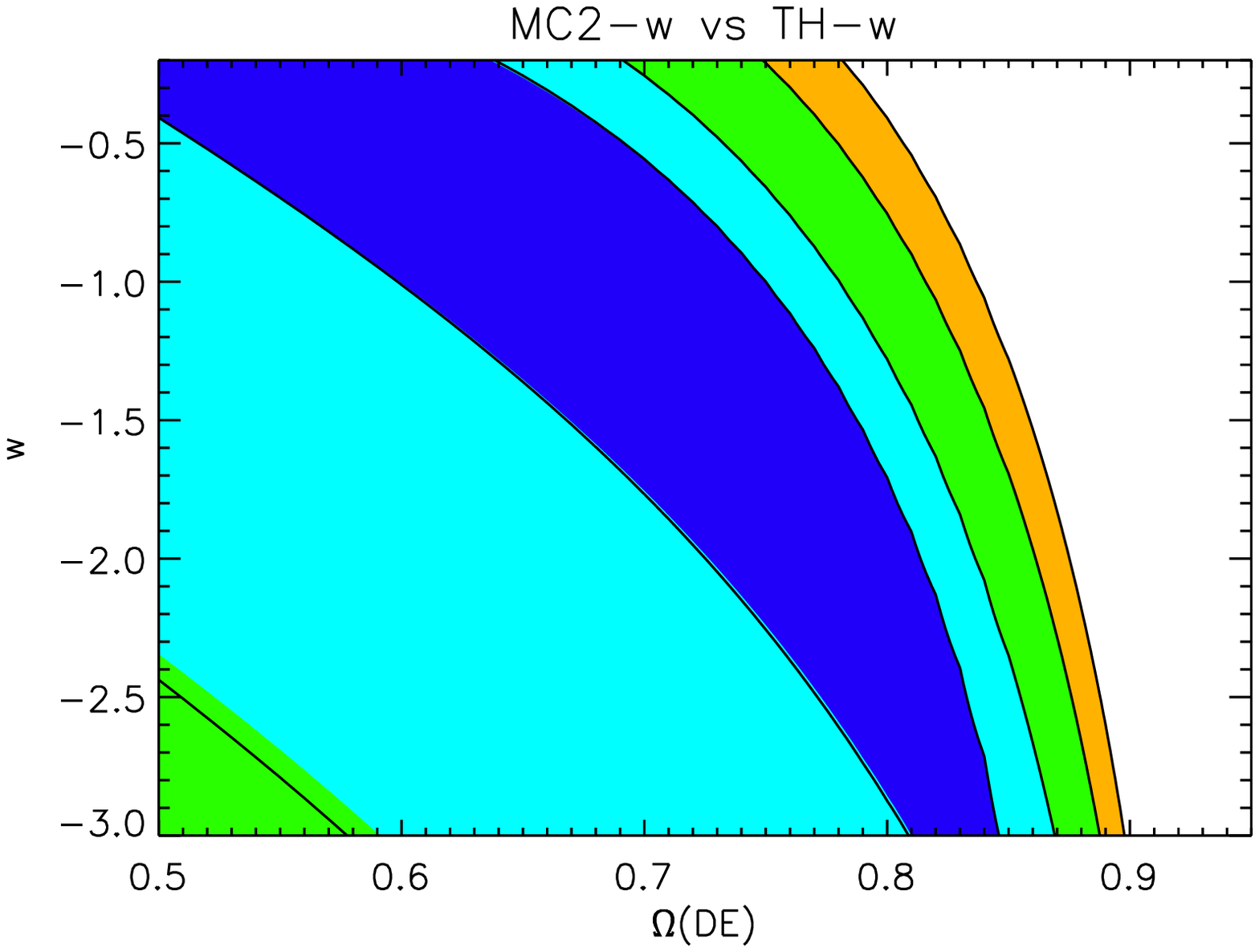}}	
	\centering
{\epsfysize=4cm \epsfbox{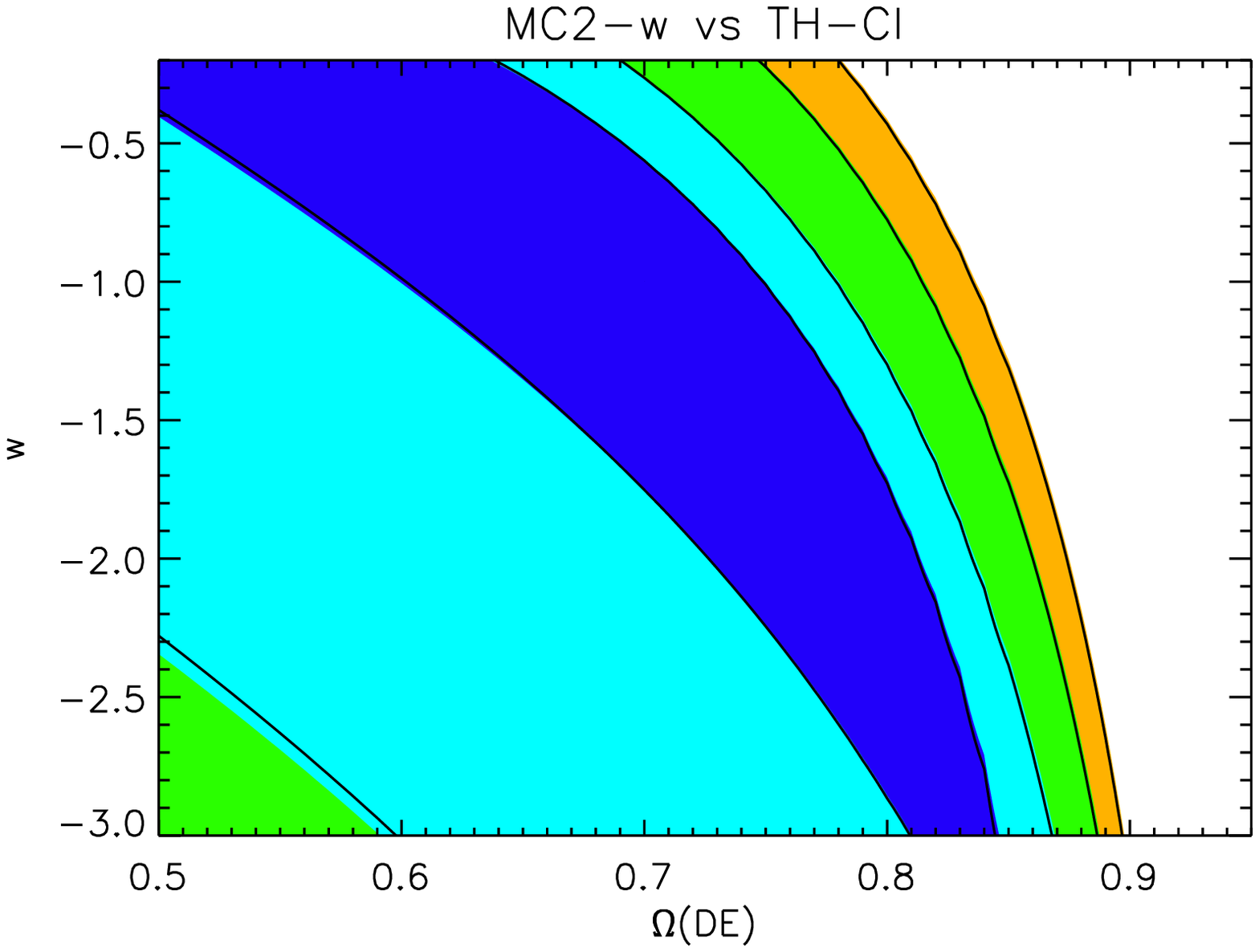}}
	\caption{$\chi^2$
	 contours from (MC2-w) simulations
	  (in color) compared to the other methods (solid line):  JK-w, MC1-w, TC-w, TH-w, TH-$\Cl$ as labeled in each panel. All cases correspond to 10\% of the sky.  The contour levels are: 0.25, 1., 4. and 9.}
	\label{fig:chi10}
\end{figure*}

\begin{figure*}
	\centering
{\epsfysize=4cm \epsfbox{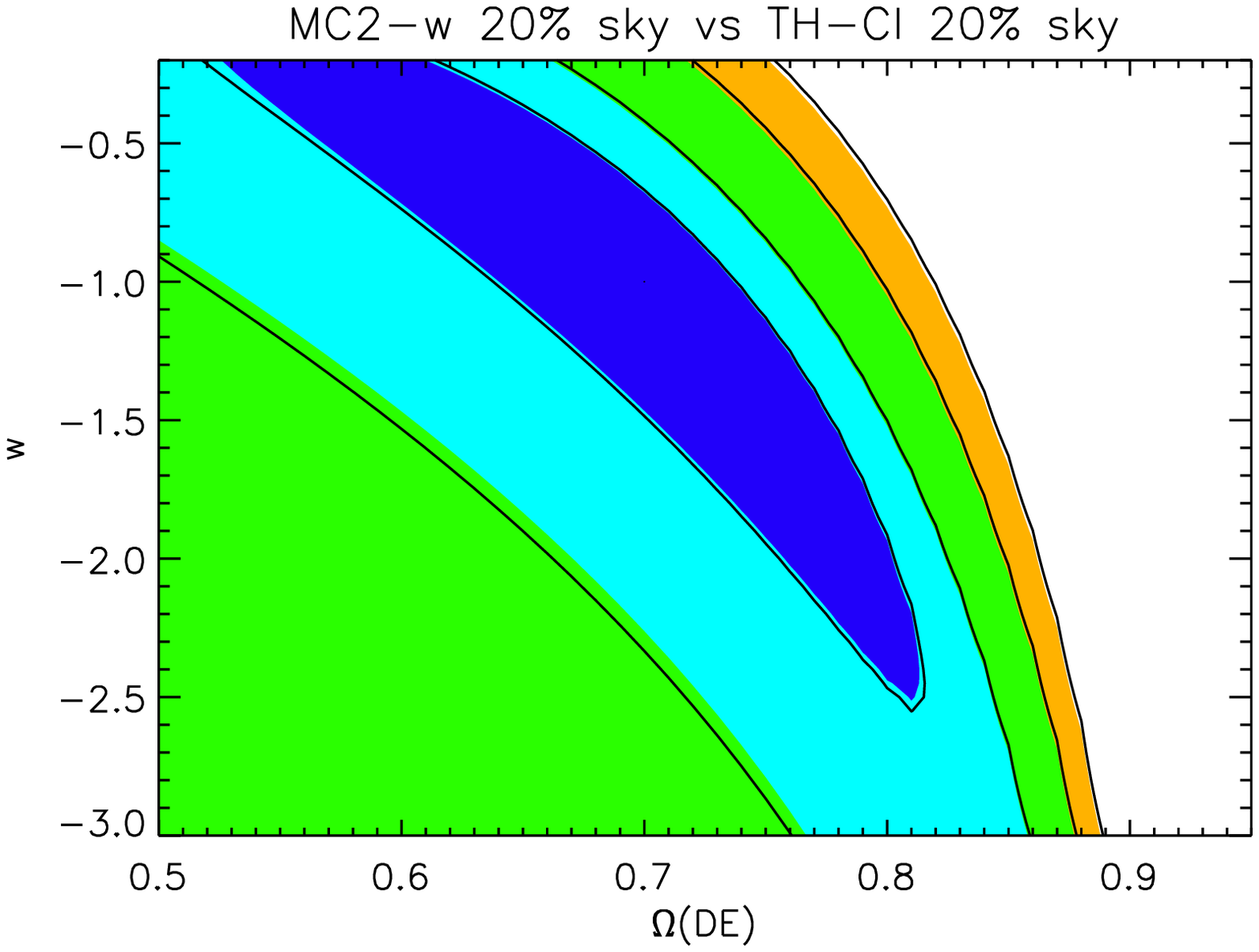}}
	\centering
{\epsfysize=4cm \epsfbox{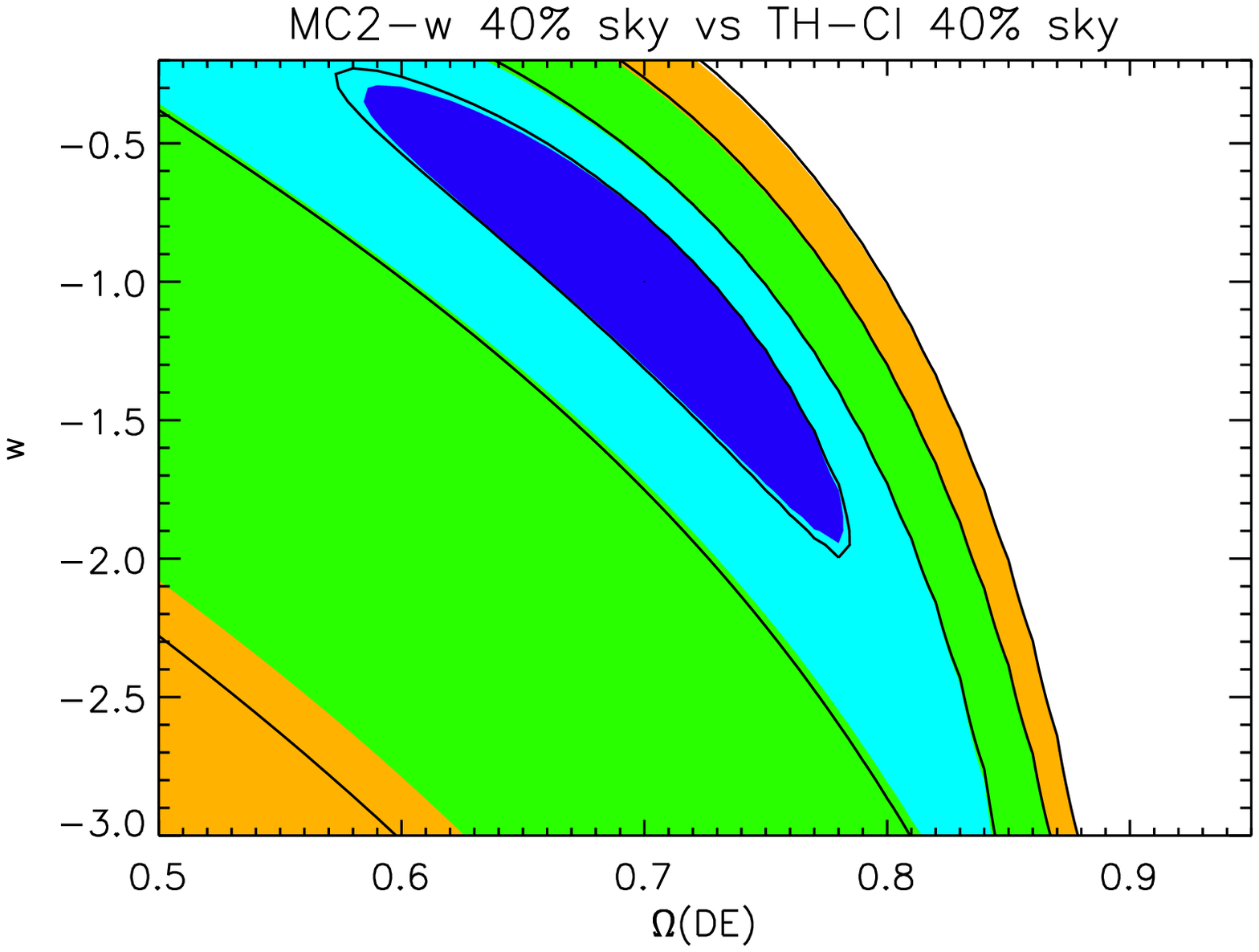}}
	\centering
{\epsfysize=4cm \epsfbox{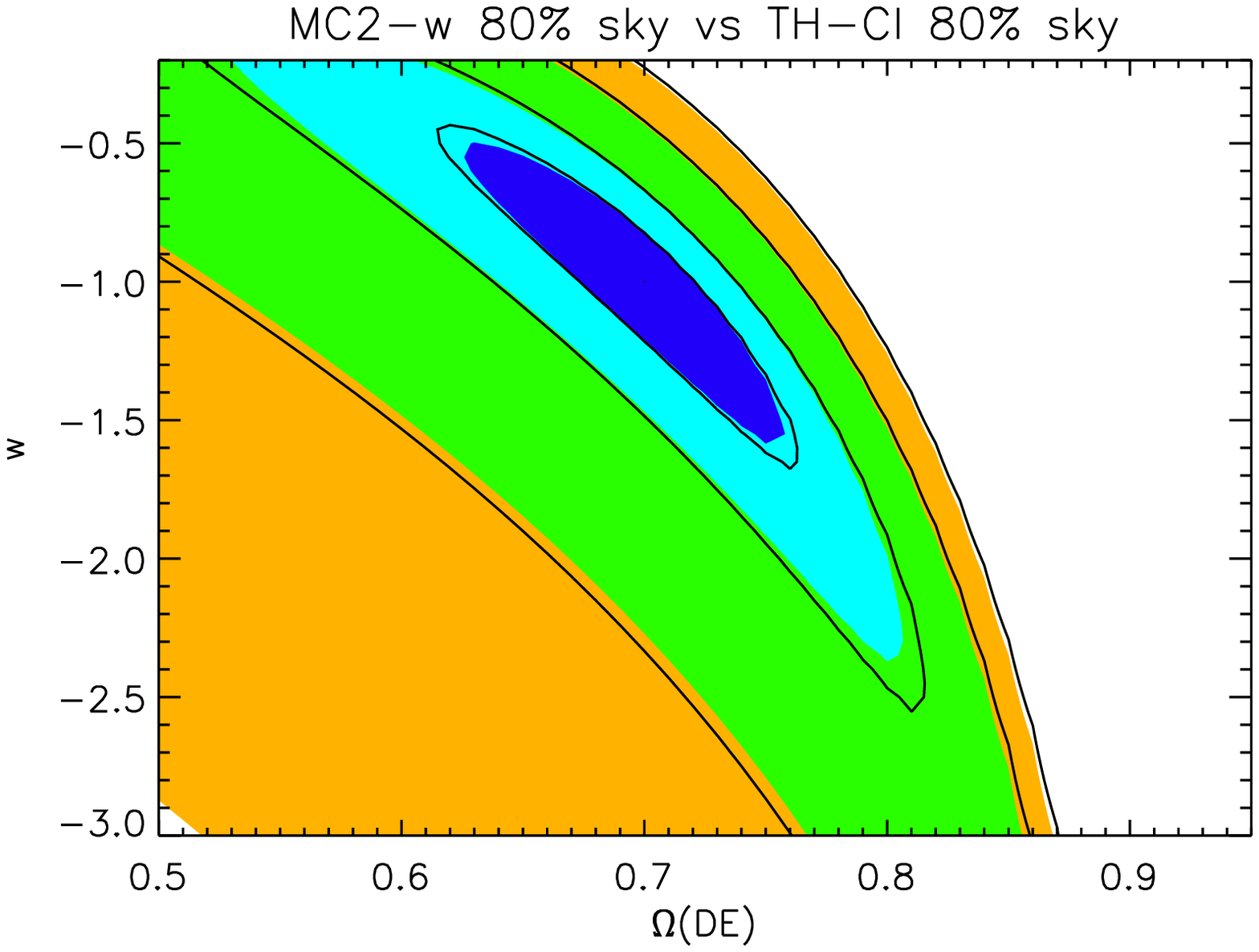}}	
	\caption{Same as Fig.\ref{fig:chi10}, but here each panel compares
	 to the method TH-$\Cl$ for 20\%,
40\% and 80\% of the sky.}
	\label{fig:chiother}
\end{figure*}

\begin{figure*}
	\centering
{\epsfysize=4cm \epsfbox{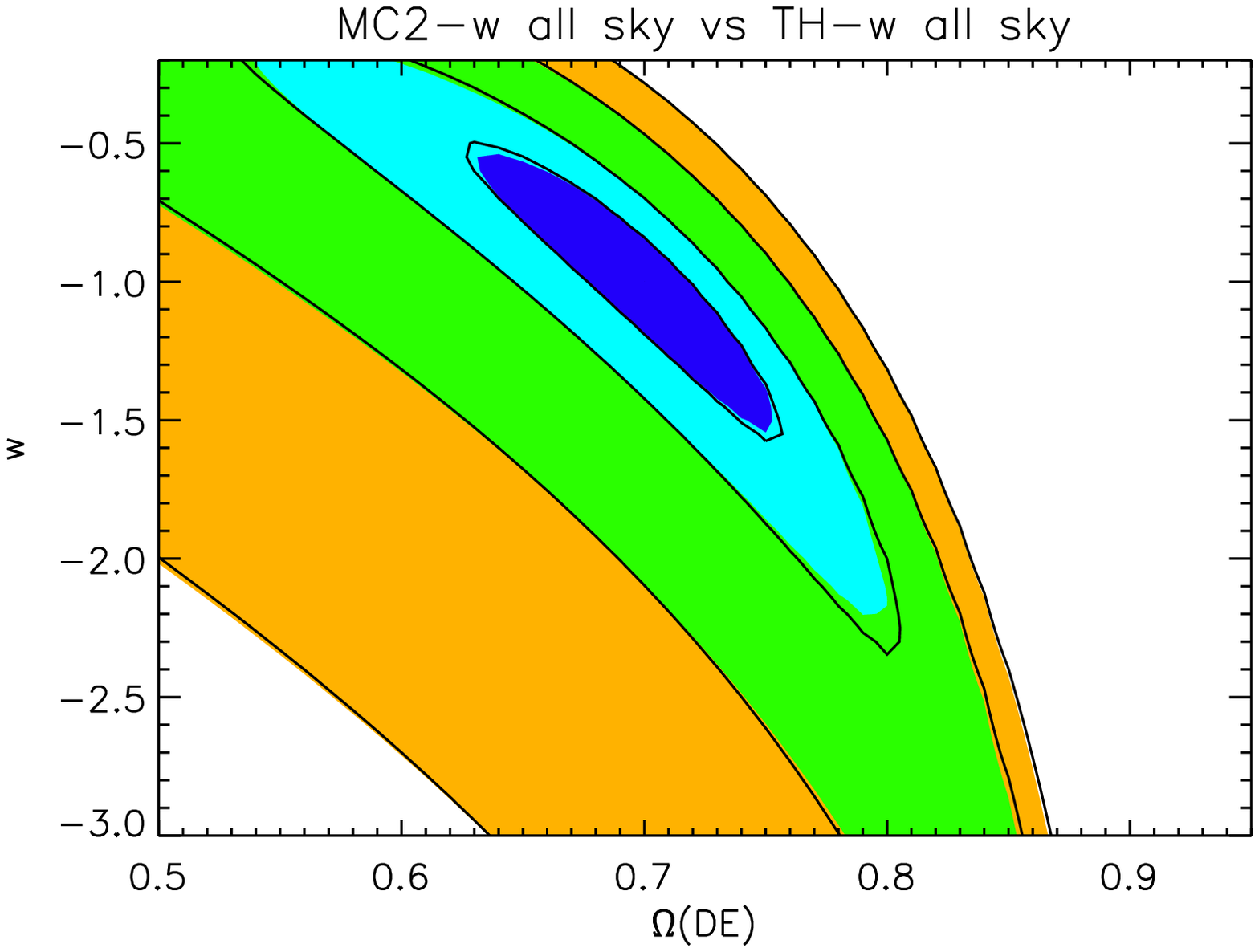}}	
	\centering
{\epsfysize=4cm \epsfbox{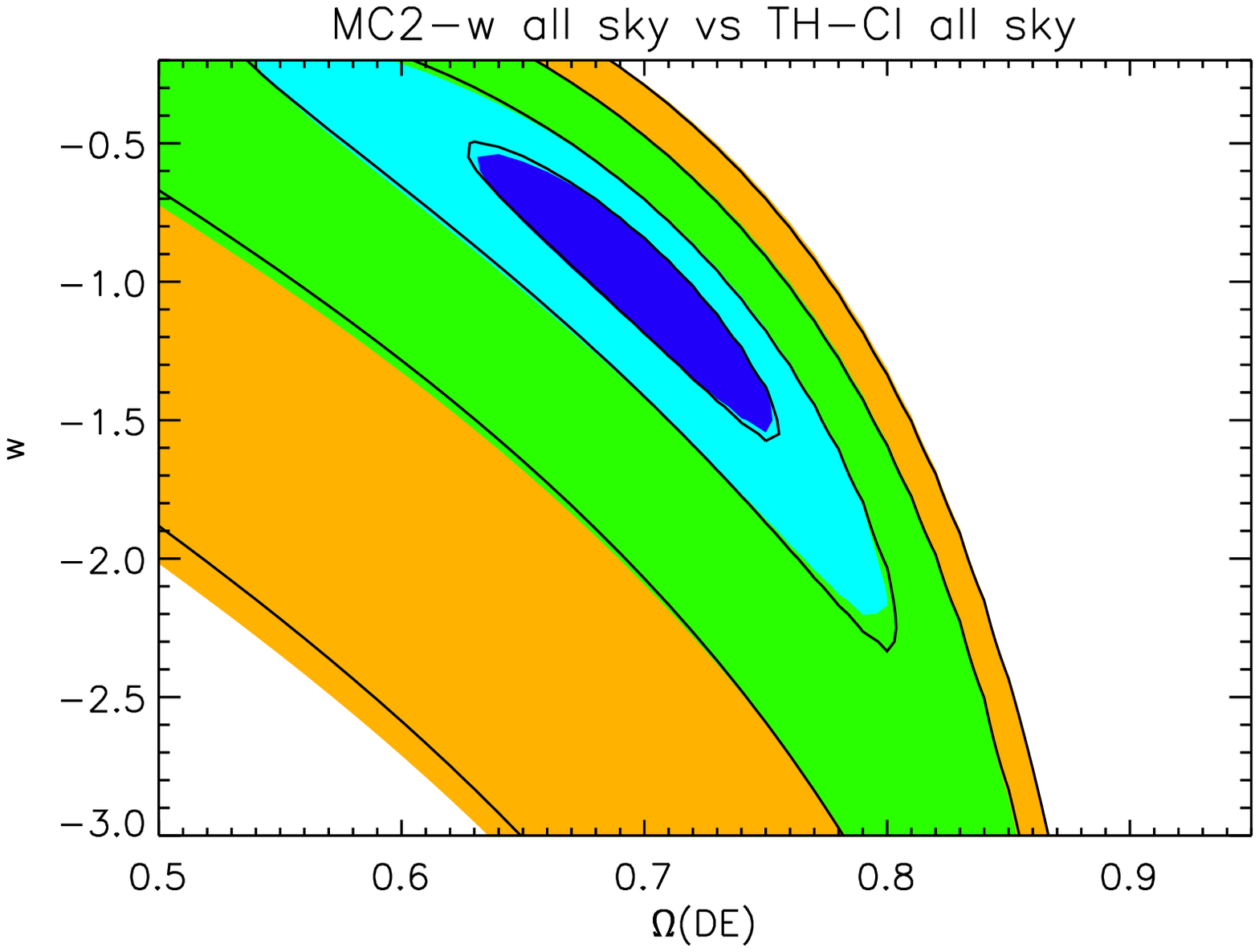}}
	\caption{Same as Fig.\ref{fig:chi10} for all sky maps. Here solid lines correspond
	 to methods TH-w (left panel) and TH-Cl (right panel).}
	\label{fig:chiall}
\end{figure*}

We shall discuss below to what extent the choice of covariance matrix estimation method
affects cosmological parameter constraints. This is specially relevant because current ISW detection
significance levels are still rather poor (i.e, at the 4-$\sigma$ level at most, see Cabre etal 2006)
and the practical implementation of methods might yield noticeably different results.

We shall compare the methods described in \S\ref{sec:cov}, whereas
the fiducial model is the one implemented in the simulations.
Our significance levels are derived from a $\chi^2$ statistic:

\begin{equation}
\label{eq:chi2}
\chi^2 = \sum_{i,j=1}^{N} \Delta_i ~ \hat{C}_{ij}^{-1} ~ \Delta_j
\end{equation}
where:
\begin{equation}
\Delta_i \equiv \frac{(w_{TG}^E(\theta_i) -
w_{TG}^M(\theta_i))}{\sigma_{TG}(\theta_i)}
\label{eq:deltai}
\end{equation}
is the difference between the
"estimation" $E$ and the model $M$. We have run models for $\Omega_{DE}$ from 0.5
to 0.9 and for $w$ from -3.0 to -0.2 and we fix the estimation $E$ to be our
fiducial model $\Omega_{DE}=0.7$ and $w=-1$ which was input in the simulation.
The size of the resulting confidence level contours depends implicitly on the
best-fit model (i.e, the fiducial model) by construction.

In each case, the error used is the one obtained from the simulations
(for cases MC2-w, MC1-w, JK-w, MC2-$\Cl$)
or from the theoretical estimator (TH-w and TH-$\Cl$) for the given
fiducial model. That is, the errors are not varied as we sample parameter space
in the $\chi^2$ estimation. This allows a direct
comparison on the contours when using different covariance matrix estimators.

Results are shown in  Fig.\ref{fig:chi10},\ref{fig:chiother} and \ref{fig:chiall}. In the
different figures we compare the real space MC (MC2-w) result
 (colored contours) with the other methods (contours traced
by solid lines). Contours from different methods agree remarkably well:
it does not depend neither on which space we compute the errors
and covariance (real or harmonic space), nor in the portion of the sky used. 
We have checked that small contour differences
are compatible once we take into account 
uncertainties in the errors, as shown in Fig.\ref{fig:varrealw}.
Moreover using a diagonal approximation for the $\Cl$ covariance matrix
to infer the covariance in real space (through the Legendre transform in Eq.\ref{eq:errorW}), works for a
small portion of the sky surprisingly well.
As explained in section $\S$\ref{sec:clvar}, when we use the theoretical error in $\Cl$ space (TH-$\Cl$)
for real data, we should use a bin of width of $\Delta l$ that varies with the portion of the sky.

\subsection{Best fit model}

\label{sec:cosm}

\begin{figure}
	\centering
{\epsfysize=6cm \epsfbox{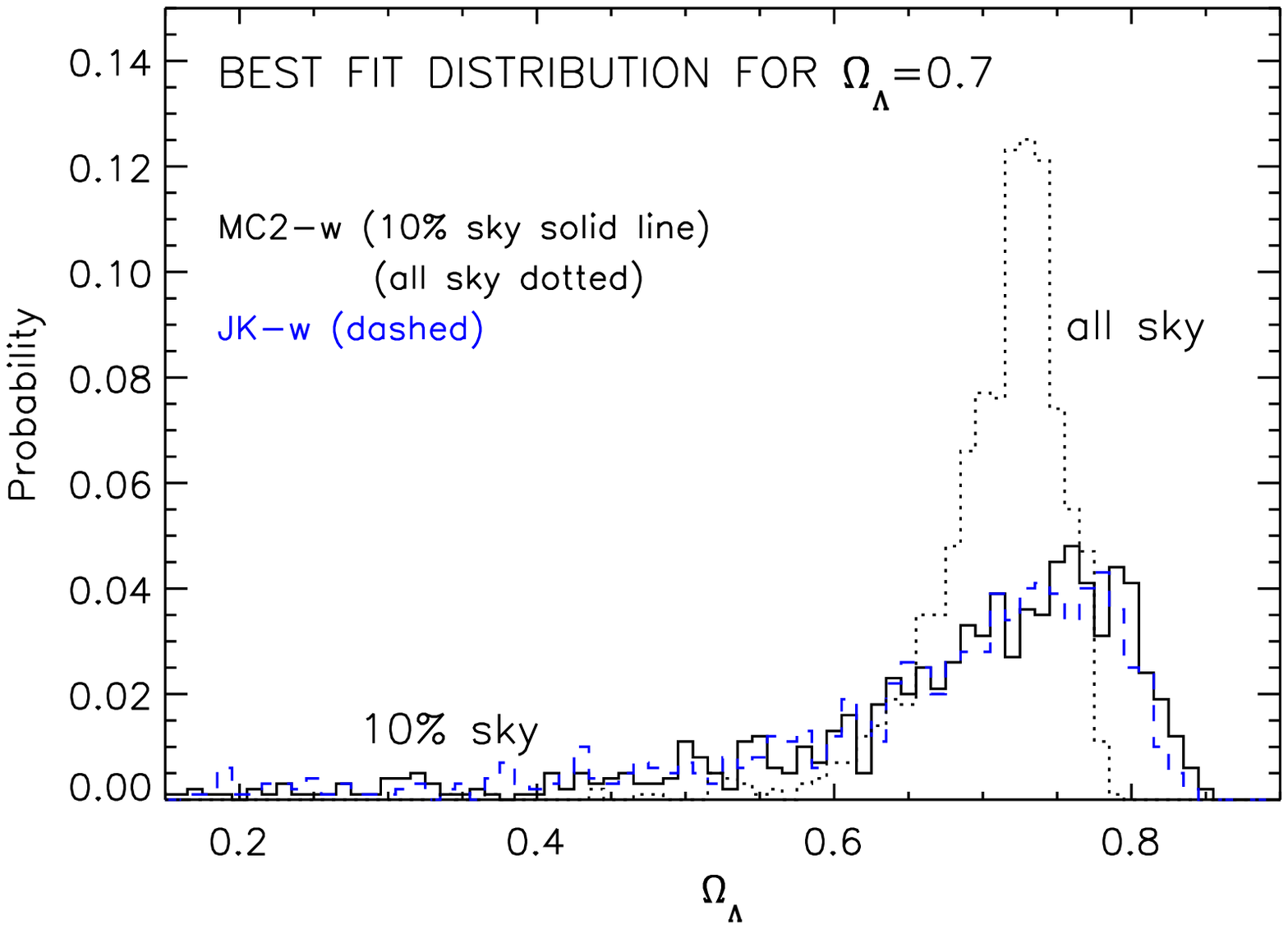}}
	\centering
{\epsfysize=6cm \epsfbox{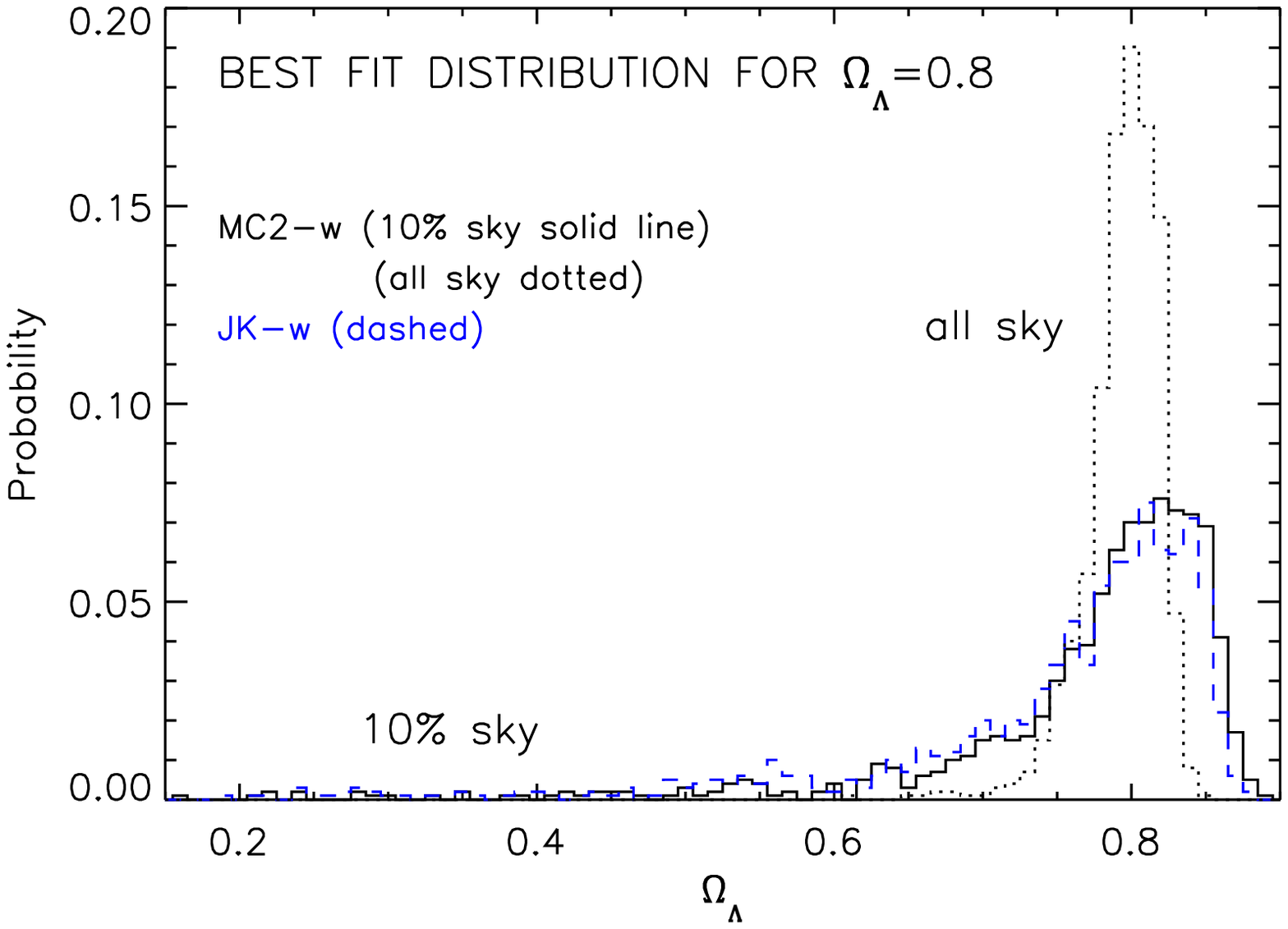}}
	\caption{Distribution of best fit values for $\Omega_{DE}$ fixing $w=-1$ and all the other parameters to the fiducial model. For JK (dotted lines), we use the JK-w error obtained for each simulation. For the MC2 simulations (solid line), we use a fixed error obtained from  the
	dispersion in the simulations.}
	\label{fig:pdfmin1}
\end{figure}

In this section we investigate how the error method used
affects  the best fit estimation of cosmological parameters. We fix all the parameters as in the fiducial model,
except for $\Omega _{DE}$. We focus on the case of the angular 2-point correlation
$w(\theta)$ and compare results for JK-w to those for MC2-w.
We do a $\chi^2$ fit of the correlation from each single simulation, which is used as
the "E" estimator in Eq.\ref{eq:deltai}. This follows what is done with real data where the
observations correspond to a single realization. We can make a distribution for all the best fit values of $\Omega _{DE}$
that we obtain from each realization, which is shown in Fig.\ref{fig:pdfmin1}.
The error and covariance used in the fit  is in one case
the JK-w obtained using this single simulation (dashed line in Fig.\ref{fig:pdfmin1}) or
the MC2-w calculated from all the
simulations (continuous line). Despite these differences there is an excellent
agreement between the JK-w and the MC2-w results.

We see that the  distribution of best fit values is biased towards higher values
than the underlying fiducial model value $\Omega _{DE}=0.7$. In partiuclar,
the distribution of best-fit values is skewed, showing a long tail of values smaller than the input model.
This is due to the fact that contours in $\chi^2$ (and in the S/N) are not symmetric. The reason for this
is the nonlinear mapping between values of  $\Omega _{DE}$  and the amplitude of $w(\theta)$.
When the errors are large, this non-linear mapping transforms an approximate
Gaussian distribution (which is a good approximation for the distribution of $w(\theta)$)
into a strongly non-Gaussian distribution in  $\Omega _{DE}$.
When the errors are smaller, as happens for larger  $\Omega_{DE}$, the
mapping between  $w(\theta)$ and  $\Omega _{DE}$, is better approximated
by a linear relation which results in a more Gaussian distribution. Thus,
if we have  small enough  errors, this bias is negligible, as we can see
for the all-sky case shown  by the dotted lines in Fig.\ref{fig:pdfmin1}.

\section{Summary \& Conclusions}   
\label{sec:discuss}   

We have run a large number of pairs of sky map simulations, that we call
MC2. Each pair  is a stochastic realization of an auto and cross correlation signal, that we
input to the simulation, what we call the fiducial model. We have
focused our attention in testing the galaxy-temperature cross-correlation,
so each pair of simulations correspond to a CMB and a galaxy map. For the
fiducial model we take the current concordance \LCDM scenario. We have
run simulations for different values of $\Omega_{DE}$ and have tested
maps with different fractions of the sky.
We have concentrated on the case $\Omega_{DE}=0.7$ and $f_{sky}=0.1$ which broadly matches
current observations and results in  large errors ($> 50\%$).

We are interested in error analysis/forecast and significance estimation.
We calculate the correlation between  maps and use the different realizations
to work out the statistics. We then
compare the results to the different approximations that have been used so
far in the literature. One of the approximations, that we call MC1, uses
montecarlo simulations for the CMB maps with a fixed (observed) galaxy map (ie with
no cross-correlation signal or sampling variance in the galaxies).
We test a popular harmonic space prediction, that we shall call
TH (Theory in Harmonic space).  We also test Jack-Knife (JK) errors which
uses sub-regions  of the actual  data  to calculate the dispersion in our estimator.
Finally,  we introduce a novel error estimator in real space, that call TC
(Theory in Configuration space). For both models and simulations  we have assumed
that the underlaying statistics in the maps is Gaussian.
Our main results can be summarized as follows:

\renewcommand{\labelenumi}{\alph{enumi}$)$ }
\begin{enumerate}

\item The number of simulations needed for numerical convergence (to within  $\simeq 5\%$ accuracy)
in the computation of the covariance matrix is about 1000 simulations (see \S\ref{sec:convergence}).

\item Diagonal errors in $w(\theta)$ are very accurate in both TH and TC
approximations for all sky maps. This is shown in the bottom lines of Fig.\ref{fig:varreal}.
For maps with different fraction of the sky
$f_{sky}<1$, the agreement is also good on small scales ($\theta <20$ deg) as can be seen in
Fig.\ref{fig:plotfsky}.

\item Even for a map as wide as $10\%$ of the sky, the survey geometry starts to be important
for errors in the cross-correlation above 10 degrees. This is shown in simulations
as a sharp inflection that begins at 30 degrees in Fig. \ref{fig:varreal} (solid line) .
Our new TC method predicts well
this inflection, while the more traditional TH method totally misses this feature.

\item If we only use one single realization for the galaxies
(MC1) the error seems to be systematically underestimated by about $10\%$ on all scales.
 This bias is expected as we have neglected the variance in the galaxy field and the
 cross-correlation signal.

\item The JK errors 
do quite well within 10\% accuracy on all scales, including the
larger scales where boundary effects start to be important
  (see triangles in  Fig.\ref{fig:varreal}).

\item The dispersion in the error estimator (error in the error) for individual realizations
is of the order $~20\%$ (see Fig.\ref{fig:varrealw}). This uncertainty is inherent to the JK
method, because one uses the observations (a single realization) to estimate errors. But
it is also implicit in other methods because our knowledge of the models is limited
by the data and can be thought of as a ``sampling variance error''.

\item S/N (see Fig.\ref{fig:s2n}) and parameter estimation (see Fig. \ref{fig:chiall}
and Table \ref{tb:ts2n}) are equivalent when we do the analysis
in configuration and harmonic space. This was expected for all sky maps, but it is not trivial
for partial sky coverage (see comments below).

\item It is possible to  propagate errors and covariances from $\Cl$ to $w(\theta)$
(harmonic to configuration space) using Eq.(\ref{eq:errorW}). Starting from a diagonal (all sky)
covariance matrix in $\Cl$,  the resulting covariance matrix in $w(\theta)$ is quite accurate
as compared to direct estimation from simulations.

\item The above  propagation also works
well for a map with a fraction $f_{sky}$ of the sky, by just scaling the $\Cl$ errors
by a factor $1/\sqrt{f_{sky}}$  respect to all the sky. This is surprising because for $f_{sky}<1$
the covariance matrix in $\Cl$ is no longer diagonal (see Fig.\ref{fig:covCl})
and the actual measured $\Cl$ errors in simulations do not simply scale with
 $1/\sqrt{f_{sky}}$ (see Fig.\ref{fig:varcl}).
Thus,  Eq.(\ref{eq:errorW}) should not be valid.
We believe that this works because the two effects compensate.
There is a transfer of power from diagonal to off-diagonal elements of the covariance matrix
which for the scales of interest (smaller than the survey area) seems to corresponds
to a rotation that somehow does not affect the final errors from Eq.(\ref{eq:errorW}).

\item If we want to use the popular TH approach in Eq.(\ref{eq:errorCl})
with $f_{sky}<1$ we need to bin the $\Cl$ data in multipole bands of width $\Delta\l$.
The binned spectrum has a diagonal covariance when $\Delta l$ is large enough
and the error in the binned spectrum approximately follows  Eq.(\ref{eq:errorCl2}).

\item When the errors are large (i.e., for partial sky coverage and \LCDM models with
not so large $\Omega_{DE}$)
there is a signficant bias in the distribution of the recovered best-fit
values of $\Omega_{DE}$, as shown in Fig.\ref{fig:pdfmin1}. This is because of the non
linear mapping between $\Omega_{DE}$  and the amplitude of $w(\theta)$.

\item S/N forecasts for future surveys, shown in Fig.\ref{fig:s2npred} and Table \ref{tb:ts2n2}, strongly depend on the fiducial model used.
For example, an all-sky survey with broadly distributed sources around a median redshift $z_m=1$ and $\Omega_{DE}=0.8$ can detect the ISW effect with a S/N$\simeq 11$.

\end{enumerate}

What method should be used when confronted with real data? Running realistic simulations seems
the best approach, but is very costly because we need of order 1000 simulations for each model
we want to explore. The theoretical modeling of errors seems quite accurate and is much faster
to implement. The main advantage of the JK approach is that the errors are obtained from the same data in a model independent way. This is important because real data
could  surprise  our prejudices and also  because, in the ISW case,  the errors are
very large and the data can accommodate different models.

As an example, consider the analysis of Cabre etal (2006) who recently
cross-correlated the SDSS-DR4 galaxy with the WMAP3 CMB anisotropies.
Using the JK approach with $w(\theta)$ they estimate  a $S/N \simeq 3.6$ for the $r=20-21$ sample, which has a mean redshift of $z_m \simeq 0.33$.
These numbers are high compared  with the values in Table \ref{tb:ts2n2}
for $z_m=0.33$ which for $f_{sky}=0.13$ gives a low $S/N \simeq 2$, even for
$\Omega_{DE}=0.8$.   The dominant contribution to the S/N in Table \ref{tb:ts2n2}
scales as  $\Cl^{TG}/\sqrt{\Cl^{GG}}$ (ie see Eq.\ref{eq:s2nforecast2})
and is therefore independent of bias, but depends on $\sigma_8$.
We have noticed that in fact the actual measured values of  $\Cl^{TG}/\sqrt{\Cl^{GG}}$
in the SDSS DR4-WMAP3 maps are almost a factor of 2 larger than the
values in the concordance $\Omega_{DE}=0.8$ ($\sigma_8=0.9$, $n=1$, $\Omega_{\nu}=0$,
$\Omega_B=0.05$, $h=0.7$) model. This explains the discrepancy in
the $S/N$ and illustrates the danger of blindly using  theoretical errors that are
model dependent. The discrepancy of the concordance model with the SDSS4-WMAP3
measured values of
$\Cl^{TG}/\sqrt{\Cl^{GG}}$ is not very significant once we account for sampling
errors (less than 3-sigma), but it could be an indication of new physics that make
the $P(k)$ normalization higher  than the concordance model, ie deviations
in $\sigma_8$, spectral index $n$,  neutrinos $\Omega_{\nu}$, etc, away
from the fiducial model we are considering.

We have also shown that it is possible to use the other theoretical models
(ie TC and TH) to make model independent error predictions from observations.
Contrary to all other methods,  the JK approach does not assume Gaussian
statistics, but its accuracy could depend on the  model or the
way it is  implemented (ie shape and number of sub-regions). We conclude that
to be safe one needs to validate the JK method with simulations, but there is
no reason apriori to expect that this method is inaccurate.

In summary, we have presented a detailed testing of  different error approximations
that have been used in the literature,   both in configuration and harmonic space.
Contrary to some claims in the literature (see Introduction), 
we show that the different errors (including the JK method)
are equivalent within the sampling uncertainties.
By this we mean not only that the error and covariance are similar but also that they
produce very similar signal-to-noise (S/N) and recovery of cosmological parameters.

\section*{Acknowledgments} 
   
We acknowledge the support from Spanish Ministerio de Ciencia y   
Tecnologia (MEC), project AYA2006-06341 with EC-FEDER funding
and research project 2005SGR00728 from  Generalitat de Catalunya.
AC and MM acknowledge support from the DURSI department of the Generalitat
de Catalunya and the European Social Fund. PF acknowledges support
from the Spanish MEC through a Ramon y Cajal fellowship. 
This work was supported by the European Commission's ALFA-II programme
through its funding of the Latin-American European Network for
Astrophysics and Cosmology (LENAC). We also would like to thank the
hospitality of Instituto Nacional de Astrofisica, Optica y Electronica (INAOE, Mexico),  
Galileo Galilei Institute for Theoretical Physics (Florence, Italy) and
the Center for Cosmology and Particle Physics (NYU, USA).


\appendix  
\onecolumn

\section{Covariance matrix and errors in configuration space}

\subsection{The estimator}

Consider two fields in the sky $A(q)$, $B(q')$ which correpond to one
realization of the universe. We want to estimate the true two point
cross-correlation function of the universe $w_{AB}$ by averaging over the sky in the
survey area $S$. The estimator is

\begin{figure*}
\includegraphics[width=50mm]{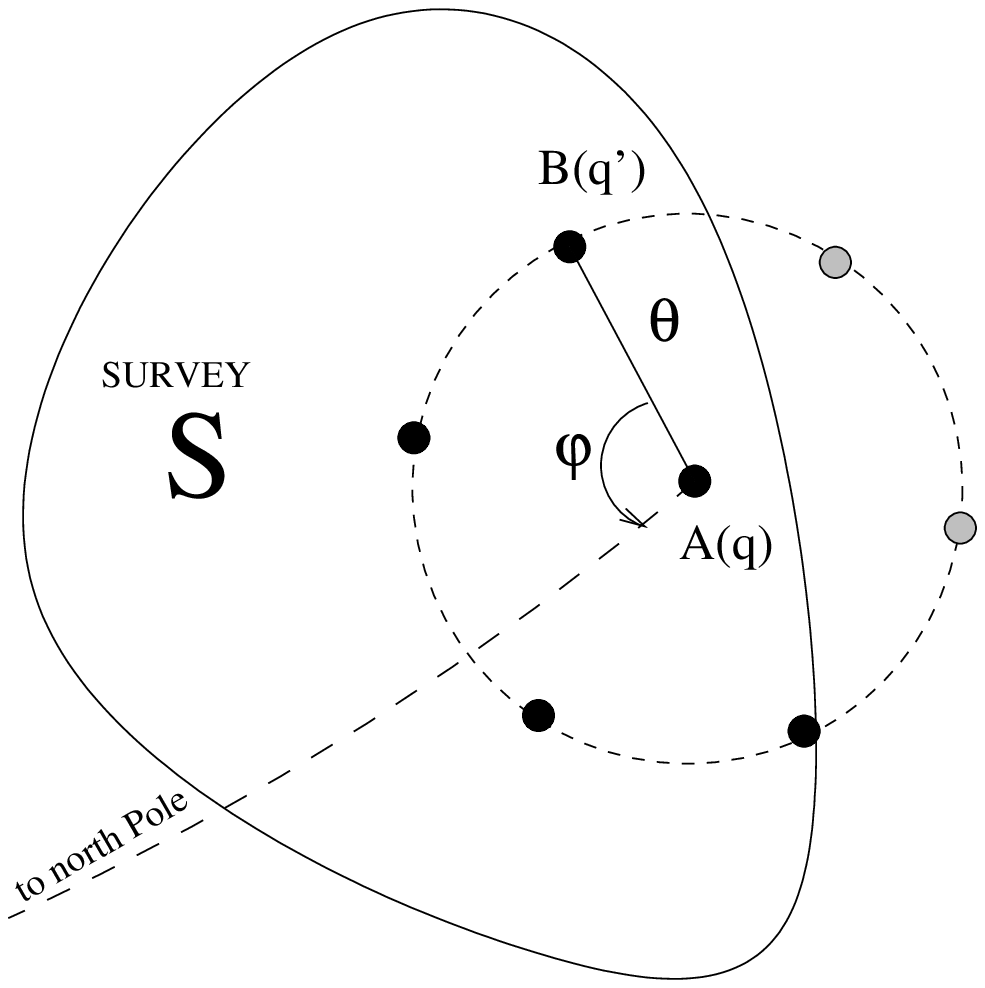}
\caption{Representation of the $\widehat{w}_{AB}(\theta)$ estimator.
The product AB is averaged in an ordered way over all pairs of
$\theta$-separated points that belong to the survey area (black points).}
\label{surveySn}
\end{figure*}

\beq{estimator}
\hw_{AB}(\theta)=<A(q)B(q')\arrowvert_{\hat{qq'}=\theta}>_{S}
\eeq
where we average over all pairs separeted by an angle $\theta$
and $\theta + \Delta \theta$ in the survey region.
This can  be put in an integral form

\beq{integralestim}
\hw_{AB}(\theta) = {1\over S_N}\int_S dq dq'\arrowvert_{\hat{qq'}=\theta} A(q) B(q')
                 = {1\over 4\pi 2\pi \sin{\theta} \Delta\theta P(\theta) }
                    \int_S dq \int_0^{2\pi} \sin{\theta} \Delta\theta d\varphi A(q) B(q+\theta(\varphi)) D(q,\theta,\varphi)
\eeq

where $S_N$ is the normalization factor and the integral is over $dq\in S$ and $d\varphi \in (0,2\pi)$.
As it is illustrated in Fig (\ref{surveySn}), we integrate all the pairs
in an ordered way. First, we fix a point $q$ and sum over all the $\theta$-separated
pairs related to this point, moving around $\varphi$. Since not
all the points in the sky $\theta$-separated from $q$ belong to the survey,
we introduce a selection function $D(q,\theta,\varphi)$ which is one if the second point
belong to the survey and zero otherwise. We perform this operation in each point
of the survey. The origin of $\varphi$ is not relevant, it could be taken for
instance as the direct angle between the $\theta$-pair and the geodesic line
between the $q$ point and the pole. The second integration is over all the points
in the survey.

The normalization factor $S_N$ is a measure of the number of $\theta$-pairs
allowed by the survey, which depends on $\theta$. For an all
sky survey, $S_N$ is $4\pi 2\pi \sin{\theta} \Delta\theta$. The geometry of
the survey is enclosed in a multiplicative  factor $P(\theta)$,
which is actually the ratio between the number of $\theta$-pairs in the survey
and the number of $\theta$-pairs in the whole sky, i.e., the probability
that when throwing a $\theta$-pair in the whole sky it falls into the survey.
When $\theta=0$, this probability is equal to the fraction of sky $f_{sky}$ covered by the survey. In an
all sky survey, $P(\theta)=1$ and also $D(q,\theta,\varphi)=1$, and
the estimator for the cross-correlation is given by:

\beq{allskyestim}
\hw_{AB}(\theta) = {1\over 4\pi 2\pi }
                    \int dq  d\varphi A(q) B(q+\theta(\varphi))
\ee

\subsection{Covariance}

In order to get the covariance, we need to relate the estimator of
the cross-correlation with the true cross-correlation value.
The true cross-correlation value $w_{AB}(\theta)$ is the average
over realizations of the estimator, (where the estimator $\hw_{AB}(\theta)$ is obtained averaging for all the $\theta$ of the sky). Due to homogeneity, $w_{AB}(\theta)$ is also
equal to average any $\theta$-pair of fixed points A and B over all the realizations.

\beq{truewab}
w_{AB}(\theta)=<\hw_{AB}(\theta)>_{realization}=<A(q_1)B(q_2)>_{realization} \quad \forall \hat{q_1q_2}=\theta
\ee
\beq{truewabprim}
\hw_{AB}(\theta)=<A(q)B(q')\arrowvert_{\hat{qq'}=\theta}>_{sky}
\ee

where $<>$ means averaging over all the realizations from now on.\\

The covariance for an arbitrary estimator is

\beq{varianca}
\begin{split}
C_{ij} \equiv  C_{\hw}(\theta_i,\theta_j)
& =  <(\hw(\theta_i)-<\hw(\theta_i)>)(\hw(\theta_j)-<\hw(\theta_j) >)>\\
& =  <(\hw(\theta_i)-w(\theta_i))(\hw(\theta_j)-w(\theta_j))>
=  <\hw(\theta_i)\hw(\theta_j)>-w(\theta_i)w(\theta_j)
\end{split}
\ee

Thus, for our cross-correlation estimator, the covariance is given by

\begin{multline}
\begin{split}
C_{ij} & =
 < {1\over S_N(\theta_i)}\int_S dq_1 dq_2  \Big|_{\hat{q_1q_2}=\theta_i} A(q_1)B(q_2)
{1\over S_N(\theta_j)}\int_S dq_3 dq_4  \Big|_{\hat{q_3q_4}=\theta_j}A(q_3) B(q_4) >
- w_{AB}(\theta_i)w_{AB}(\theta_j) \\
& =\int_S {dq_1 dq_2 dq_3 dq_4\over S_N(\theta_i) S_N(\theta_j)}
\bigg|_{\hat{q_1q_2}=\theta_i,\hat{q_3q_4}=\theta_j}
< A(q_1)B(q_2) A(q_3) B(q_4) > - w_{AB}(\theta_i)w_{AB}(\theta_j)
\end{split}
\label{variance}
\end{multline}

\begin{figure*}
\includegraphics[width=120mm]{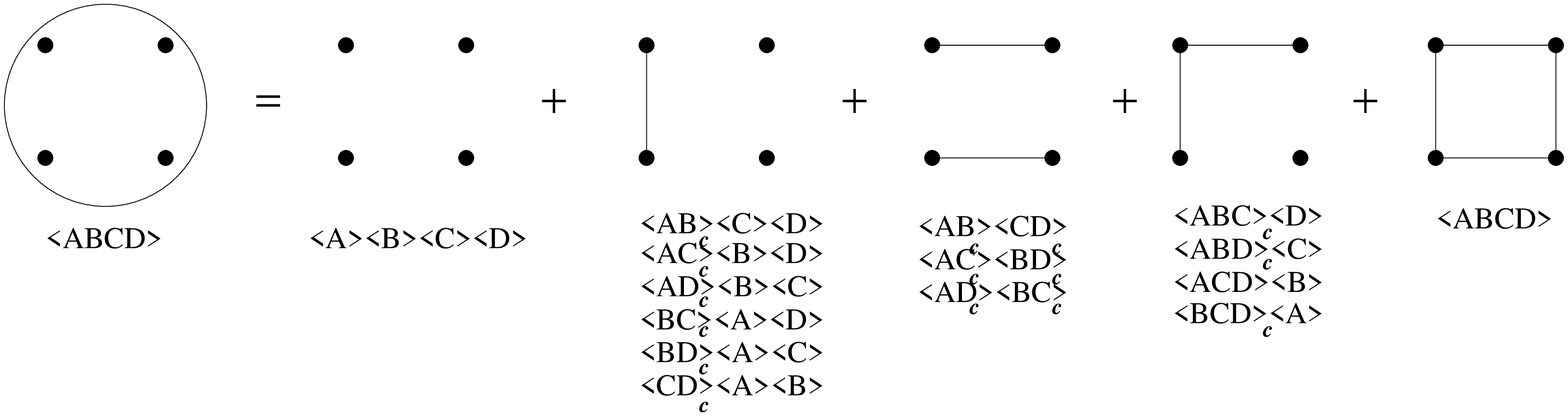}
\caption{Writing the four points moment in 15 terms of connected parts.
For a gaussian field only the two point terms will remain.}
\label{4points}
\end{figure*}

What we are doing in eq (\ref{variance}) is fixing four points
in the sky (two $\theta$-separated pairs) and average this
fixed configuration over realizations of the universe. Then we integrate over all 4-points allowed
configurations. The realization average over the four fixed points can be simplified and
expressed as a function of two-fields-correlations (shown in Fig \ref{4points})

\begin{equation}
\begin{split}
<A(1)B(2)A(3)B(4)>  = &  <A(1)B(2)><A(3)B(4)> + \\
 & + <A(1)A(3)><B(2)B(4)> + <A(1)B(4)><A(3)B(2)>
\end{split}
\label{twopairs}
\end{equation}

where $A(i)=A(q_i)$ and $B(j)=B(q_j)$,
under these two conditions:

\begin{itemize}
\item $<A(k)>=<B(l)>=0$
\item $<A(1)B(2)A(3)B(4)>_c=0$
\end{itemize}

Those are very soft requeriments. Regarding the first condition,
we can always modify a field
with non zero average to one with zero average just by substracting
its mean (sky averaged)
\footnote{here the sky average mean is the estimator for the true mean}
value at each point.  The second condition is that the fourth connected
moment is zero. This is true for a gaussian
statistics and always a very good approximation for almost gaussian fields. Note that for
fields with zero mean, the second moments and the second-connected moments are equal.

Focus for a moment in the first term of equation (\ref{twopairs}). This term has
two $\theta$-pairs that are uncoupled. The
average over realizations will give, for each pair, the cross-correlation value at
the corresponding $\theta$, i.e., $<A(1)B(2)>$ $<A(3)B(4)>=w_{AB}(\theta_i)w_{AB}(\theta_j)$.
This value is constant for each 4-point
configuration, thus when integrating this term, we still get the same result.
This uncoupled term will cancel the last term in equation \ref{variance}.
Therefore we only have to calculate two terms:

\beq{twoterms}
\begin{split}
C_{ij} & =
\int_S {dq_1 dq_2 dq_3 dq_4\over S_N(\theta_i) S_N(\theta_j)}
\bigg|_{\hat{q_3q_4}=\theta_j,\hat{q_1q_2}=\theta_i}  <A(1)A(3)><B(2)B(4)>  \\
& + \int_S {dq_1 dq_2 dq_3 dq_4\over S_N(\theta_i) S_N(\theta_j)}
\bigg|_{\hat{q_3q_4}=\theta_j,\hat{q_1q_2}=\theta_i} <A(1)B(4)><A(3)B(2)>
\end{split}
\ee

\begin{figure*}
\includegraphics[width=120mm]{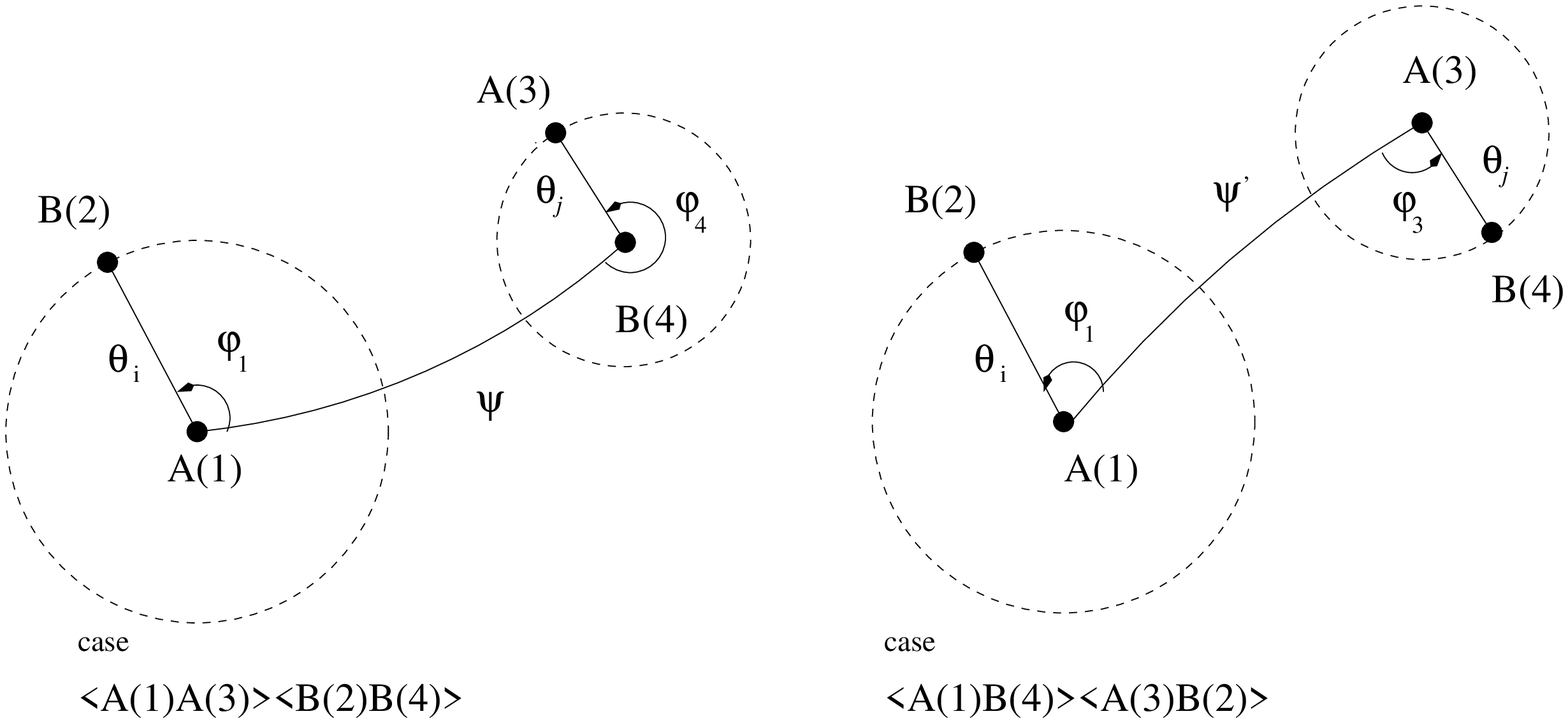}
\caption{Variables for integrating equation \ref{twoterms}.}
\label{fig_esquema}
\end{figure*}

We have to choose convenient variables to integrate, which will differ slightly for the
first and second integral. In Fig \ref{fig_esquema} is shown how to choose the variables.
The idea is the following. Let's stay in the first case where we have to integrate
$<A(1)A(3)><B(2)B(4)>$. First, we fix one point in the sky, $A(1)$.
Second, we notice that $A(1)$ is related
to $B(2)$ because they are a $\theta$-pair and is related to $A(3)$ because they have to
be cross-correlated over realizations.
Then we decide to fix distance $\psi$ to the fourth point $B(4)$.
This is the adequate way to desacoplate the integrations partly. Now the cross-correlations pairs only
depend on two angles $\psi$ and one $\varphi_i$ ,i.e, $<A(1)A(3)>=w_{AA}(\phi(\psi,\varphi_4,\theta_j))$
and $<B(2)B(4)>=w_{BB}(\phi'(\psi,\varphi_1,\theta_i))$. Here $\phi$ and $\phi'$ are the angles
between $\hat{1,3}$ and  $\hat{2,4}$ respectively.
The same idea about which variables to use in the integration is applied over the
$<A(1)B(4)><A(3)B(2)>$ term.

In order to make our deduction clearer we will follow our explanation for an all sky survey.
Afterwards we will comment on the case when only a fraction of the sky area is allowed.
For all sky survey, we easily separate the two $\theta$-pairs having

\beq{novesvariables}
\begin{split}
C_{ij}  & =
{1 \over (4\pi 2\pi)^2}  \int_{4 \pi} dq_1 \int_0^\pi 2\pi sin(\psi) d\psi
\left[ \int_0^{2\pi} d\varphi_4 w_{AA}(\phi(\psi,\varphi_4,\theta_j))\right]
\left[\int_0^{2\pi} d\varphi_1 w_{BB}(\phi'(\psi,\varphi_1,\theta_i))\right] \\
& + {1 \over (4\pi 2\pi)^2} \int_{4 \pi} dq_1 \int_0^\pi 2\pi sin(\psi) d\psi
\left[ \int_0^{2\pi} d\varphi_3 w_{AB}(\phi(\psi,\varphi_3,\theta_j))\right]
\left[\int_0^{2\pi} d\varphi_1 w_{AB}(\phi'(\psi,\varphi_1,\theta_i))\right]
\end{split}
\ee

When doing the average over realizations we have lost the dependence on the
position of the 4-points configuration and only the distances
between points remain important. We will get $4 \pi$ from the $dq_1$ integration.
Also, if preferred, due to the symmetry in $\varphi$,
$\int_0^{2\pi} d\varphi \rightarrow 2 \int_0^\pi d\varphi$.

Using spherical trigonometry, we can relate the angular distance $\phi$ for the
cross-correlation $w_X({\phi})$ with their related angles $\varphi$, $\psi$ and $\theta$.
The relation is given by the cosinus law in spherical trigonometry

%
\beq{cosinuslaw}
cos(\phi)=cos(\psi)cos(\theta)+sin(\psi)sin(\theta)cos(\varphi)
\ee

We arrive to the following equations:

\beq{resultB}
 C_{ij} ={1\over 8 \pi^2}
\int_0^{\pi}sin\psi d\psi[W_{AA}(\theta_j,\psi)W_{BB}(\theta_i,\psi)+
W_{AB}(\theta_j,\psi)W_{AB}(\theta_i,\psi)]
\ee

\beq{WWWB}
W_X(\theta,\psi)=\int_0^{2\pi}w_X(\phi) \big|_{cos\phi=cos\theta cos\psi+sin\theta sin\psi cos\varphi} d\varphi
\ee

where $X$ stands for any two field combination $AA$,$AB$,$BB$. When estimating
the covariance, the true value of $w_X$ has to be substituted by its estimated value.

By construction, the covariance is symmetric in its arguments, i.e,
$C_{ij}=C_{ji}$.
This symmetry still remains in equation (\ref{resultB}) but it is hidden.
It remains because we integrated over all four points configurations.
It is hidden because of the chosen coordinates for the integration.
When integrating, we priviledge some points over others. We separate
the integral by fixing two $\psi$-separeted points and integrating
over $\varphi$ angles. If the points chosen to be $\psi$-separated
were $B(2)$ and $A(3)$ instead of $A(1)$ and $B(4)$ we would have
ended by equation (\ref{resultB}) with $\theta_i\leftrightarrow\theta_j$.
Although the symmetry exists, we find convenient to put it more explicitly.
In equation (\ref{resultB}) we change the kernel to

\beq{kernelKB}
K[\theta_i,\theta_j,\psi]={1\over 2}[W_{AA}(\theta_i,\psi)W_{BB}(\theta_j,\psi)+W_{AA}(\theta_j,\psi)W_{BB}(\theta_i,\psi)]
+W_{AB}(\theta_i,\psi)W_{AB}(\theta_j,\psi)
\ee

\subsection{Partial sky survey}

\subsubsection{Probability considerations}
\label{sec:procon}

In most cases, our survey only have a restricted area of the sky to
estimate the cross-correlation signal. If we throw a point, the
probability to fall into the survey area is the fraction
of the sky covered by the survey area $f_{sky}$.
We define $P(\theta)$ as the probability that a randomly
thrown $\theta$-pair in the sky falls (both points) into the
survey area. This probability can also be understood as the ratio
between the number of $\theta$-pairs of the survey and the
total $\theta$-pairs in the sky.
The conditional probability $P(\theta / 1p) $ that both
points fall into the survey, once
we know that one is already inside, is given by

\beq{prob2cond}
P(\theta / 1p ) = \frac{P(\theta)}{f_{sky}}
\ee

We define $P(\psi,\theta,\phi)$ as the probability that the
triangle of sides $\psi,\theta,\phi$ falls (all) inside the survey area when thrown randomly in the sky .
The conditional probability that the third point of a triangle
falls into the survey, when we know that the other two points,
$\psi$-separeted, are already in, is

\beq{prob3cond}
P(\psi,\theta,\phi / \psi) = \frac{P(\psi,\theta,\phi)}{P(\psi)}
\ee



It is also useful to remember that $ S_N(\theta)= 4\pi 2\pi \sin{\theta} \Delta\theta P(\theta)$
is the normalization factor of the estimator.


Probabilities for $P(\theta)$ and $P(\psi,\theta,\phi)$ have to be
computed for each survey geometry. In Appendix B we compute
those probabilities for a polar cap survey, i.e., which
contains all points with distances less that $r$ to a given
point (the pole). A polar cap geometry is a very useful approximation
for most cosmological surveys, which are compact and extend
to a wide area. For those surveys, we can use the probabilites
$P(\theta)$ and $P(\psi,\theta,\phi)$ in Appendix B.

\subsubsection{The covariance integration}

We are in the case of limited area of the sky, where we have to integrate only the 4-point configuration allowed in this area to compute the covariance. We focus in the first term of the equation
(\ref{twoterms}) that we named $I_1$.
We replace the integration over the survey configuration
by the integration over all configurations convolved with a delta-selection
function $D$ which selects the configurations in the survey.

\begin{equation}
\begin{split}
I_1 & = \int_{4\pi} {dq_1 dq_2 dq_3 dq_4\over S_N(\theta_1) S_N(\theta_2)}
D(q_1,q_2,q_3,q_4)  <A(1)A(3)><B(2)B(4)> \\
& =  {1 \over  S_N(\theta_1) S_N(\theta_2)}
\int_{4 \pi} dq_1 D(q_1) \int_0^\pi d\psi  \int_0^{2\pi} ~d\alpha~  sin(\psi) D(q_4))\\
& \quad \quad \quad \quad
\left[ \int_0^{2\pi} d\varphi_4 \sin{\theta_2} \Delta\theta_2 w_{AA}(\phi(\psi,\varphi_4,\theta_2))
D(q_3)\right]
\left[\int_0^{2\pi} d\varphi_1 \sin{\theta_1} \Delta\theta_1 w_{BB}(\phi'(\psi,\varphi_1,\theta_1))
D(q_2)\right]
\end{split}
\end{equation}
where $\alpha$ is the angle between $\psi$-pair and the line from $q_1$ to the pole.
The other angles are as in figure (\ref{fig_esquema}). In the expression above we have
formally split $D(q_1,q_2,q_3,q_4)$ into four parts $D(q_1)D(q_2) D(q_3) D(q_4)$ which
in fact have the same meaning: they are unity only  when all the 4 points are inside the survey
and they are zero otherwise.

This is an exact result for the $I_1$ term. The key point here is to approximate
the integrals over $D$'s by replacing those selection functions $D$ by a
convenient probability. Here we are throwing all 4 points in an ordered way.
We throw the first point at $q_1$, the selection function $D(q_1)$ will select if this point falls into
the survey area. The substitution $D(q_1)\rightarrow f_{sky}$ applies here. Next point
to be thrown is $q_4$ which is $\psi$-related to $q_1$. We substitute
$D(q_4)$ by $P(\psi / 1p)$, i.e, the probability that once a point is inside the survey area,
a second point $\psi$-separated also falls into. The next two points $q_2$ $q_3$ are
not related between them but they are related to the two previous points already thrown.
$D(q_2)$ and $D(q_3)$ have to be substituted by the probability that,
given two points $\psi$-separated inside the survey,
a third point is also in the survey at distances $\theta$ and $\phi$ to those previous points.
$D(q_3)$ is substituted by $P(\psi,\theta_2,\phi / \psi)$, and $D(q_2)$ by
$P(\psi,\theta_1,\phi' / \psi)$. When expanding
the normalization factors and doing the integrals we get:

\begin{equation}
\begin{split}
I_1 & = {1 \over 4\pi 2\pi P(\theta_i) P(\theta_j)}
f_{sky}  \int_0^\pi d\psi  sin(\psi) P(\psi / 1p)\\
& \quad \quad \quad \quad
\left[\int_0^{2\pi} d\varphi_4 w_{AA}(\phi(\psi,\varphi_4,\theta_j)) P(\psi,\theta_j,\phi /\psi) \right]
\left[\int_0^{2\pi} d\varphi_1 w_{BB}(\phi'(\psi,\varphi_1,\theta_i))
P(\psi,\theta_i,\phi' / \psi)\right]
\end{split}
\end{equation}

Unfortunately, by replacing $D\rightarrow P$ we (slightly) break the symmetry
$\theta_i \leftrightarrow \theta_j$. Thus, for the covariance, we will use the
kernel in equation (\ref{kernelKB}) which will recover this symmetry.
Replacing the conditional probabilities calculated in \S\ref{sec:procon} with
the non-conditional ones we arrive to the final result

\beq{resultfinal}
C_{ij}={1\over 8 \pi^2 P(\theta_i) P(\theta_j)}
\int_0^{\pi} \frac{K[W]}{P(\psi)} sin\psi d\psi
\ee

\beq{kernel}
K[W]={1\over 2}[W_{AA}(\theta_i)W_{BB}(\theta_j)+W_{AA}(\theta_j)W_{BB}(\theta_i)]+W_{AB}(\theta_i)W_{AB}(\theta_j)
\ee

\beq{WW}
W_X(\theta)=\int_0^{2\pi} P(\psi,\theta,\phi) w_X(\phi) \big|_{cos\phi=cos\theta cos\psi+sin\theta sin\psi cos\varphi} d\varphi
\ee

\subsubsection{Small angle  approximation}

In the small angle approximation, $\theta_i$ and $\theta_j$ are small. We
can consider that when one point of the $\theta$-pair falls into
the survey, the other also does.
It corresponds to $P(\psi,\theta,\phi)\rightarrow P(\psi)$,
while $P(\theta) \rightarrow  f_{sky}$. We then have:
\beq{fsky}
C_{ij} = {{C_{ij}[all ~sky]}\over{f_{sky}}}
\ee
which gives the popular approximation of error scaling as $ \propto \frac{1}{\sqrt{f_{sky}}}$
which is used in the TH method in \S\ref{sec:TH} and Eq.\ref{eq:errorCl}. To the same level
of accuracy,  we can also chose to use the equation above together with Eq.\ref{resultB}
and avoid further calculations. As we want to go one step further we will give some
prescriptions below for more realistic situations. This will not only improve the accuracy of the
calculation but will also allow us to study when the small angle approximation is good enough
in a given  situation.

\section{Probabilities of finding pairs, triangles, and polynomials in a polar cap survey}

Let it be a given $\theta$-pair or an spherical triangle $\Delta(\theta,\psi,\phi)$
randomly thrown in a sphere. In this section we compute the probabilities
of finding them inside a polar cap survey of area A.

A polar cap of radius $r$ is the union of all points of an sphere
with (spherical) distances less than $r$ to a given point (the pole).
The Area of a polar cap is:

\beq{polarcaparea}
A=2\pi R^2 [1-\cos(r)]
\ee
where $R$ is the radius of the sphere. We set it equal to one
as usual in spherical trigonometry.

The probability of a N-points polygon to be thrown inside a polar cap
of radius $r$ is equal to the intersection area of N circles of radius
$r$, each one centered in one of the polygon (vertex) points, normalized to
(divided by) the total area of the sphere, i.e., $4\pi$ .

How is it so? The probability for a given poglygon to be thrown inside
a circle of radius $r$ (polar cap) already in the sphere is the same as
the probability of first drawing the polygon in the sphere and then
throwing the circle and finding it encompassing all N-points (vertex).
For this to happen, the center of the circle must be at a distance less
than $r$ for any of the polygon points. Only those points in the
area intersected by N circles of radius $r$, one from each vertex, hold this
condition.
Then, the probability of finding the polygon inside a polar cap survey of
radius $r$ is that area divided by the total area of the sphere.

\subsection{Probability for a $\theta$-pair: $ P(\theta)$}

As we have seen, the probability for a $\theta$-pair thrown randomly
into a sphere to fall inside a circle of radius r (polar cap survey) is:

\beq{prob2}
P(\theta)=\frac{ { \mbox{intersection area of two cirecles radius $r$ } \choose
\mbox{ separated a distance of $\theta$}} }
{4\pi}
\ee

\begin{figure}
\begin{center}
\includegraphics[width=80mm]{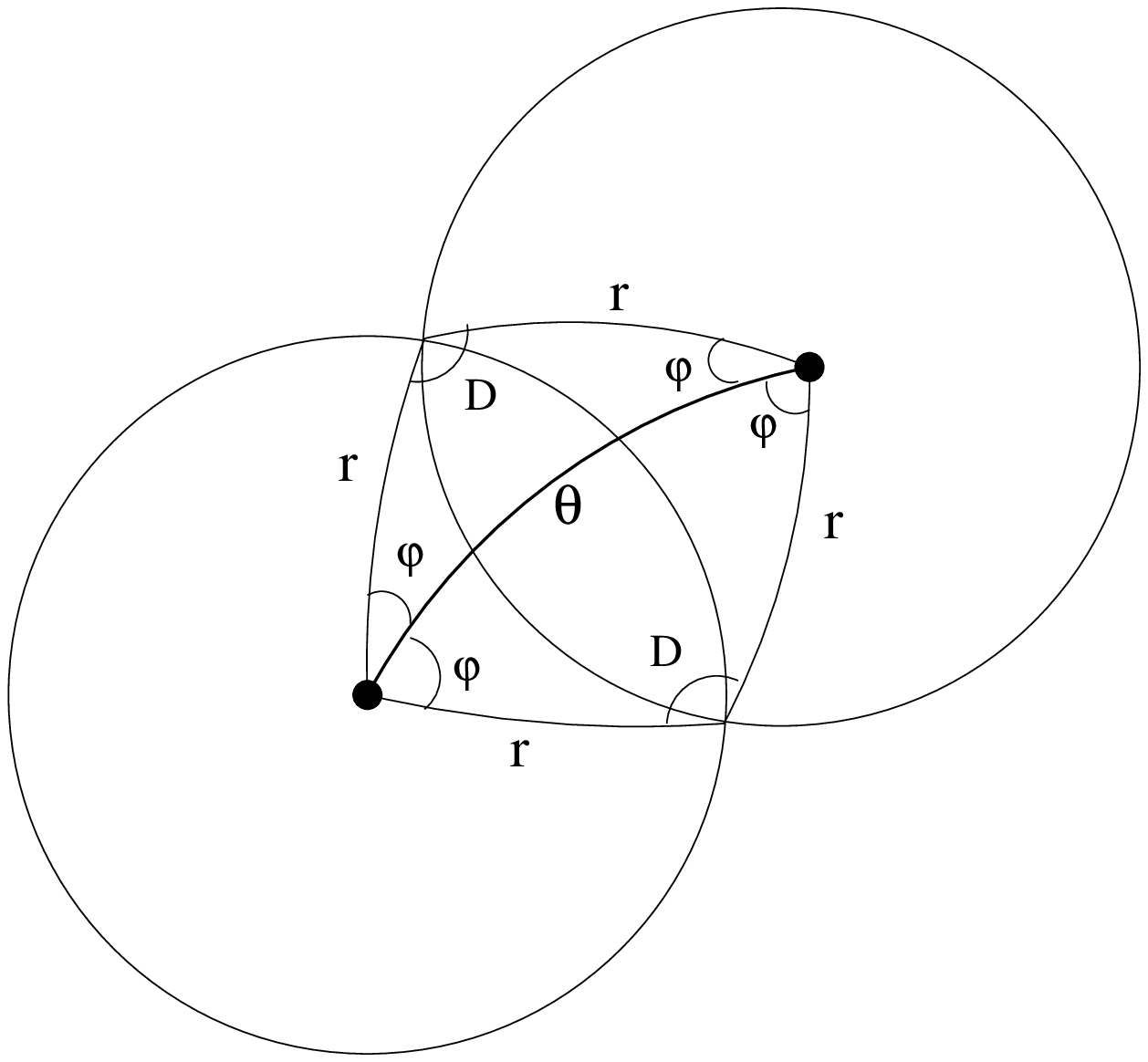}
\caption{Geometry for the intersection of two spherical circles. It is
needed for $P(\theta)$ determination.}
\end{center}
\label{fig:fprob2p}
\end{figure}

\begin{figure}
\begin{center}
\includegraphics[width=90mm]{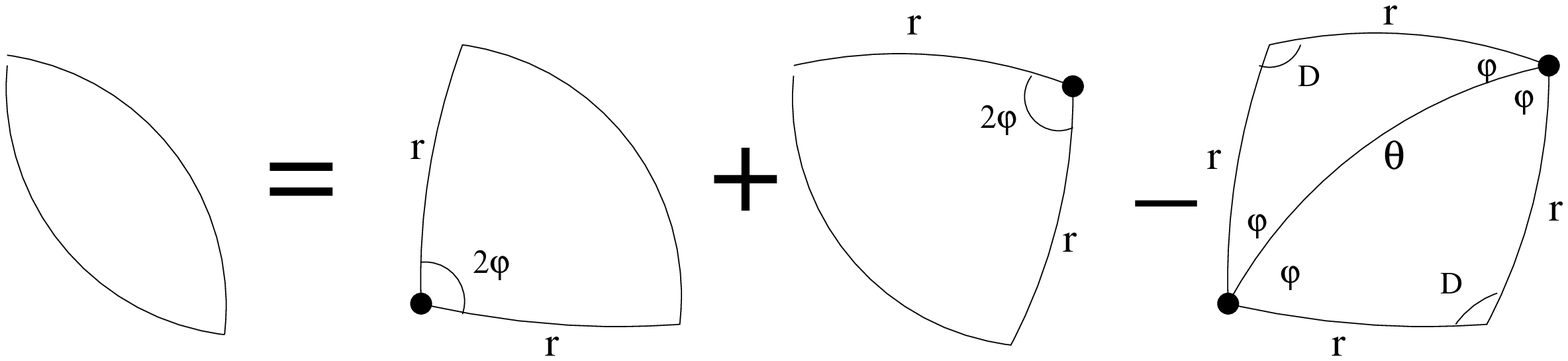}
\caption{Representation of how to obtain the intersection area of two spherical
circles from sectors of spherical circles and spherical triangles}
\end{center}
\label{fig:twopsuma}
\end{figure}

In order to compute the area of the intersection $A$, we make use of the figures \ref{fig:fprob2p} and \ref{fig:twopsuma}
as well as spherical trigonometry formulae. When $\theta > 2r$ there is no intersection and the probability
$P(\theta)$ is zero. When $\theta < 2r$ and $ 2r+\theta < 2\pi$
we contruct two symmetrical triangles $\Delta(\theta,r,r)$, as shown in figure \ref{fig:fprob2p}.
The area of the intersection is given by the sum of two sectors of spherical circle minus
the area of those triangles. In spherical trigonometry, the area of a triangle is given
by the sum of the angles between its sides minus $\pi$. Thus,

\begin{equation}
A  = [ 2 \varphi (1-\cos(r)) ] + [2 \varphi (1-\cos(r))] - [ 2 ( \varphi + \varphi + D - \pi ) ]
   = 2\pi - 2 D - 4 \varphi \cos(r)
\end{equation}

where $D$ and $\varphi$ are given by the cosinus law and semiperimeter
half angle formulaes


\begin{eqnarray}
\cos(D) &= &\frac{\cos(\theta)-\cos^2(r)}{\sin^2(r)} \\
\tan(\frac{\varphi}{2}) & = &\sqrt{\frac{\sin(s-r) \sin(s-d)}{\sin(s)\sin(s-r)}}
=\sqrt{\frac{\sin(r-\theta/2)}{\sin(r+\theta/2)}} \quad , \quad\quad  s=\frac{\theta+r+r}{2}
\end{eqnarray}

When $\theta < 2r $ but  $2r+\theta < 2\pi$ (and therefore $r> \pi/2$)
the two r-circumferences do not intersect each other although
the two circles area still overlap. The area of the intersection is all the sphere
exept the area of the two complementary circles, i.e.,

\begin{equation}
A =  4\pi-[4\pi- 2\pi(1-cos(r))]-[4\pi-2\pi(1-cos(r))]=-4\pi cos(r)
\end{equation}

\subsection{Probability for a triangle $\Delta(\psi,\theta,\phi)$ : $P(\psi,\theta,\phi)$}

As we have seen, the probability for a triangle $\Delta(\psi,\theta,\phi)$  thrown randomly
in a sphere to fall inside a circle of radius r (polar cap survey) is:

\beq{prob3}
P(\psi,\theta,\phi)=\frac{ { \mbox{intersection area of three circles radius $r$}
\choose \mbox{centered at the vertex}}}
{4\pi}
\ee

\begin{figure}
\begin{center}
\begin{tabular}{lr}
\includegraphics[height=80mm]{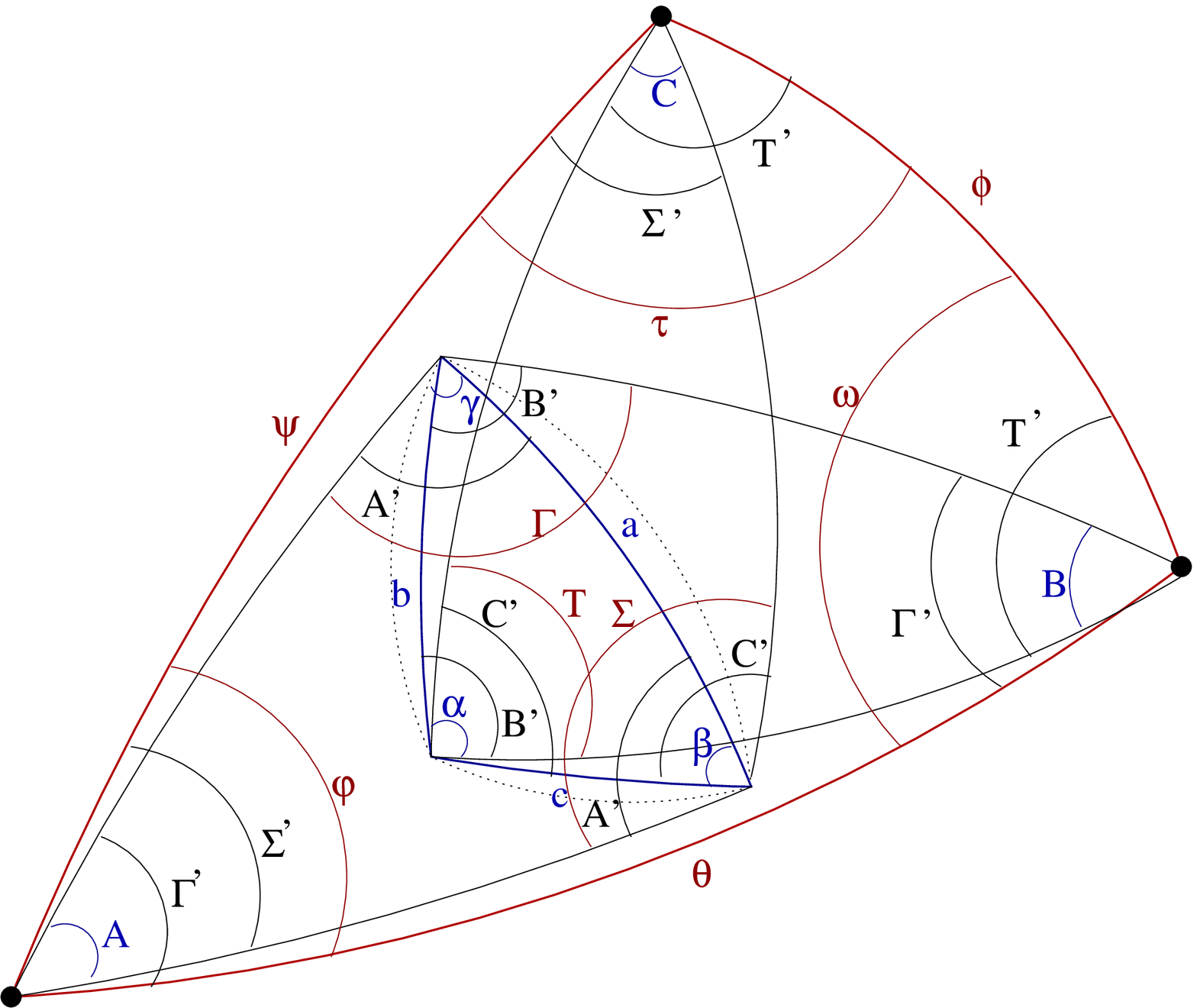} & \includegraphics[width=50mm]{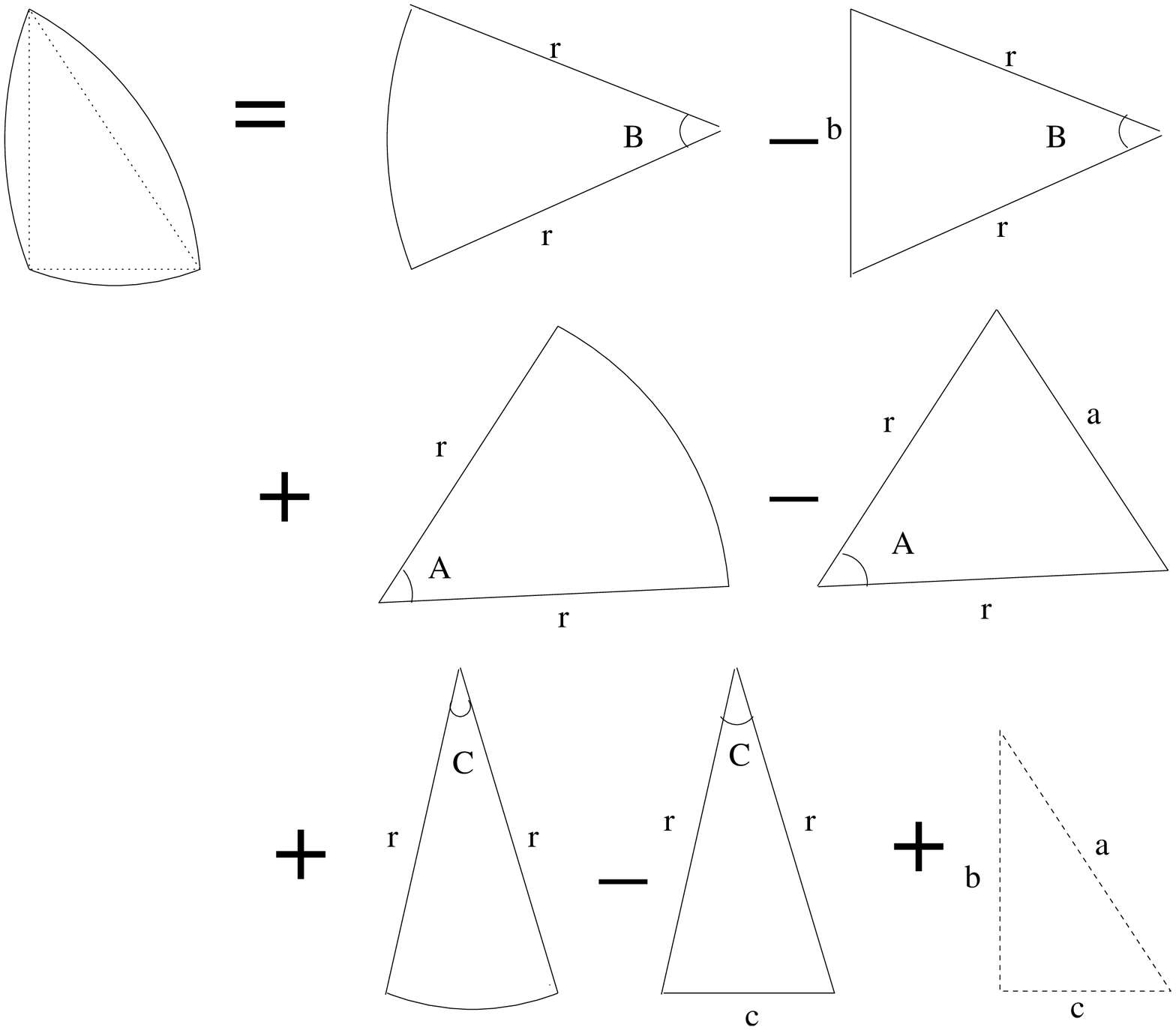} \\
\end{tabular}
\caption{Left: Geometry for the intersection area of three circles of radius r
centered at the three black points. Angles shown in the figure
are used in equation \ref{area3p}  to compute the intersection area, which
is directly related to $P(\psi,\theta,\phi)$. Right: Representation of how
to obtain the intersection area from sectors of spherical circles and spherical triangles.}
\end{center}
\label{fig:fprob3p}
\end{figure}


When the intersection does not exist $P(\psi,\theta,\phi)=0$. This is the case if $r < \rho_o$,
where $\rho_o$ is the radius of the spherical circumference that circumscribes the triangle
$\Delta(\psi,\theta,\phi)$. It can be shown \footnote{$\rho_o$ can be deduced by relating
the spherical and the Euclidean triangle with the same points for the vertex. The
radius of a circumscribed circumference is well known for an Euclidean case.} that,

\begin{equation}
sin(\rho_o)=2 \frac{\sin(\psi/2) \sin(\theta/2) \sin(\phi/2)}
   {\sqrt{(S(S-\sin(\psi/2)(S-\sin(\theta/2)(S-\sin(\phi/2)}}
\end{equation}

where $S=\sin(\psi/2)+ \sin(\theta/2)+ \sin(\phi/2)$
When $\rho_o < r$ and  $\rho_o + r < \pi$, the intersection area, $A$, exist, and
to compute it
we can make use of the figures in \ref{fig:fprob3p}
as well as spherical trigonometry formulae.
The intersection area $A$ is delimited by three arcs of a circle of radius $r$.
Three points mark the intersection of those arcs in the limiting region. One can
contruct an spherical triangle $\Delta(a,b,c)$ having those points as vertex. This
is the blue triangle in figure \ref{fig:fprob3p}, which shows all necessary angles for this section.
Figure 
\ref{fig:fprob3p} also
shows how one can get the area $A$ from the triangles an sectors
of circles. Following this figure the area is

\begin{eqnarray}
\label{area3p}
A & = &[ B ( 1- \cos(r))] - [ B'+B'+B-\pi] + [ A ( 1- \cos(r))] - [ A'+A'+A-\pi]   \\ \nonumber
  & + & [ C ( 1- \cos(r))] - [ C'+C'+C-\pi]  + [ \alpha+\beta+\gamma-\pi]           \\ \nonumber
  & = & 2 \pi - \cos(r) ( A+B+C)- \Sigma-\Gamma-T
    = 2 \pi - \cos(r) (2\Sigma'-2\Gamma'-2T'- \varphi-\omega-\tau) -\Sigma-\Gamma-T \\ \nonumber
\end{eqnarray}

where we have applied that angles $\alpha,\beta,\gamma,A,B,C$ can be expressed
as a sum of other angles. The angles left can be obtained by the spherical cosinus law.
We write only three angles here, but the others can be computed in a similar way.

\bea
\cos(\varphi) & = &\frac{\cos(\phi)-\cos(\psi)\cos(\theta)}{\sin(\psi)\sin(\theta)} \\ \nonumber
\cos(\Sigma)  & = &\frac{\cos(\psi)-\cos^2(r)}{\sin^2(r)}             \\ \nonumber
\cos(\Sigma') & = &\frac{\cos(r)-\cos(\psi)\cos(r)}{\sin(\psi)\sin(r)} \\ \nonumber
\eea


When $\rho_o < r$ but $\rho_o+r>\pi$ the intersection of the three circles exist but we
can not contruct such a triangle as in \ref{fig:fprob3p}.
The area is given by all sky exept the sum of
the intersection areas of the two points complementary circles, i.e.,

\begin{equation}
A=4\pi-4\pi(1-P(\psi))-4\pi(1-P(\theta))-4\pi(1-P(\phi))=4\pi[P(\psi)+P(\theta)+P(\phi)-2]
\end{equation}


\end{document}